\documentclass{article}

\usepackage[utf8]{inputenc}
\usepackage[fleqn]{amsmath}
\usepackage{amsfonts, amssymb}

\usepackage[pdftex,backref,colorlinks]{hyperref}
  \hypersetup{
     colorlinks,
     linkcolor={red!50!blue},
     citecolor={blue!50!green},
     urlcolor={blue!80!black}
 }

\usepackage[ntheorem]{empheq}
\usepackage[hyperref]{ntheorem}

\newtheorem{theorem}{Theorem}

\newtheorem{definition}{Definition}

\usepackage{extarrows}                        \usepackage{units} \usepackage{bm} \usepackage{bbold}                    \usepackage{tensor}                     \usepackage{soul} \usepackage{cancel} \usepackage[retainorgcmds]{IEEEtrantools} \IEEEeqnarraydefcolsep{0}{\leftmargini} 
\usepackage{graphicx}
\usepackage{fixltx2e} 
\usepackage{array} 
\newcolumntype{C}[1]{>{\centering\let\newline\\\arraybackslash\hspace{0pt}}m{#1}}

\usepackage[numbers,sort&compress]{natbib} 
\usepackage{color}
\usepackage{realboxes}
\usepackage{framed}
\usepackage{array} \usepackage{mathtools} \usepackage[version=3]{mhchem}

\usepackage{enumerate} 
\usepackage{bookmark}                       \usepackage{float}                    \usepackage{rotating}

\usepackage{ytableau} 
\usepackage{youngtab}
\usepackage{adjustbox}

\ytableausetup{mathmode, centertableaux}
\usepackage{xargs}
\usepackage{letltxmacro}                          
\let\oldytableau\ytableau
\let\endoldytableau\endytableau
\renewenvironment{ytableau}{\begin{adjustbox}{scale=.78}\oldytableau}{\endoldytableau\end{adjustbox}}

\usepackage{xcolor}
\usepackage{tikz}
\usepackage{tkz-euclide}
\usetikzlibrary{matrix,fit,arrows,decorations}
\usepackage{pbox}

\newcommand{\qed}{\hfill\tikz{\draw[draw=black,line width=0.6pt] (0,0) rectangle (2.8mm,2.8mm);}\bigskip}

\usepackage[a4paper,left=2.0cm]{geometry}
\newcommand{\ud}{\mathrm{d}}
\newcommand{\bra}[1]{\langle #1 \vert}
\newcommand{\ket}[1]{\vert #1 \rangle}
\newcommand{\Tr}[1]{\text{tr}\left(#1\right)}

\newcommand{\FPic}[2][{}]{\hspace{-0.25mm}\pbox{\textwidth}{\includegraphics[#1]{{#2}}}\hspace{-0.25mm}}

\newcommand{\ybox}[1]{\begin{ytableau} #1 \end{ytableau}}

\newcommand{\SUN}{\mathsf{SU}(N)}
\newcommand{\MixedPow}[2]{V^{\otimes
    #1}\otimes\left(V^*\right)^{\otimes #2}}
\newcommand{\Pow}[1]{V^{\otimes #1}}

\newcommand{\Lin}[1]{\mathrm{Lin}\left( #1 \right)}
\newcommand{\API}[1]{\mathsf{API}\left( #1 \right)}

\def\smath#1{\text{\scalebox{.8}{$#1$}}}
\def\sfrac#1#2{\smath{\frac{#1}{#2}}}

\makeatletter
\newcommand{\vast}{\bBigg@{3}}
\newcommand{\Vast}{\bBigg@{4}}
\makeatother

\tikzstyle basiclabel=[draw=none,fill=none,shape=rectangle,inner sep=2pt,scale=.8]
\tikzstyle leftlabel=[basiclabel,anchor=east]
\tikzstyle rightlabel=[basiclabel,anchor=west]

\begingroup\makeatletter\ifx\SetFigFont\undefined\gdef\SetFigFont#1#2#3#4#5{  \reset@font\fontsize{#1}{#2pt}  \fontfamily{#3}\fontseries{#4}\fontshape{#5}  \selectfont}\fi\endgroup
\newcommand{\blockmatrix}[9]{
  \draw[draw=#4,fill=#5,every node/.style={inner sep=0,outer sep=0}] (0,0) rectangle( #1,#2);
  \ifthenelse{\equal{#6}{true}}
  {
    \draw[draw=#7,fill=#8] (0,#2) -- (#9,#2) -- ( #1,#9) -- ( #1,0) -- ( #1 - #9,0) -- (0,#2 -#9) -- cycle;
  }
  {}
  \draw ( #1/2, #2/2) node[inner sep=0,outer sep=0]{ #3};
}

\newcommand{\mblockmatrix}[4][none]{
  \begin{tikzpicture} 
  \ifthenelse{\equal{#1}{none}}
  {
    \blockmatrix{#2}{#3}{#4}{none}{none}{false}{none}{none}{0.0}
  }
  {
    \definecolor{fillcolor}{rgb}{#1}
    \blockmatrix{#2}{#3}{#4}{none}{fillcolor}{false}{none}{none}{0.0}
  }
  \end{tikzpicture}}

\newcommand{\fblockmatrix}[4][none]{  \begin{tikzpicture}[outer sep=0,inner sep=0] 
  \ifthenelse{\equal{#1}{none}}{\blockmatrix{#2}{#3}{#4}{black}{none}{false}{none}{none}{0.0}}{\definecolor{fillcolor}{rgb}{#1}\blockmatrix{#2}{#3}{#4}{black}{fillcolor}{false}{none}{none}{0.0}}\end{tikzpicture}}

\newcommand{\dblockmatrix}[4][none]{
  \begin{tikzpicture} 
  \ifthenelse{\equal{#1}{none}}
  {\blockmatrix{#2}{#3}{#4}{black}{none}{true}{black}{none}{0.35cm}}{\definecolor{fillcolor}{rgb}{#1}\blockmatrix{#2}{#3}{#4}{black}{none}{true}{black}{fillcolor}{0.35cm}}\end{tikzpicture}}

\newcommand{\diagonalblockmatrix}[5][none]{
  \begin{tikzpicture} 

  \ifthenelse{\equal{#1}{none}}
  {
    \blockmatrix{#2}{#3}{#4}{black}{none}{true}{black}{none}{#5}
  }
  {
    \definecolor{fillcolor}{rgb}{#1}
    \blockmatrix{#2}{#3}{#4}{black}{none}{true}{black}{fillcolor}{#5}
  }

  \end{tikzpicture}}

\renewenvironment{abstract}{\centering\begin{minipage}{.95\textwidth}
\sffamily{\bf Abstract:}}
{\end{minipage}\vskip 3em}
\makeatletter
\renewcommand\@maketitle{\hfill
\begin{minipage}{\textwidth}
\vskip 2em
\let\footnote\thanks 
{\LARGE \bf \@title \par }
\vskip 1.5em
{\large \@author \par}
\end{minipage}
\vskip 3em \par
}
\makeatother
\usepackage{authblk}

\title{Transition Operators}
\author[1]{J. Alcock-Zeilinger}
\author[1]{H. Weigert}
\affil[1]{\small University of Cape Town; Dept. of Physics, Private Bag X3, Rondebosch 7701, South Africa}
\date{October 2016}

\begin{document}
\maketitle

\begin{abstract}
  \noindent In this paper, we give a generic algorithm of the
  transition operators between Hermitian Young projection operators
  corresponding to equivalent irreducible representations of $\SUN$,
  using the compact expressions of Hermitian Young projection
  operators derived
  in~\cite{Alcock-Zeilinger:2016sxc}. We show that the
  Hermitian Young projection operators together with their transition
  operators constitute a \emph{fully orthogonal} basis for the algebra
  of invariants of $\Pow{m}$ that exhibits a systematically simplified
  multiplication table. We discuss the full algebra of invariants over
  $\Pow{3}$ and $\Pow{4}$ as explicit examples. In our presentation we
  make use of various standard concepts such as Young projection
  operators, Clebsch-Gordan operators, and invariants (in birdtrack
  notation). We tie these perspectives together and use them to shed
  light on each other.
\end{abstract}

\setlength{\parskip}{2pt plus 2pt minus 1pt}
\tableofcontents
\setlength{\parskip}{7pt plus 2pt minus 1pt}

\section{Introduction}\label{sec:Introduction}

Applications of representation theory generally are concerned with
irreducible representations of the group under scrutiny. Physics
applications in particular are generally aimed at finding all
irreducible representations in an $m$-particle Fock space. The
textbook example here is of course angular momentum and spin with the
group $\mathsf{SU}(2)$ and the construction of the periodic table in
quantum mechanics via the irreducible multiplets for $m$-electrons in
an atom with $m$ protons that classify its orbital configuration, its
spectral and chemical properties. In quantum chromodynamics we meet
flavor symmetry (flavor $\mathsf{SU}_f(2)$ or $\mathsf{SU}_f(3)$) as
an approximate symmetry that guide us through interpreting the mesons
and baryons as members of irreducible representation of the flavor
group in the \emph{eight-fold
  way}~\cite{GellMann:1961ky,GellMann:1964xy}. Gauge invariance and
confinement force the same particles into singlets of the color gauge
group $\mathsf{SU}_c(3)$. The latter are of particular interest in the
color glass condensate which dominates QCD in high energy
applications, i.e. in modern collider experiments. In this set of
applications Wilson line correlators appear whose color structures are
of central importance and the presently existing techniques are
limited to explicit calculations at a given order of
complexity. In~\cite{Alcock-Zeilinger:2016sxc}, we have established an
efficient algorithm to construct a full set of Hermitian projection
operators for the decoposition of a product of $m$ fundamental
representations of $\SUN$ as a subset of the associated algebra of
invariants. Here we aim to find a complete basis for the algebra of
invariants that is fully shaped by irreducible representations these
operators represent and identify the missing basis elements as
transition operators between \emph{equivalent} representation
contained in the product. In a future
paper~\cite{AlcockZeilinger2016Singlets}, this information will be
used to give a full account of all singlets (i.e. all one dimensional
representations that remain invariant under the group action)
constrained in a product of $m$ fundamental representations with $m'$
anti-fundamental representations. In physics parlance, this gives
access to the gauge invariant part of the Fock space of $m$ particles
and $m'$ anti-particles.

There are, of course, several different technologies on the market to
address these questions, the most familiar to the practising
physicist being the construction of Clebsch-Gordan
coefficients~\cite{Clebsch1866theorie}
(see~\cite{Tung:1985na,Fulton:2004,Peskin:1995ev} for textbook
introductions), Eli{\'e} Cartan's method of roots and
weights~\cite{Cartan1894structure} and Alfred Young's combinatorial
method of classifying the idempotents on the algebra of
permutations~\cite{Young:1928}. The \emph{Schur-Weyl
  duality},~\cite{Weyl:1946} relates these idempotents to the
irreducible representations of compact, semi-simple Lie groups and
thus to $\SUN$. This duality is based on the \emph{theory of
  invariants},~\cite{Cvitanovic:2008zz,Weyl:1946}, which exploits the
invariants (in particular the \emph{primitive invariants}) of a Lie
group $G$ and constructs projection operators corresponding to the
irreducible representations of $G$ from the invariants of that
group. It is this method that allows us to carry $N$ as a parameter
throughout, which has important advantages in applications in QCD we
are ultimately interested in. The core part of our discussion will
deal with a product representations of $\SUN$ constructed from the
fundamental representation of a Lie group $G$ on a given vector space
$V$ with $\text{dim}(V) = N$, whose action will simply be denoted by
$v\mapsto U v$ for all $U\in\SUN$ and $v\in V$. Choosing a basis
$\{e_{(i)} |i = 1,\ldots,\text{dim}(V)\}$ such that $v = v^i e_{(i)}$ this
becomes $v^i \mapsto \tensor{U}{^i_j} v^j$. This immediately induces 
product representations of $\SUN$, representations on $\Pow{m}$, if
one uses this basis of $V$ to induce a basis on $\Pow{m}$ so that a
general element $\bm v\in\Pow{m}$ takes the form $\bm v = v^{i_1\ldots
  i_m} e_{(i_1)}\otimes\cdots\otimes e_{(i_m)}$:
\begin{align}
  \label{eq:SUN-Action}
  U\circ\bm v 
  = 
  U\circ v^{i_1\ldots i_m} e_{(i_1)}\otimes\cdots\otimes e_{(i_m)}
  :=
  \tensor{U}{^{i_1}_{j_1}} 
  \cdots 
  \tensor{U}{^{i_m}_{j_m}} 
  v^{j_1\ldots j_m} e_{(i_1)}\otimes\cdots\otimes e_{(i_m)}
\end{align} 
Since all the factors in $\Pow{m}$ are identical, the notion of
permuting the factors is a natural one and leads to a linear map on
$\Pow{m}$ according to
\begin{align}
  \label{eq:perm-facs-def}
  \rho\circ\bm v 
  = 
  \rho\circ v^{i_1\ldots i_m} e_{(i_1)}\otimes\cdots\otimes e_{(i_m)}
  :=
  v^{\rho(i_1)\ldots \rho(i_m)} e_{(i_1)}\otimes\cdots\otimes e_{(i_m)}
\end{align}
where $\rho$ is an element of $S_m$, the group of permutations of $m$
objects.\footnote{Permuting the basis vectors instead involves
  $\rho^{-1}$: $v^{\rho(i_1)\ldots \rho(i_m)} e_{(i_1)} \otimes \cdots
  \otimes e_{(i_m)} = v^{i_1\ldots i_m} e_{(\rho^{-1}(i_1))} \otimes
  \cdots \otimes e_{(\rho^{-1}(i_m))}$}{} From the
definitions~\eqref{eq:SUN-Action} and~\eqref{eq:perm-facs-def} one
immediately infers that the product representation commutes with all
permutations on any $\bm v\in\Pow{m}$:
\begin{equation}
\label{eq:InvariantsIntro1}
  U \circ \rho \circ \bm v  = \rho \circ U\circ  \bm v
  \ .
\end{equation}
In other words, any such permutation $\rho$ is an \emph{invariant} of
$\SUN$ (or in fact any Lie group $G$ acting on $V$):
\begin{align}
  \label{eq:per-invariant}
  U\circ\rho\circ U^{-1} = \rho
  \ .
\end{align}
It can further be shown that these permutations in fact span the space
of all linear invariants of $\SUN$ over
$\Pow{m}$~\cite{Cvitanovic:2008zz}. The permutations are thus referred
to as the \emph{primitive invariants} of $\SUN$ over $\Pow{m}$. The
full space of linear invariants is then given by
\begin{align}
  \label{eq:PI-def}
  \API{\SUN,\Pow{m}} 
  := 
  \Bigl\{ 
  \sum_{\sigma\in S_m} \alpha_\sigma \sigma \Big| 
  \alpha_\sigma\in \mathbb R, \sigma\in S_m \Bigr\}
  \subset
  \Lin{\Pow{m}}
  \ .
\end{align}

As defined in~\eqref{eq:PI-def}, $\API{\SUN,\Pow{m}}$ is a real vector
space and becomes a real algebra with the product induced by the
product of permutations. It will become obvious from our presentation
that this space is large enough to encompass all group-theoretically
interesting objects, namely
\begin{enumerate}
\item Hermitian projectors onto irreducible representations (see~\cite{Keppeler:2013yla,Alcock-Zeilinger:2016sxc}), and
\item any transition operators associated with equivalent
  representations.
\end{enumerate}
We will show that these operators do not only fit into it, they in
fact span the whole space and form an orthogonal basis for
$\API{\SUN,\Pow{m}}$. We will call this the projector basis for
$\API{\SUN,\Pow{m}}$ in the remainder of this paper and discuss in
detail its unique structures and the freedom of choice still left
open.

Naively, since the number of permutations in $S_m$ is $m!$, one might
expect the dimension of $\API{\SUN,\Pow{m}}$ to be simply $m!$ and
indeed this is the maximal dimension of the algebra. However, this is
realized only if $N = \text{dim}(V) \ge m$. Failing that, not all
permutations remain linerarly indepdendent (as elements of the vector
space $\Lin{\Pow{m}}$). In the projector basis we construct this is
particularly clearly exhibited: a number of clearly distinguished
basis elements will explicitly become null operators, all others will
remain non-zero and thus form a basis for the now smaller space of
invariants (see appendix~\ref{sec:VanishingReps}). It is this feature
that allows us to use $N$ as a parameter in calculations.

We begin with a short presentation of some background material needed
to fully appreciate the arguments made in this paper,
sec.~\ref{sec:Background}. This section begins by introducing the
birdtrack formalism~\cite{Cvitanovic:2008zz}, which is particularly
suited for dealing with the objects discussed in this paper. We then
proceed by summarizing classic textbook material on Young tableaux and
their corresponding projection operators, see for
example~\cite{Cvitanovic:2008zz,Fulton:1997,Fulton:2004,Tung:1985na,Sagan:2000}. Lastly,
we state some cancellation rules for birdtrack
operators~\cite{Alcock-Zeilinger:2016bss}.

Section~\ref{sec:YoungBasis} provides the first new results: we show
that the Young projection operators can be augmented by what we choose
to call transition operators to give an alternative basis for the
algebra of invariants $\API{\SUN,\Pow{m}}$ for $m\leq4$, and proceed to give the full basis for
$\API{\SUN,\Pow{3}}$ (a diagram depicting the full basis up to $m=4$
is given in Figure~\ref{fig:YT-hierarchy-4}). Since orthogonality of Young projection
operators breaks down beyond $m=4$, the Young basis cannot be generalized to
larger $m$. This motivates a basis in terms of Hermitian 
projection operators and their unitary transition operators:

Section~\ref{sec:Orth-Proj-Basis} discusses such a basis through
Clebsch-Gordan operators for all $m$. As it turns out, Hermiticity and
unitarity of these operators automatically guarantee mutual
orthogonality of the basis elements with respect to the inner product
$\langle A, B \rangle := \text{tr} \left( A^{\dagger} B
\right)$.
Since this method requires the construction of $N^m$ normalized states
to find $m!$ basis elements for $\API{\SUN,\Pow{m}}$ we then proceed
to present an more efficient algorithm to reach this goal. Our method
is based on streamlined methods to construct Hermitian Young
projection operators~\cite{Alcock-Zeilinger:2016sxc} (themselves based
on earlier work by Keppeler and
Sj{\"o}dahl~\cite{Keppeler:2013yla}). These Hermitian Young projection
operators are complemented with unitary transition operators to
provide a full basis for $\API{\SUN,\Pow{m}}$ for all $m$ in
section~\ref{sec:HermitianYoungProjectorsSection},
Theorem~\ref{thm:TransitionElement}. This construction algorithm for
transition operators serves as a starting point for a more efficient
method, Theorem~\ref{thm:TransitionCompact}, yielding much shorter
expressions of the transition operators. The proof of
Theorem~\ref{thm:TransitionCompact} can be found in
appendix~\ref{sec:ProofsCompactTransition}.

We close with some examples: We give the basis of $\API{\SUN,\Pow{m}}$
in terms of Hermitian Young projection operators and unitary
transition operators for both $m=3$ and $m=4$ in
section~\ref{sec:Herm-Algebra-Example}. Figures~\ref{fig:YT-hierarchy-4}
and~\ref{fig:hierarchy-4} summarize the most important aspects of
Young and Hermitian Young decompositions of $\API{\SUN,\Pow{m}}$ for
all $m \le 4$.

\section{Background: Birdtracks, Young tableaux, notations and
  conventions}\label{sec:Background}

\subsection{Birdtracks, scalar products, and Hermiticity}

In the 1970's Penrose devised a graphical method of dealing with
primitive invariants of Lie groups including Young projection
operators,~\cite{Penrose1971Mom,Penrose1971Com}, which was
subsequently applied in a collaboration with
MacCallum,~\cite{Penrose:1972ia}. This graphical method, now dubbed
the \emph{birdtrack formalism}, was modernized and further developed
by Cvitanovi{\'c},~\cite{Cvitanovic:2008zz}, in recent years. The
immense benefit of the birdtrack formalism is that it makes the
actions of the operators visually accessible and thus more intuitive.
For illustration, we give as an example the permutations of $S_3$
written both in their cycle notation (see~\cite{Tung:1985na} for a
textbook introduction) as well as birdtracks:
\begin{equation}
\label{eq:S3-Birdtracks}
S_3 = \Bigl\{
  \underbrace{\FPic{3ArrLeft}\FPic{3IdSN}\FPic{3ArrRight}}_{\mathrm{id}}
  \; , \quad 
  \underbrace{\FPic{3ArrLeft}\FPic{3s12SN}\FPic{3ArrRight}}_{(12)}
  \; , \quad 
  \underbrace{\FPic{3ArrLeft}\FPic{3s13SN}\FPic{3ArrRight}}_{(13)}
  \; , \quad 
  \underbrace{\FPic{3ArrLeft}\FPic{3s23SN}\FPic{3ArrRight}}_{(23)}
  \; , \quad 
  \underbrace{\FPic{3ArrLeft}\FPic{3s123SN}\FPic{3ArrRight}}_{(123)}
  \; , \quad
  \underbrace{\FPic{3ArrLeft}\FPic{3s132SN}\FPic{3ArrRight}}_{(132)}
  \Bigr\}\; .
\end{equation}
The action of each of the above permutations on a tensor product
$v_1\otimes v_2\otimes v_3$ is clear, for example
\begin{equation}
  (123) \left(v_1\otimes v_2\otimes v_3\right) = v_3\otimes v_1\otimes v_2.
\end{equation}
In the birdtrack formalism, this equation is written as
\begin{equation}
  \FPic{3ArrLeft}\FPic{3s123SN}\FPic{3ArrRight}\FPic{3v123Labels} \; = \; \FPic{3v312Labels} \; ,
\end{equation}
where each term in the product $v_1\otimes v_2\otimes v_3$ (written as
a tower $\FPic{3v123Labels}$) can be thought of as being moved along
the lines of
$\FPic{3ArrLeft}\FPic{3s123SN}\FPic{3ArrRight}$. Birdtracks are thus
naturally read from right to left as is also indicated by the arrows
on the legs.

The graphical structure faithfully represents the multiplication table
of $S_m$ by impolementing a ``glue and follow the lines
prescription'' along the lines of
\begin{equation}
  \label{eq:BT-product}
  \FPic{3ArrLeft}\FPic{3s12SN}\FPic{3ArrRight}
  \cdot
  \FPic{3ArrLeft}\FPic{3s13SN}\FPic{3ArrRight}
  :=
  \FPic{3ArrLeft}\FPic{3s12SN}\hspace{-0.2mm}\FPic{3s13SN}\FPic{3ArrRight}
  =
  \FPic{3ArrLeft}\FPic{3s132SN}\FPic{3ArrRight}
  \ .
\end{equation}
Selecting a set of integers $\{a_1,\ldots,a_n\}$ we can introduce two
prominent types of elements of these algebras: symmetrizers
\begin{equation}
  \label{eq:symmetrizer-def}
  \bm S_{a_1,\ldots,a_n} := \frac1{n!} \sum\limits_{\sigma\in S_n} \sigma_{a_1,\ldots,a_n}
\end{equation}
where $\sigma_{a_1,\ldots,a_n}$ denotes a permutation in $S_n$ over
(any subset of) the letters ${a_1,\ldots,a_n}$, and antisymmetrizers
\begin{equation}
  \label{eq:anti-symmetrizer-def}
  \bm A_{a_1,\ldots,a_n} := \frac1{n!} \sum\limits_{\sigma\in S_n}
  \text{sign}(\sigma) \sigma_{a_1,\ldots,a_n}
\ .
\end{equation}
These may act on subsets of $n$ factors in $\Pow{m}$. Both
symmetrizers and antisymmetrizers are by definition idempotent,
\begin{equation}
  \label{eq:A2AS2S}
  \bm S_{a_1,\ldots,a_n}^2 = \bm S_{a_1,\ldots,a_n}
  \hspace{1cm}\text{ and }\hspace{1cm}
  \bm A_{a_1,\ldots,a_n}^2 = \bm A_{a_1,\ldots,a_n}
  \ ,
\end{equation}
they are projection operators.

All of these
have birdtrack representations in which symmetrizers (resp. antisymmetrizers) are shown as unfilled (filled) boxes covering the lines to be symmetrized (resp. antisymmetrized). Take for example $\bm S_{134}\in
\API{\Pow{5}}$ and $\bm A_{35}\in \API{\Pow{5}}$, which take the form
\begin{equation}
  \label{eq:SBTexample}
  \bm S_{134} = \FPic{5ArrLeft}\FPic{5s234N}\FPic{5Sym123N}\FPic{5s243N}\FPic{5ArrRight} \in \API{\SUN,\Pow{5}}
  \hspace{1cm}\text{ and }\hspace{1cm}
  \bm A_{35} = \FPic{5ArrLeft}\FPic{5s45N}\FPic{5ASym34N}\FPic{5s45N}\FPic{5ArrRight}
  \in \API{\SUN,\Pow{5}}
\ .
\end{equation}
We note in passing that Hermitian conjugation for birdtracks (in the
sense of linear maps on $\Pow{m}$ with the scalar product inherited
from $V$) is achieved by reflection around a vertical axis, followed
by a reversal of the arrows,~\cite{Cvitanovic:2008zz}. As an example take
\begin{align}
  \label{eq:conj-example}
  \FPic{3ArrLeft}\FPic{3s123SN}\FPic{3ArrRight}
  \xrightarrow{\text{reflect}} 
  \reflectbox{\FPic{3ArrLeft}\FPic{3s123SN}\FPic{3ArrRight} }
  \xrightarrow{\text{rev. arr.}}
  \FPic{3ArrLeft}\FPic{3s132SN}\FPic{3ArrRight}
  \hspace{1cm}\text{i.e} \hspace{1cm}
  \left(\FPic{3ArrLeft}\FPic{3s123SN}\FPic{3ArrRight}\right)^\dagger = \FPic{3ArrLeft}\FPic{3s132SN}\FPic{3ArrRight}
\ .
\end{align}
This implies that all symmetrizers and antisymmetrizers as defined
above are Hermitian
\begin{equation}
  \label{eq:ASHermitian}
  \bm S_{a_1,\ldots,a_n}^\dagger = \bm S_{a_1,\ldots,a_n}
  \hspace{1cm}\text{ and }\hspace{1cm}
  \bm A_{a_1,\ldots,a_n}^\dagger = \bm A_{a_1,\ldots,a_n}
\end{equation}
and that all permutations are unitary:
\begin{equation}
  \label{eq:perm-unitary}
  \sigma^{-1} = \sigma^\dagger \hspace{1cm}\text{ for all $\sigma\in S_m$.}
\end{equation}
The direction of arrows on the legs also allows
  us to account for complex conjugation,
  \emph{c.f.}~\cite{Cvitanovic:2008zz}. In this paper, we will
  exclusively be working with real operators and thus suppress the
  direction of the arrows, for example
  \begin{equation}
    \label{eq:suppress-Arrows}
\FPic{3s132SN} \quad \text{will refer to} \quad
\FPic{3ArrLeft}\FPic{3s132SN}\FPic{3ArrRight}
\ .
  \end{equation}

Lastly, if a \emph{Hermitian} projection operator $A$ projects onto a
subspace completely contained in the image of a projection
operator $B$, then we denote this as $A\subset B $, transferring the
familiar notation of sets to the associated projection operators. In
particular, $A\subset B$ if and only if
\begin{equation}
  \label{eq:OperatorInclusion1}
  A \cdot B = B \cdot A = A
\end{equation}
for the following reason: If the subspaces obtained by consecutively
applying the operators $A$ and $B$ in any order is the same as that
obtained by merely applying $A$, then not only need the subspaces that
$A$ and $B$ project onto overlap (as otherwise $A\cdot B=B\cdot A=0$),
but the subspace corresponding to $A$ must be completely contained in
the subspace of $B$ - otherwise the last equality
of~\eqref{eq:OperatorInclusion1} would not hold. Hermiticity is
crucial for these statements -- they thus do not apply to most Young
projection operators on $\Pow{m}$ if $m \ge 3$.\footnote{As can be
  explicitly verified by an example.} A familiar example for this
situation is the relation between symmetrizers of different length: a
symmetrizer $\bm{S}_{\mathcal{N}}$ can be absorbed into a symmetrizer
$\bm{S}_{\mathcal{N'}}$, as long as the index set $\mathcal{N}$ is a
subset of $\mathcal{N'}$, and the same statement holds for
antisymmetrizers,~\cite{Cvitanovic:2008zz}. For example,
\begin{equation}
\label{eq:Absorb-syms}
  \FPic{3Sym12SN}\FPic{3Sym123SN} 
  \; = \;
  \FPic{3Sym123SN} 
  \; = \; 
  \FPic{3Sym123SN}\FPic{3Sym12SN}
\ .
\end{equation}
Thus, by the above notation,
$\bm{S}_{\mathcal{N'}}\subset\bm{S}_{\mathcal{N}}$ if
$\mathcal{N}\subset\mathcal{N'}$. Or, as in our example,
\begin{equation}
  \FPic{3Sym123SN} 
  \; \subset \; 
  \FPic{3Sym12SN}
\ .
\end{equation}

$\API{\SUN,\Pow{m}}$ itself is equipped with a scalar product for linear
maps that is consistent with the scalar product on $\Pow{m}$ and
simply given by a trace
\begin{equation}
  \label{eq:APItr}
  \langle\ ,\ \rangle: \API{\SUN,\Pow{m}}\otimes\API{\SUN,\Pow{m}}  \to \mathbb R, 
  \hspace{1cm}
  \langle A, B\rangle := \text{tr}(A^\dagger B)
\ .
\end{equation}

In birdtrack notation, the trace $\text{tr}$ merely connects each
line exiting $A^{\dagger}B$ on the left with the line entering
$A^{\dagger}B$ on the right that is on the same level,
\begin{equation}
  \label{eq:trace-def}
\text{tr} \left( A^{\dagger}B \right) = \;
 \scalebox{0.8}{\begin{tikzpicture}[anchor=base, baseline=-8pt, yscale=0.25, xscale=0.25,
  shift={(0,-0.8)}]
\coordinate(LL1)at(-1.6,2.5);
\coordinate(RR1)at(1.6,-2.5);
\coordinate(L1)at(0,1.8);
\coordinate(L2)at(0,0.8);
\coordinate(L3)at(0,-0.2);
\coordinate(L5)at(0,-1.8);
\node(DL)[leftlabel]at(-1.8,-0.7){$\vdots$};
\node(DR)[rightlabel]at(1.8,-0.7){$\vdots$};
\draw[line width=0.75pt](L1)arc(270:-90:2.5 and 0.8);
\draw[line width=0.75pt](L2)arc(270:-90:3.5 and 1.8);
\draw[line width=0.75pt](L3)arc(270:-90:4.5 and 2.8);
\draw[line width=0.75pt](L5)arc(90:450:2.5 and 0.8);
\draw[fill=gray!30](LL1)rectangle(RR1);
\node(Rho)[label]at(0,-0.5){\large$A^{\dagger}B$};
\end{tikzpicture}}.
\end{equation}
For example, 
\begin{equation}
  \text{tr} \left( \left(\FPic{3s12SN}\right)^{\dagger} \FPic{3s123SN}
  \right) = \text{tr} \left( \FPic{3s12SN}\hspace{-0.2mm}\FPic{3s123SN} \right) =
  \text{tr} \left( \FPic{3s23SN} \right) = \FPic{3FTrs23Red} = N^2,
\end{equation}
where we have drawn the lines originating from the trace in red for
visual clarity. Each closed loop in the trace yields a factor of $N$
(the dimension of the fundamental representation), so that the scalar
product~\eqref{eq:APItr} will always yield a polynomial of $N$. In
particular, it is clear that this polynomial will not be identically
$0$ if $A,B\in S_m$ (and hence $A^{\dagger}\in S_m$), since $S_m$ is a
group and thus $A^{\dagger}B\in S_m$. The cyclic property of the trace
becomes very apparent in birdtrack notation, as the operator $A_1$ in
\begin{equation}
  \text{tr} \left( A_1 A_2 \ldots A_n \right)
\end{equation}
can just be ``pulled'' to the right of $A_n$ along the lines induced
by the trace.

\subsection{The hierarchichal nature of Young tableaux}

\begin{figure}[h]
  \centering
  \includegraphics[width=\linewidth]{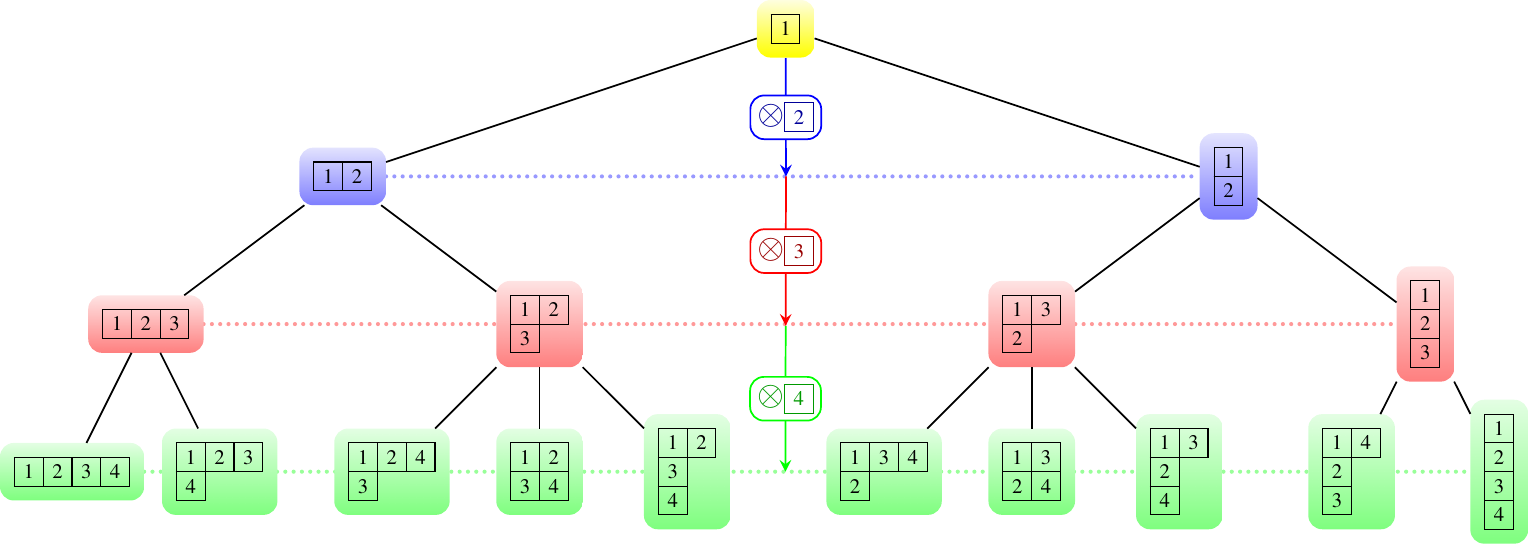}
  \caption{Branching tree of Young tableaux from its root to the $4^{th}$
    generation}
  \label{fig:YTbranching}
\end{figure}

A Young tableau is a conglomerate of numbered boxes with shape and
ordering restrictions imposed. The shape and ordering restrictions
automatically emerge if we construct these tableaux iteratively,
starting from a single box \begin{ytableau} 1 \end{ytableau} in a
process governed by branching rules (see for example~\cite{Fulton:1997,Sagan:2000}). The second box \begin{ytableau}
  2 \end{ytableau} and all further boxes are attached (in order) to
the right of or below an existing box in all possible ways that lead
to a set of boxes in which no row is longer than that above it and no
column is longer than the one left of it. The process results in a
tree whose first few branchings are displayed in
Fig~\ref{fig:YTbranching}.

We denote the set of Young tableaux with $n$ boxes by $\mathcal Y_n$
and note that in each branching step every Young tableau $\Theta \in
\mathcal Y_{n-1}$ creates a whole set of descendant tableaux in
$\mathcal Y_n$. We will refer to  this set by \ytableausetup{mathmode,
  boxsize=1.2em}$\{\Theta\otimes \begin{ytableau}
  n 
\end{ytableau} \}$ and notice that it has no overlap with the descendants of any
other element of $\mathcal Y_{n-1}$. Any $\mathcal Y_n$ is the
disjunct union of descendant sets: For example,
\ytableausetup{mathmode, boxsize=normal}
\begin{equation}
  \mathcal{Y}_3 := \vast\lbrace
  \begin{ytableau}
    1 & 2 & 3
  \end{ytableau}, \quad
  \begin{ytableau}
    1 & 2 \\
    3
  \end{ytableau}, \quad
  \begin{ytableau}
    1 & 3 \\
    2
  \end{ytableau}, \quad
  \begin{ytableau}
    1 \\
    2 \\
    3
  \end{ytableau}
\vast\rbrace = \Biggl\lbrace
  \begin{ytableau}
    1 & 2
  \end{ytableau} \otimes \ybox{3} \Biggr\rbrace \cup \Biggl\lbrace
  \begin{ytableau}
    1 \\
    2
  \end{ytableau} \otimes \ybox{3} \Biggr\rbrace.
\end{equation}

\ytableausetup{mathmode, boxsize=normal}

Traversing the tree of Fig.~\ref{fig:YTbranching} downwards is a
branching operation, in which each desendant has a well defined
ancestry chain: Starting at a Yound tableau and taking away the box
with the highest entry is a map in the mathematical sense, it yields
a unique tableau. We call this map the parent map and denote it by
$\pi$. $\pi$ can then repeatedly be applied to the resulting tableau
generating the ancestry chain for a given tableau. An example for part
of such a chain is
\begin{equation}
  \ldots\xlongrightarrow{\pi} \quad
  \begin{ytableau}
    1 & 3 & *(magenta!45) 6 \\
    2 & *(magenta!30) 5 \\
    *(magenta!15) 4
  \end{ytableau} \quad \xlongrightarrow{\pi} \quad
  \begin{ytableau}
    1 & 3 \\
    2 & *(magenta!30) 5 \\
    *(magenta!15) 4
  \end{ytableau} \quad \xlongrightarrow{\pi} \quad
  \begin{ytableau}
    1 & 3 \\
    2 \\
    *(magenta!15) 4
  \end{ytableau} \quad \xlongrightarrow{\pi} \quad
  \begin{ytableau}
    1 & 3 \\
    2 
  \end{ytableau}
  \quad \xlongrightarrow{\pi} \ldots
\ .
\end{equation}
This idea can obviously be formalized in a way that provides some
useful notation:
\begin{definition}[parent map and ancestor tableaux]
\label{ParentMap}
Let $\Theta \in \mathcal{Y}_n$ be a Young tableau. We define its
\emph{parent tableau} $\Theta_{(1)} \in \mathcal{Y}_{n-1}$ to be the
tableau obtained from $\Theta$ by removing the box \ybox{n} of
$\Theta$. Furthermore, we will define a \emph{parent map} $\pi$ from
$\mathcal{Y}_n$ to $\mathcal{Y}_{n-1}$, for a particular $n$,
\begin{equation}
   \label{eq:ParentMap1}
  \pi: \mathcal{Y}_n \rightarrow \mathcal{Y}_{n-1},
\end{equation}
which acts on $\Theta$ by removing the box \ybox{n} from
$\Theta$,\footnote{We note that the tableau $\Theta_{(1)}$ is always a
  Young tableau if $\Theta$ was a Young tableau, since removing the
  box with the highest entry cannot possibly destroy the properties of
  $\Theta$ (and thus $\Theta_{(1)}$) that make it a Young tableau.}
\begin{equation}
   \label{eq:ParentMap2}
   \pi: \Theta \mapsto \Theta_{(1)}.
\end{equation}
In general, we define the successive action of the parent map $\pi$ by
\begin{equation}
  \label{eq:ParentMapSucc1}
  \mathcal{Y}_n \; \xlongrightarrow[]{\pi} \; \mathcal{Y}_{n-1} \;
  \xlongrightarrow[]{\pi} \; \mathcal{Y}_{n-2} \;
  \xlongrightarrow[]{\pi} \; \ldots \; \xlongrightarrow[]{\pi} \; \mathcal{Y}_{n-m},
\end{equation}
and denote it by $\pi^m$,
\begin{equation}
  \label{eq:ParentMapSucc2}
  \pi^m: \mathcal{Y}_n \rightarrow \mathcal{Y}_{n-m}, \qquad  \pi^m:= \mathcal{Y}_n \; \xlongrightarrow[]{\pi} \; \mathcal{Y}_{n-1} \;
  \xlongrightarrow[]{\pi} \; \mathcal{Y}_{n-2} \;
  \xlongrightarrow[]{\pi} \; \ldots \; \xlongrightarrow[]{\pi} \; \mathcal{Y}_{n-m}
\end{equation}
We will further denote
the tableau obtained from $\Theta$ by applying the map $\pi$ $m$
times, $\pi^m(\Theta)$, by $\Theta_{(m)}$, and refer to it as the
\emph{ancestor tableau} of $\Theta$ $m$ generations back. Applying the
map $\pi^m$ to a Young tableau $\Theta$ then yields the \emph{unique}
tableau $\Theta_{(m)}$,
\begin{equation}
  \pi^m: \Theta \mapsto \Theta_{(m)}.\
\end{equation}
\end{definition}

We now reverse direction again and return to thinking about adding
boxes. As we keep adding more and more boxes we encounter more and
more tableaux that share their overall shape, they only differ by the
ordering of entries. The shape (represented by the boxes with the
entries deleted) is commonly referred to as a Young
\emph{diagram}. The reordering required to relate two tableaux of the
same shape defines a tableau permutation:
\begin{definition}[tableau permutation]\label{thm:TableauPermutation}
  Consider two Young tableaux $\Theta,\Theta'\in\mathcal{Y}_n$ with
  the same shape. Then, $\Theta'$ can be obtained from $\Theta$ by
  permuting the numbers of $\Theta$; clearly, the permutation needed
  to obtain $\Theta'$ from $\Theta$ is unique. We denote this permutation
  mapping $\Theta'$ into $\Theta$ by $\rho_{\Theta\Theta'}$,
\begin{equation}
  \Theta = \rho_{\Theta\Theta'} (\Theta')
\qquad \Longleftrightarrow \qquad 
  \Theta' = \rho_{\Theta\Theta'}^{-1} (\Theta) 
    = \rho_{\Theta'\Theta} (\Theta)
\ .
\end{equation}
\end{definition}

\subsection{From Young tableaux to Young operators}

The literature (see for example~\cite{Tung:1985na}) provides a
standard manner in which each Young tableau is associated with a Young
projection operator that is constructed from symmetrizers and
antisymmetrizers -- symmetrizers for the rows and antisymmetrizers for
the columns. (For completeness we assign the identity permutation for
rows or columns of length one. If \emph{all} rows or
columns are exactly of length one, we refer to this as the
symmetrizers or antisymmetrizers becoming trivial.) As such, the Young
projection operators corresponding to tableaux in $\mathcal{Y}_m$ are
elements of the (real) algebra of invariants $\API{\SUN,\Pow{m}}$.

Take, for example
\begin{equation}
  \label{eq:SymsEx1}
  \Theta =
  \begin{ytableau}
    1 & 3 & 4 \\
    2 & 5
  \end{ytableau} \; .
\end{equation}
The Young projection operator corresponding to this tableau,
$Y_{\Theta}$, is given by
\begin{equation}
  \label{eq:YT-example}
  Y_{\resizebox{!}{3mm}{\ytableausetup{mathmode, boxsize=1em} \begin{ytableau}
    1 & 3 & 4 \\
    2 & 5
  \end{ytableau}}}
  =
  2\cdot\bm{S}_{134}\bm{S}_{25}\bm{A}_{12}\bm{A}_{35}
  \ ,
\end{equation}
where the constant $2$ ensures idempotency of $Y_{\resizebox{!}{3mm}{\ytableausetup{mathmode, boxsize=1em} \begin{ytableau}
    1 & 3 & 4 \\
    2 & 5
  \end{ytableau}}}$, \emph{c.f.} eq.~\eqref{eq:alpha-Def} . (This is a
textbook topic. For a reminder on how to construct Young projection
operators from Young tableaux, readers are referred
to~\cite{Fulton:1997, Sagan:2000, Tung:1985na}.) All the symmetrizers
of a tableau commute with each other and so do the antisymmetrizers,
since no number appears more than once in any tableau. Thus, when
constructing the birdtrack corresponding to
$Y_{\resizebox{!}{3mm}{\ytableausetup{mathmode,
      boxsize=1em} \begin{ytableau}
      1 & 3 & 4 \\
      2 & 5
  \end{ytableau}}}$, we are able to draw the two symmetrizers appearing in it
underneath each other
(since they are disjoint), and similarly for the two
antisymmetrizers, 
\begin{equation}
  Y_{\resizebox{!}{3mm}{\ytableausetup{mathmode, boxsize=1em} \begin{ytableau}
    1 & 3 & 4 \\
    2 & 5
  \end{ytableau}}} = 2 \cdot
    \FPic{5s234N}\FPic{5Sym123Sym45N}    \FPic{5s2453N}\FPic{5ASym12ASym34N}    \FPic{5s45N}
  \; .
\end{equation}

We denote the set (or product -- it does not really matter as they
mutually commute) of symmetrizers associated with the tableau $\Theta$
by $\mathbf S_\Theta$ and the set (or product) of antisymmetrizers by
$\mathbf A_\Theta$. However, the symmetrizers of a tableau do not
commute with its antisymmetrizers (unless one or both are trivial):
\begin{equation}
  \label{eq:AScomm}
  [\mathbf S_\Theta,\mathbf
A_\Theta]\neq 0
\ .
\end{equation}
Therefore their relative order matters in the general
definition of a Young projector\footnote{Placing the antisymmetrizers to the right of the symmetrizers is a \emph{choice} of convention -- the reverse order leads to equivalent strutural results, it only matters to stay consistent.} 
\begin{equation}
  \label{eq:YTheta}
  Y_\Theta 
:= \alpha_\Theta \mathbf S_\Theta \mathbf A_\Theta
\ ,
\end{equation}
where $\alpha_\Theta\in\mathbb R$ is defined as
\begin{equation}
\label{eq:alpha-Def} \alpha_{\Theta}:=\frac{\mathcal{H}_{\Theta}}{\prod_{\mathcal{R}}\vert\text{length}(\mathcal{R})\vert!
  \prod_{\mathcal{C}}\vert\text{length}(\mathcal{C})\vert!}
\ ;
\end{equation}
the products in the denominator run over every row $\mathcal{R}$
respectively over every column $\mathcal{C}$ of
$\Theta$~\cite{Cvitanovic:2008zz} and $\mathcal{H}_{\Theta}$ is the
hook length of the tableau $\Theta$~\cite{Fulton:1997,Sagan:2000}: For
a particular Young tableau $\Theta$, form its underlying Young diagram
$\mathbf{Y}_{\Theta}$ by deleting all entries and then re-fill each
box $c$ in $\mathbf{Y}_{\Theta}$ with the integer counting all boxes
to the right and below it (including itself), called the \emph{hook}
of $c$ $\mathcal{H}_c$, for example\ytableausetup{mathmode, boxsize=normal}\begin{equation}
  \label{eq:Hook-length-example}
\Theta = \;
  \begin{ytableau}
    1 & 3 & 5 & 8 \\
    2 & 6 & 7 \\
    4
  \end{ytableau}
\; \xrightarrow{\text{delete entries}} \;
\ydiagram{4,3,1}
\; \xrightarrow{\text{hooks $\mathcal{H}_c$}} \;
  \begin{ytableau}
    6 & 4 & 3 & 1 \\
    4 & 2 & 1 \\
    1
  \end{ytableau}
\ .
\end{equation}
The hook length of $\Theta$, $\mathcal{H}_\Theta$, (equivalently the
hook length of $\mathbf{Y}_{\Theta}$, $\mathcal{H}_{\mathbf{Y}_{\Theta}}$) is defined to be the
product of all the hooks $\mathcal{H}_c$,
\begin{equation}
  \label{eq:Hook-length-def}
\mathcal{H}_\Theta 
=
\mathcal{H}_{\mathbf{Y}_{\Theta}} 
:= \prod_{c\in\Theta} \mathcal{H}_c
\ ;
\end{equation}
for the tableau in~\eqref{eq:Hook-length-example}, the hook length is
$\mathcal{H}_\Theta=6\cdot4^2\cdot3\cdot2=576$.

The Young projection operators are nonzero precisely if all their
columns are at most of length $N$, otherwise they vanish identially --
we refer to this as being dimensionally zero (see
app.~\ref{sec:VanishingReps}).

 From~\eqref{eq:YTheta} one infers
that $Y_\Theta$ is not Hermitian (unless at least one of the sets is trivial):
\begin{equation}
  \label{eq:YT-not-herm}
  Y_\Theta^\dagger 
  = \alpha_\Theta (\mathbf S_\Theta \mathbf A_\Theta)^\dagger 
  = \alpha_\Theta\mathbf A_\Theta \mathbf S_\Theta
  \neq  Y_\Theta
\ .
\end{equation}

Hermiticity (or the lack thereof) in birdtrack notation is best
judged after expanding in primitive invariants, for example
\ytableausetup{mathmode, boxsize=0.7em}
\begin{alignat}{5}
  \label{eq:Y13-2}
  &  Y_{\resizebox{!}{3mm}{\ytableausetup{mathmode, boxsize=1em} \begin{ytableau}
    1 & 3\\
    2
  \end{ytableau}}}\ 
  && 
  = \frac43 \cdot \FPic{3Sym13ASym12} 
&& \quad \Rightarrow \quad && 
  Y^{\dagger}_{\resizebox{!}{3mm}{\ytableausetup{mathmode, boxsize=1em} \begin{ytableau}
    1 & 3\\
    2
  \end{ytableau}}}\
  && = \frac43 \cdot \FPic{3ASym12Sym13}  \\
&  &&= \frac{1}{3} \left(
  \FPic{3IdSN} -
  \FPic{3s12SN} +
  \FPic{3s13SN} -
  \FPic{3s123SN} \right)
&& &&
  && = \frac{1}{3} \left(
  \FPic{3IdSN} -
  \FPic{3s12SN} +
  \FPic{3s13SN} -
  \FPic{3s132SN} \right) \ . \nonumber
\end{alignat}

The definition of $Y_{\resizebox{!}{3mm}{\ytableausetup{mathmode, boxsize=1em} \begin{ytableau}
    1 & 3 & 4 \\
    2 & 5
  \end{ytableau}}}$ in~\eqref{eq:YT-example} speaks of a linear map on
$\Pow{5}$, i.e. an element of $\Lin{\Pow{5}}$, or with equal validity
of a linear map on a larger space $\Pow{m}$, $m \ge 5$, in which the
factors beyond the first five remain unaffected. We speak of this case
as the canonical embedding of $\Lin{\Pow{n}} \hookrightarrow
\Lin{\Pow{m}}$ (with $m \ge n$). For a given tableau $\Theta \in
\mathcal Y_n$ we will, in a slight abuse of notation, employ the same
notation, $Y_\Theta$, to talk both about the original case or any of
the embeddings. The idea of an embedding in birdtrack terms requires
to explicitly draw the ``unaffected lines'', for example the operator
$\Bar Y_{\resizebox{!}{3mm}{\ytableausetup{mathmode,
      boxsize=1em} \begin{ytableau}
      1 & 2\\
      3
  \end{ytableau}}}$ is canonically embedded into
$\mathrm{Lin}\left(V^{\otimes 5}\right)$ as
\begin{equation}
  \label{eq:CanonicalEmbedding}
  \FPic{3Sym12ASym13} 
  \; \hookrightarrow \;   
\FPic{5IdN}\FPic{5Sym12N}\FPic{5s23N}\FPic{5ASym12N}\FPic{5s23N}
\ . 
\end{equation}

Furthermore, for any operator $O$, the symbol $\bar{O}$ will refer to
a product of symmetrizers and antisymmetrizers without any
\emph{additional} scalar factors. For example,
\begin{equation}
\label{eq:Bar-Notation-Def}
  Y_{\Theta} := \underbrace{\frac{4}{3}}_{=: \alpha_{\Theta}} \;
  \underbrace{\FPic{3Sym13ASym12}}_{    =: \bar{Y}_{\Theta}}.
\end{equation}
The benefit of this notation is that it allows us to ignore all
additional scalar factors; in particular, for a non-zero scalar
$\omega$, 
\begin{equation}
\label{eq:Bar-Notation-Benefit}
  \omega \cdot O \neq O \qquad \text{but} \qquad \omega \cdot \Bar O =
  \Bar O
\ .
\end{equation}

Tableau permutations can be represented as birdtracks. A graphical
procedure is probably the most efficient mean to obtain this
representation:
\begin{definition}[tableau permutations as birdtracks]
\label{thm:TableauPermutation-birdtrack}
To construct the birtrack form of the tableau permutation
$\rho_{\Theta\Phi}$ between tableaux of the same shape (\emph{c.f.}
Def.~\ref{thm:TableauPermutation}) explicitly, write the Young
tableau $\Theta$ and $\Phi$ next to each other, such that $\Theta$ is
to the left of $\Phi$ and then connect the boxes in the corresponding
position of the two diagrams, such as
\begin{equation}\label{eq:SchematicTransitionPermutation}
\Theta \rightarrow \;
\raisebox{-26pt}{\scalebox{0.5}{
\begin{picture}(0,0)\includegraphics{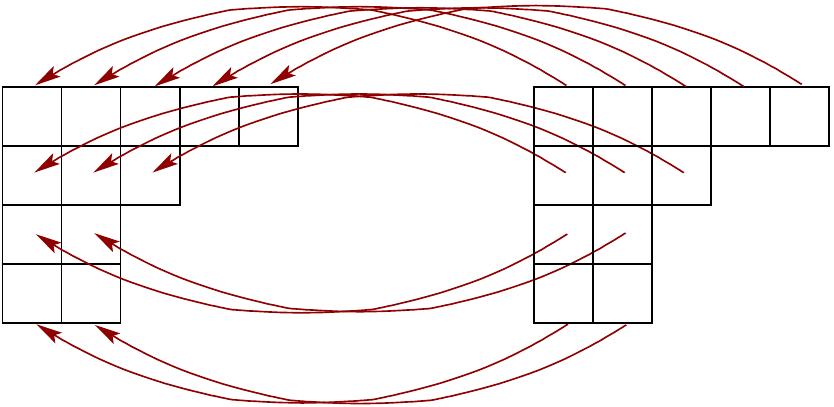}\end{picture}\setlength{\unitlength}{4144sp}\begin{picture}(3804,1844)(169,-712)
\end{picture}}} \leftarrow
\Phi.
\end{equation}
Then, write two columns of numbers from $1$ to $n$ next to each
other in descending order; the left column represents the entries of
$\Theta$ and the right column represents the entries of
$\Phi$. Lastly, connect the entries in the left and the right
column in correspondence to
\eqref{eq:SchematicTransitionPermutation}. The resulting tangle of
lines is the birdtrack corresponding to $\rho_{\Theta\Phi}$ and
thus determines the permutation.
\end{definition}

As a birdtrack, $\rho_{\Theta\Phi}$ immediately becomes a linear map
in $\API{\Pow{m}}$ and as such directly relates the associated Young
projectors:
\begin{equation}
  \label{eq:permute-tableaux2}
  Y_{\Theta} = 
\rho_{\Theta\Phi} Y_{\Phi} \rho_{\Theta\Phi}^{-1} 
= 
\rho_{\Theta\Phi} Y_{\Phi} \rho_{\Phi\Theta} 
\ .
\end{equation}
This property is in fact part and parcel of the very definition of
Young projectors in~\cite[def. 5.4]{Tung:1985na}.
Eq.~\eqref{eq:permute-tableaux2} demonstrates that tableaux of
the same shape correspond to equivalent representations and 
$\rho_{\Theta\Phi}$ is the isomorphism that seals the equivalence.
For the Hermitian projection operators
(\emph{c.f.} sec.~\ref{sec:Three-Hermitian-Ops}), eq.~\eqref{eq:permute-tableaux2}
breaks down, as is exemplified in appendix~\ref{sec:Illustration-Rho}.

  It may help to illustrate this with an example: Take the equivalence
  pair corresponding to the Young tableaux
\ytableausetup{mathmode, boxsize=normal}
\begin{equation}
  \label{eq:Equiv3quarks}
  \Theta :=
  \begin{ytableau}
    1 & 2 \\
    3
  \end{ytableau}
\qquad \text{and} \qquad \Phi :=
\begin{ytableau}
  1 & 3 \\
  2
\end{ytableau}
\end{equation}
with
\begin{equation}
  \label{eq:young3}
  Y_{\Theta} = \sfrac{4}{3}\cdot\FPic{3Sym12ASym13} \qquad
  \text{and} \qquad Y_{\Phi} = \sfrac{4}{3}\cdot\FPic{3Sym13ASym12}
\ .
\end{equation}
To find the permutation $\rho_{\Theta \Phi}$, we connect boxes
between $\Theta$ and $\Phi$
\begin{equation}
\label{eq:3-Rho-ThetaThetap}
\FPic[scale=.8]{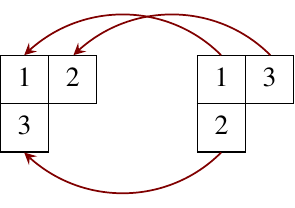}
\end{equation}
and identify $\rho_{\Theta \Phi}$ as \FPic{3s23SN}. Evidently
eq.~\eqref{eq:permute-tableaux2} holds, since
\begin{align}
  \label{eq:Yequiv}
  \sfrac{4}{3}\cdot\FPic{3Sym12ASym13}  =
  \sfrac{4}{3}\cdot\FPic{3s23N}\FPic{3Sym13ASym12}\FPic{3s23N}
\ .
\end{align}

\subsection{Cancellation rules}\label{sec:CancellationRules}

In~\cite{Alcock-Zeilinger:2016bss}, we established
various rules designed to easily manipulate birdtrack operators
comprised of symmetrizers and antisymmetrizers. Since all operators
considered in this paper are of this form, the simplification rules
of~\cite{Alcock-Zeilinger:2016bss} are immediately
applicable here. One of them plays a crucial role throughout
this paper so we recall the result without repeating the proof:

\begin{theorem}[cancellation of parts of the
  operator~\cite{Alcock-Zeilinger:2016bss}]\label{thm:CancelMultipleSets}
  Let $\Theta\in\mathcal{Y}_n$ be a Young tableau and
  $M\in\API{\SUN,\Pow{n}}$ be an algebra element. Then,
  there exists a (possibly vanishing) constant $\lambda$ such that
  \begin{equation}
    \label{eq:Cancel-General-O}
O := \mathbf{S}_{\Theta} \; M \; \mathbf{A}_{\Theta} =
\lambda \cdot Y_{\Theta}
\ .
  \end{equation}
Note that if the operator $O$ is non-zero, then $\lambda\neq 0$. 
\end{theorem}

To benefit of this statement we provide some crucial criteria that
allow us to identify particularly important cases of nonzero $O$ in
$\API{\SUN,\Pow{n}}\subset\Lin{\Pow{n}}$:
\begin{enumerate}
\item\label{itm:nonzero-O-1} Let
  $\mathbf{A}_{\Phi_i}\supset\mathbf{A}_{\Theta}$ and
  $\mathbf{S}_{\Phi_j}\supset\mathbf{S}_{\Theta}$ be (anti-)
  symmetrizers that can be absorbed into $\mathbf{A}_{\Theta}$ and
  $\mathbf{S}_{\Theta}$ for every $i\in\lbrace 1,3,\ldots k-1\rbrace$
  and for every $j\in\lbrace 2,4,\ldots k\rbrace$. If $M$
  in~\eqref{eq:Cancel-General-O} is of the form
\begin{equation}
  \label{eq:CancelWedgedParent1}
  M = \; \mathbf{A}_{\Phi_1} \;
  \mathbf{S}_{\Phi_2} \; \mathbf{A}_{\Phi_3} \;
  \mathbf{S}_{\Phi_4} \; \cdots \; \mathbf{A}_{\Phi_{k-1}} \;
  \mathbf{S}_{\Phi_k}
\ ,
\end{equation}
then $O$ is non-zero unless $Y_\Theta$ is dimensionally zero.
\item\label{itm:nonzero-O-2} Let $\Theta,\Phi\in\mathcal{Y}_n$ be two
  Young tableaux with the same shape and construct the permutations
  $\rho_{\Theta\Phi}$ and $\rho_{\Phi\Theta}$ between the two tableaux
  according to Definition~\ref{thm:TableauPermutation-birdtrack}. Furthermore,
  let $\mathcal{D}_{\Phi}$ be a product of symmetrizers and
  antisymmetrizers which can be absorbed into $\mathbf{S}_{\Phi}$ and
  $\mathbf{A}_{\Phi}$ respectively.  If $M$
  in~\eqref{eq:Cancel-General-O} is of the form
  \begin{equation}
  \label{eq:CancelWedgedParent2}
    M = \rho_{\Theta\Phi} \; \mathcal{D}_{\Phi} \; \rho_{\Phi\Theta}
    \ ,
  \end{equation}
then $O$ is non-zero unless $Y_\Theta$ is dimensionally zero.
\item \label{itm:nonzero-O-3} If $M$ is a product of expresstions of
  the form of~\eqref{eq:CancelWedgedParent1}
  and~\eqref{eq:CancelWedgedParent2}, then $O$ is non-zero unless
  $Y_\Theta$ is dimensionally zero.
\end{enumerate}
The general proof of these statements can again be found
in~\cite{Alcock-Zeilinger:2016bss}, but it is apparent that under the
conditions listed for the ingredients
of~\eqref{eq:CancelWedgedParent1} and~\eqref{eq:CancelWedgedParent2},
any dimensional zero of $O$ manifests itself as a dimensional zero of
$Y_\Theta$ since $\mathbf A_\Theta$ automatically contains the longest
antisymmetrizer in $O$.  \ytableausetup{mathmode, boxsize=normal}

As an example, consider the operator
\begin{equation}
  O := \; \FPic{5s354N}\FPic{5Sym123Sym45N}\FPic{5s243N}\FPic{5ASym12N}\FPic{5s23s45N}\FPic{5Sym12Sym34N}\FPic{5s23N}\FPic{5ASym12ASym34N}\FPic{5s23N}
\;  = \; \lbrace\bm{S}_{125},\bm{S}_{34}\rbrace \cdot
  \lbrace\bm{A}_{13}\rbrace \cdot \lbrace\bm{S}_{12},\bm{S}_{34}\rbrace \cdot \lbrace\bm{A}_{13},\bm{A}_{24}\rbrace.
\end{equation}
This operator meets all conditions of the above Theorem
\ref{thm:CancelMultipleSets}: the sets
$\lbrace\bm{S}_{125},\bm{S}_{34}\rbrace$ and
$\lbrace\bm{A}_{13},\bm{A}_{24}\rbrace$ together constitute the birdtrack of
a Young projection operator $\bar{Y}_{\Theta}$ corresponding to the
tableau
\begin{equation}
  \Theta :=
  \begin{ytableau}
    1 & 2 & 5 \\
    3 & 4
  \end{ytableau}
  \ .
\end{equation}
The set $\lbrace\bm{A}_{13}\rbrace$ corresponds to the ancestor
tableau $\Theta_{(2)}$, and the set
$\lbrace\bm{S}_{12},\bm{S}_{34}\rbrace$ corresponds to the ancestor tableau
$\Theta_{(1)}$ and thus can be absorbed into $\mathbf{A}_{\Theta}$
resp. $\mathbf{S}_{\Theta}$, \emph{c.f.} eq.~\eqref{eq:Absorb-syms}. Hence $O$ can be written as
\begin{equation}
  O = \; \mathbf{S}_{\Theta} \; \mathbf{A}_{\Theta_{(2)}} \;
  \mathbf{S}_{\Theta_{(1)}} \; \mathbf{A}_{\Theta}.
\end{equation}
According to the Cancellation-Theorem~\ref{thm:CancelMultipleSets}, we may cancel the wedged ancestor sets
$\mathbf{A}_{\Theta_{(2)}}$ and $\mathbf{S}_{\Theta_{(1)}}$ at the
cost of a non-zero constant $\lambda$,
\begin{equation}
 O = \lambda \cdot
\underbrace{
\FPic{5s354SN}\FPic{5Sym123Sym45N}\FPic{5s2453N}\FPic{5ASym12ASym34N}\FPic{5s23N}
}_{\Bar
O = \bar{Y}_{\resizebox{!}{2mm}{\ytableausetup{mathmode, boxsize=1em} \begin{ytableau}
    1 & 2 & 5 \\
    3 & 4
  \end{ytableau}}}}
\ ,
\end{equation}
which is proportional to $Y_{\resizebox{!}{3mm}{\ytableausetup{mathmode, boxsize=1em} \begin{ytableau}
    1 & 2 & 5 \\
    3 & 4
  \end{ytableau}}}$.

\section{Young projection and transition operators over
  \texorpdfstring{$\Pow{m}$}{Vm} for small \texorpdfstring{$m$}{m}: an
  inspiration for a multiplet adapted basis for \texorpdfstring{$\API{\SUN,\Pow{m}}$}{API(SU(N),Vm)}}\label{sec:YoungBasis}

The group theoretical interest in Young operators is that they project
onto irreducible representations. They satisfy the following three properties:

\begin{enumerate}
\item\label{itm:YoungIntro3} Young projection operators are
  idempotent, that is they satisfy\footnote{This is surprisingly hard
    to demonstrate unless you have access to the cancellation rule
    Theorem~\ref{thm:CancelMultipleSets}.}
\begin{subequations}
  \label{eq:Ymult}
  \begin{equation}
    \label{eq:YoungIntro3}
    Y_\Theta\cdot Y_\Theta=Y_\Theta
    \ .
  \end{equation}
They are mutually orthogonal as projectors: if $\Theta$ and $\Phi$ are
two distinct Young tableaux in $\mathcal{Y}_m$, then
  \begin{equation}
    \label{eq:YoungIntro2}
    Y_\Theta\cdot Y_\Phi = 0 
\hspace{1cm} 
\text{for $m=1,2,3,4$}
  \end{equation}
and for all $m$ if $\Theta$ and $\Phi$ have a different shape.

\item\label{itm:YoungIntro1} The set of   Young projection
  operators for $\SUN$ over $\Pow{m}$ sum up to the identity element
  of $\Pow{m}$:
  \begin{equation}
    \label{eq:Yidpowm}
    \sum\limits_{\Theta\in\mathcal Y_m} Y_\Theta = \text{id}_{\Pow{m}}
\hspace{1cm}
\text{for $m=1,2,3,4$}
\ .
  \end{equation}
\end{subequations}
  In physics parlance this constitutes a completeness relation.
\end{enumerate}
The Young tableaux underlying the projection operators fully classify
the irreducible representations of $\SUN$ over $\Pow{m}$ for any
$m$. If $m\leq4$, the Young projectors split the space $\Pow{m}$ into
mutually orthogonal subspaces, which can be shown to be
irreducible~\cite{Tung:1985na}. For $m\geq4$, generalizations of
Young projectors take over this role. Such generalizations include
subtracted operators~\cite{Littlewood:1950,Schensted:1976} and
Hermitian Young projection
operators~\cite{Keppeler:2013yla,Alcock-Zeilinger:2016sxc}. In this
paper, we will focus on the latter.

Besides Young projection operators not being pairwise orthogonal for
$m\geq5$ as linear maps, they are not orthogonal with
respect to the scalar product on $\API{\Pow{m}}$ even for smaller $m$,
for example
\ytableausetup{mathmode, boxsize=0.7em}
\begin{equation}
  \label{eq:not-orth}
 \text{tr}\Bigl(
   Y_{\begin{ytableau} \scriptstyle 1 & \scriptstyle 2 \\
      \scriptstyle 3 \end{ytableau}}^\dagger  
  Y_{\begin{ytableau} \scriptstyle 1 & \scriptstyle 3 \\
      \scriptstyle 2 \end{ytableau}}
  \Bigr)
  \neq 0
\end{equation}
as emerges quickly from an explicit calculation:
\begin{align*}
   \text{tr}\Bigl(
   Y_{\begin{ytableau} \scriptstyle 1 & \scriptstyle 2 \\
      \scriptstyle 3 \end{ytableau}}^\dagger  
  Y_{\begin{ytableau} \scriptstyle 1 & \scriptstyle 3 \\
      \scriptstyle 2 \end{ytableau}}  \Bigr)
& = 
\text{Tr} 
 \left( 
\left(\frac{4}{3}\right)^2 
\FPic{3ASym13Sym12}\cdot\FPic{3Sym13ASym12} 
 \;  \right) \\
& = \frac{1}{9} \text{Tr} 
\Bigl( \;
- \ \FPic{3IdSN} + \FPic{3s12SN} -2\cdot \FPic{3s23SN} + \FPic{3s13SN} 
  +2\cdot \FPic{3s123SN} - \FPic{3s132SN}
  \; \Bigr) \\
 & = - \frac{N_c^3}{9} + \frac{N_c}{9} \neq 0 \ .
\end{align*}
This mishap is only possible since the Young
projectors are not Hermitian -- otherwise their orthogonality as
projectors would create a zero automatically at least for $m\leq4$.\footnote{This is truly
  an issue with Hermiticity, not a consequence of choosing an
  unsuitable scalar product: The sets of left and right eigenvectors
  differ if $Y_\Theta$ is not Hermitian.}
\ytableausetup{mathmode, boxsize=normal}

Nevertheless the trace of $Y_\Theta$ corresponding to $\Theta\in\mathcal{Y}_m$, normalized as a projector,
uncovers the dimension of the associated irreducible representation
(see~\cite[app. B4.]{Cvitanovic:2008zz} for a textbook exposition):
\begin{equation}
  \label{eq:dimTheta}
  \text{tr} (Y_\Theta) = \text{dim}(\Theta) 
\hspace{1cm}
\text{for all $m$}
\ .
\end{equation}

As is evident from Fig.~\ref{fig:YTbranching}, $|\mathcal Y_m|$, the
number of Young tableaux in $\mathcal Y_m$, is smaller than the
dimension of $\API{\SUN,\Pow{m}}$, which is $m!$ (up to dimensional zeros).

\subsection{Transition operators for Young projectors over
  \texorpdfstring{$\Pow{m}$}{Vm} for
  \texorpdfstring{$m\leq4$}{mless4}}\label{sec:Young-trans-ops}

Generally, the established goal of representation theory is to find a
set of operators that satisfy idempotency, orhthogonality and
decopmosition of unity, nothing more, nothing less, and for $m\leq4$,
Young projection operators do just that.

We step beyond this point by noting that there is an additional set of
linearly independent operators in $\API{\SUN,\Pow{m}}$ for $m\leq4$
that are closely related to the set of Young projectors and that complete it 
to a basis of the full algebra in a transparent way. Recall that for
any pair of equivalent representations corresponding to tableaux
$\Theta$ and $\Phi$ in $\mathcal{Y}_m$, there exist
a unique tableau permutations $\rho_{\Theta\Phi}$ such that
$Y_\Theta = \rho_{\Theta\Phi} Y_{\Phi} \rho_{\Phi\Theta}$
(c.f. eq.~\eqref{eq:permute-tableaux2}). From this define
\emph{transition operators}
\begin{equation}
  \label{eq:Ytrans}
  T_{\Theta\Phi} 
  := 
  \rho_{\Theta\Phi}Y_{\Phi} 
  = 
  Y_\Theta \rho_{\Theta\Phi}
  =
  Y_\Theta \rho_{\Theta\Phi}Y_{\Phi}
\hspace{1cm}
\text{for $m=1,2,3,4$}
\end{equation}
and observe that they seamlessly extend the multiplication table of
the Young projectors:
\begin{subequations}
\label{eq:YtransMult}
\begin{align}
  \label{eq:nHTranns-Op-Def1}
  & Y_{\Theta}T_{\Theta\Phi}=T_{\Theta\Phi}=T_{\Theta\Phi}Y_{\Phi} \\
  \label{eq:nHTranns-Op-Def2}
  & T_{\Theta\Phi}T_{\Phi\Theta}=Y_{\Theta} \\
  \label{eq:nHTranns-Op-Def3}
  & T_{\Phi\Theta}T_{\Theta\Phi}=Y_{\Phi} 
\end{align}
\end{subequations}
We see that $T_{\Theta\Phi}$ maps the image $Y_{\Phi}(\Pow{m})$
bijectlively onto $Y_{\Theta}(\Pow{m})$ ($m\leq4$). The inverse on the images is
$T_{\Phi\Theta}$. The $T_{\Theta\Phi}$ are \emph{transition
  operators} between the irreducible representations. An example of
all Young projection and transition operators over $\Pow{3}$ is given
in section~\ref{sec:3qAlgebraSectionYoung}.

Note that in general the transition operators between Young projectors
are not unitary on the subspaces,
\begin{equation}
  \label{eq:Herm-Conj-nHTransition-Ops}
\left(T_{\Theta\Phi}\right)^{\dagger} = Y^\dagger_\Phi \rho_{\Phi\Theta} Y^\dagger_{\Theta} \neq T_{\Phi\Theta} \ ,
\end{equation}
since the Young projection operators are not Hermitian (for an explicit example see app.~\ref{sec:cons-non-herm}).

\subsection{A multiplet adapted basis for \texorpdfstring{$\API{\SUN,\Pow{m}}$}{API(SU(N)Vm)}}\label{sec:Multiplet-basis-for-API}

In this section we prove that the set of all mutually orthogonal
projection operators corresponding to irreducible representations of
$\SUN$ over $\Pow{m}$ and their transition operators -- call this set
$\mathfrak{S}_m$ -- spans the algebra of invariants
$\API{\SUN,\Pow{m}}$. This proof holds for all $m$ allowing us to
construct $\mathfrak{S}_m$: for Young projection operators, this means
that $m\leq4$. Later on (section~\ref{sec:Orth-Proj-Basis}), we see that
$\mathfrak{S}_m$ can be constructed \emph{for all $m$} if \emph{Hermitian}
Young projection operators are used.

The projection operators corresponding to irreducible representations
of $\SUN$ over $\Pow{m}$ project onto equivalent irreducible
representations \emph{if and only if} the corresponding Young tableaux
have the same shape,~\cite{Cvitanovic:2008zz,Tung:1985na}, and thus
correspond to the same underlying Young diagram. Suppose now that a
particular Young diagram ${\mathbf{Y}}$ gives rise to $l$ Young
tableaux. Then, the set of all projection operators corresponding to
these $l$ tableaux and all transition operators between them -- let us
denote this set by $\mathfrak{S}_{{\mathbf{Y}}}$ -- will be of size $l^2$,
\begin{equation}
  \vert \mathfrak{S}_{{\mathbf{Y}}} \vert = l^2
\ ,
\end{equation}
since one may always arrange the elements of
$\mathfrak{S}_{{\mathbf{Y}}}$ into an $l\times l$ matrix which has the
projection operators on the diagonal and each off-diagonal element in
position $ij$ is the transition operator between the diagonal elements
$ii$ and $jj$.  Fortunately, there is a way of counting how many Young
tableaux can be obtained from a Young diagram with a particular shape,
namely via the hook length
$\mathcal{H}_{{\mathbf{Y}}}$~\cite{Fulton:1997,Sagan:2000},\footnote{Note
  that one often finds the statement that ``the number of tableaux
  corresponding to a diagram is given by the hook length.'' It would
  be less misleading to state that it is a function of the hook
  length.}  \emph{c.f.}  eq.~\eqref{eq:Hook-length-def}: If
${\mathbf{Y}}$ is a particular Young diagram, then the set
$\mathfrak{S}_{{\mathbf{Y}}}$ has size
$(m!/\mathcal{H}_{{\mathbf{Y}}})^2$~\cite{Glass:2004},
\begin{equation}
  \vert \mathfrak{S}_{{\mathbf{Y}}} \vert = \left(\frac{m!}{\mathcal{H}_{{\mathbf{Y}}}}\right)^2
\ .
\end{equation}
If we sum the $\vert \mathfrak{S}_{{\mathbf{Y}}}\vert$ over all
Young diagrams ${\mathbf{Y}}$ consisting of $m$ boxes, we obtain the
aggregate number of all projection and transition
operators associated with $\SUN$ over $\Pow{m}$ $\vert\mathfrak{S}_m\vert$,
\begin{equation}
  \label{eq:FullAlgebra0}
  \vert \mathfrak{S}_m \vert 
= \sum_{{\mathbf{Y}}} \vert  \mathfrak{S}_{\mathbf{Y}} \vert 
= \sum_{{\mathbf{Y}}}
\left(\frac{m!}{\mathcal{H}_{{\mathbf{Y}}}}\right)^2
\ .
\end{equation}

To proceed further, we need to use some well established facts of the
representation theory of the permutation group of $m$ elements, $S_m$,
which can be found in many standard textbooks such
as~\cite{Artin:2011}. To this end, let us briefly recapitulate: Each
irreducible representation of $S_m$ corresponds to a Young
\emph{diagram} ${\mathbf{Y}}$ (not a Young tableau!), and the
multiplicity of each representation in the regular representation of
the symmetric group is given 
by $m!/\mathcal{H}_{{\mathbf{Y}}}$. From
representation theory of finite groups (such as $S_m$), it is further
known that the sum of the square of the multiplicities of all
irreducible representations of a finite group $G$ is equal to the size
of the group. In particular, for the finite group $S_m$, this means
that
\begin{equation}
  \label{eq:FullAlgebra1}
  \vert S_m \vert = \sum_{{\mathbf{Y}}}
  \left(\frac{m!}{\mathcal{H}_{{\mathbf{Y}}}}\right)^2
\ ,
\end{equation}
where we sum over all Young tableaux ${\mathbf{Y}}$ consisting of $m$
boxes. (For a bijective proof of eq.~\eqref{eq:FullAlgebra1}
see~\cite{Sagan:2000}.) However,~\eqref{eq:FullAlgebra0} tells us that
the sum on the right hand side of equation~\eqref{eq:FullAlgebra1}
also represents the aggregate number of all Hermitian Young projection
and transition operators of $\SUN$ over $\Pow{m}$, so that
\begin{equation}
  \label{FullAlgBasisSize}
  \vert S_m \vert = \vert \mathfrak{S}_m \vert.
\end{equation}
Provided that $N \ge m$ so that dimensional zeroes are absent, the
projection and transition operators in $\mathfrak{S}_m$ are all
linearly independent (see app.~\ref{sec:VanishingReps} for the general
case). It follows that these operators span the algebra of invariants
over $\Pow{m}$, and thus constitute an alternative basis of this
algebra,
\begin{equation}
  \label{eq:SM-span}
  \API{\SUN,\Pow{m}} = \Bigl\{ \alpha_k \mathfrak{s}_k |\alpha_k\in\mathbb R, \mathfrak{s}_k\in\mathfrak{S}_m \Bigr\}
\ .
\end{equation}

\subsection{An example: the full algebra over \texorpdfstring{$\Pow{3}$}{V3}
  in a Young projector basis}\label{sec:3qAlgebraSectionYoung}
\ytableausetup{mathmode, boxsize=normal}

 $\API{\SUN,\Pow{3}}$ is spanned by the primitive invariants
\begin{equation}
  \Big\lbrace \FPic{3IdSN}, \; \FPic{3s12SN}, \; \FPic{3s13SN}, \; \FPic{3s23SN},
  \; \FPic{3s123SN}, \; \FPic{3s132SN} \Big\rbrace \subset \Lin{\Pow{3}}.
\end{equation}
If $N\ge 3$, its dimension is $3!=6$.  There are $3$ Young diagrams
consisting of $3$ boxes, which give rise to a total of $4$ Young
tableaux, \ytableausetup {mathmode, boxsize=1.2em}
\begin{equation}\label{S3YTableaux}
  \raisebox{-0.5\height}{\begin{tikzpicture}[level distance=1.4cm,
level 1/.style={sibling distance=1.2cm}]
\tikzstyle{every node}=[rectangle, inner sep = 4pt]
\node{$\ydiagram{3}$}
        child{node{$\begin{ytableau}
            \scriptstyle 1 & \scriptstyle 2 & \scriptstyle 3
          \end{ytableau}$}};
\end{tikzpicture}} \qquad
\raisebox{-0.5\height}{\begin{tikzpicture}[level distance=1.4cm,
level 1/.style={sibling distance=1.2cm}]
\tikzstyle{every node}=[rectangle, inner sep = 4pt]
 \node{$\ydiagram{2,1}$}
        child{node{$\begin{ytableau}
            \scriptstyle 1 & \scriptstyle 2 \\
            \scriptstyle 3
          \end{ytableau}$}}
        child{node{$\begin{ytableau}
            \scriptstyle 1 & \scriptstyle 3\\
            \scriptstyle 2
          \end{ytableau}$}};
\end{tikzpicture}} \qquad
\raisebox{-0.5\height}{\begin{tikzpicture}[level distance=1.4cm,
level 1/.style={sibling distance=1.2cm}]
\tikzstyle{every node}=[rectangle, inner sep = 4pt]
\node{$\ydiagram{1,1,1}$}
        child{node{$\begin{ytableau}
            \scriptstyle 1 \\
            \scriptstyle 2 \\
            \scriptstyle 3 
          \end{ytableau}$}};
\end{tikzpicture}}.
\end{equation}
Indeed, we find that the sum of the squares of 
$\nicefrac{3!}{\mathcal{H}_{{\mathbf{Y}_i}}}$ corresponding to each diagram ${\mathbf{Y}_i}$ is equal
to the size of the group $S_3$,
\ytableausetup
{mathmode, boxsize=normal}
\begin{equation}
  3! = \vert S_3 \vert = \sum_{{\mathbf{Y}_i}} \left(\frac{3!}{\mathcal{H}_{{\mathbf{Y}_i}}}\right)^2 = 1^2 +
  2^2 + 1^2.
\end{equation}
The first and last Young tableaux have a unique shape and thus project
onto unique irreducible representations of $\SUN$. The corresponding 
Young projection operators are given by
\begin{equation}
\label{eq:3Herm-Youngs}
  Y_1 = \FPic{3Sym123} \quad \text{and} \quad Y_4 = \FPic{3ASym123}
\ ,
\end{equation}
where $Y_i$ corresponds to the $i^{th}$ tableau (read from left to right) in~\eqref{S3YTableaux}. The central two
tableaux of~\eqref{S3YTableaux} stem from the same Young diagram. Thus, their corresponding
projection operators project onto equivalent irreducible
representations; there must therefore exist two transition operators
between them. The projection operators $Y_2$ and $Y_3$ are
\begin{equation}
\label{eq:3nonHerm-Youngs}
  Y_2 = \sfrac{4}{3} \; \FPic{3Sym12ASym13} \quad \text{and}
  \quad Y_3 = \sfrac{4}{3} \; \FPic{3Sym13ASym12}
\ ,
\end{equation}
and the transition operators $T_{i j}$ between $Y_i$ and $Y_j$
are
\begin{equation}
  T_{23} 
= Y_2 \rho_{23} 
= \sfrac{4}{3} \; \FPic{3Sym12s23ASym12} 
\quad  \text{and} \quad 
T_{32} 
= Y_3 \rho_{32} 
= \sfrac{4}{3} \; \FPic{3s23Sym12s23ASym12s23}
\end{equation}
in accordance with eq.~\eqref{eq:Ytrans}.  Arranging all projection and transition operators in a matrix $\mathfrak{M}$ where
the diagonal elements $\mathfrak{m}_{ii}$ are projection operators and
each off-diagonal element $\mathfrak{m}_{ij}$ is the transition
operator between $\mathfrak{m}_{ii}$ and $\mathfrak{m}_{ij}$, one obtains the following matrix of
\emph{operators},
\begin{equation}
\label{eq:YMatrixRepAlgS3}
\mathfrak{M}= \;
\begin{pmatrix}
\colorbox{blue!25}{$\FPic{3Sym123SN}$} & 0 & 0 & 0 \\%\\
0 & \colorbox{blue!25}{$\frac{4}{3} \; \FPic{3Sym12ASym13}$} & \frac{4}{3} \; \FPic{3Sym12s23ASym12} & 0 \\%\\
0 & \frac{4}{3} \; \FPic{3s23Sym12s23ASym12s23} & \colorbox{blue!25}{$\frac{4}{3} \; \FPic{3Sym13ASym12}$} & 0 \\%\\
0 & 0 & 0 & \colorbox{blue!25}{$\FPic{3ASym123SN}$}
\end{pmatrix}
\ ,
\end{equation}
where all projection operators are highlighted in blue.  

Notice that the operator in the $1 \times 1$-block in the bottom right
corner contains an antisymmetrizer of length $3$. Thus, for $N\leq 2$,
this operator is a null-operator and only the remaining two blocks are
non-trivial in the above matrix, \emph{c.f.}
app.~\ref{sec:VanishingReps}. If $N\leq 1$, also the central $2 \times
2$-block vanishes as it contains antisymmetrizers of length $2$.

\section{Orthogonal projector bases}\label{sec:Orth-Proj-Basis}

As we have seen in section~\ref{sec:YoungBasis}, the Young projection
and transition operators provide a basis for the algebra of invariants
of $\SUN$ over $\Pow{m}$ provided $m\leq4$. Due to a lack of pairwise
orthogonality and completeness of the Young projection operators
beyond this point, \emph{c.f.} eqs.~\eqref{eq:YoungIntro2}
and~\eqref{eq:Yidpowm}, the Young basis cannot be generalized to
larger $m$. This motivates a basis in terms of \emph{Hermitian} Young
projection operators, since these are orthogonal and complete for all
values of $m$~\cite{Keppeler:2013yla}.

We re-state the most important aspects of Hermitian Young projection
operators in section~\ref{sec:Herm-Proj-Ops}, before discussing
transition operators between Hermitian projections in
sections~\ref{sec:ClebschBasis} (in terms of Clebsch-Gordan operators)
and~\ref{sec:HermitianYoungProjectorsSection} (between Hermitian Young
projection operators). 

Section~\ref{sec:MultiplicationTable} discusses the mutliplication
table of the basis of the algebra of invariants of $\SUN$ over
$\Pow{m}$ in terms of Hermitian projectors and their corresponding
transition operators.

\subsection{Hermitean projection operators} 
\label{sec:Herm-Proj-Ops}

If we replace the Young projectors $Y_{\Theta}$ by their (more complicated)
Hermitian counterparts $P_\Theta$, either following Keppeler and
Sjödahl~\cite{Keppeler:2013yla,Sjodahl:2013hra,Keppeler:2012ih} or our
own improved versions
thereof~\cite{Alcock-Zeilinger:2016bss,Alcock-Zeilinger:2016sxc}, the group theoretically
important features of Young projection operators now apply \emph{for all
values of $m$}~\cite{Keppeler:2013yla}:\footnote{The constructions used in this
  paper are summarized in section~\ref{sec:Three-Hermitian-Ops}.}

\begin{subequations}
  \label{eq:PTmult}
\begin{enumerate}
\item\label{itm:PTproj} The Hermitian Young projection operators are
  \emph{idempotent} and \emph{mutually orthogonal as projectors}: for
  any two Young tableaux $\Theta$ and $\Phi$ in $\mathcal{Y}_m$ they
  satisfy
  \begin{equation}
    \label{eq:PTproj}
    P_\Theta\cdot P_\Theta= \delta_{\Theta\Phi} P_\Theta
\hspace{1cm}
\text{for all $m$}
    \ .
  \end{equation}
\item They provide a complete decomposition of unity on $\Pow{m}$
\begin{equation}
    \label{eq:Pidpowm}
    \sum\limits_{\Theta\in\mathcal Y_m} P_\Theta = \text{id}_{\Pow{m}}
\hspace{1cm}
\text{for all $m$}
  \end{equation}
  into irreducible representations.
\item\label{itm:PTherm}  Unlike their Young counterparts, they \emph{are} Hermitian
  \begin{equation}
    \label{eq:PTherm}
    P_\Theta^\dagger = P_\Theta
    \ .
  \end{equation}
\end{enumerate}
Due to this new property (Hermiticity) several new features
appear~\cite{Alcock-Zeilinger:2016sxc}:
\ytableausetup{mathmode, boxsize=1.3em}
\begin{enumerate}\setcounter{enumi}{3}
\item\label{itm:Pdecomp} The projectors in the descendant set
  $\lbrace\Theta\otimes\ybox{m}\rbrace$ of any
  $\Theta\in\mathcal Y_{m-1}$ sum to the parent projector thus
  augmenting the single identity~\eqref{eq:Pidpowm} by a whole nested
  set of inclusion sums or partial completness
  statements~\cite{Alcock-Zeilinger:2016sxc}:
\ytableausetup{mathmode, boxsize=1em}
  \begin{equation}
    \label{eq:Pdecomp}
\sum
\limits_{\Phi\in \lbrace\Theta\otimes\ybox{m}\rbrace} 
P_\Phi 
= 
P_\Theta
\ .
  \end{equation}
  This can be generalized further,
  \begin{equation}
\sum
\limits_{\Phi\in
\lbrace\Theta\otimes\ybox{k}\otimes\cdots \otimes\ybox{m}\rbrace
}  
P_\Phi 
= 
P_\Theta
\hspace{1cm} 
\text{for $\Theta\in\mathcal{Y}_{k-1}$, $\Phi\in\mathcal{Y}_m$
  and $k<m$}
\ .
  \end{equation}
\item Unlike their conventional Young counterparts $Y_\Theta$ (or the
  corrected Littlewood-Young operators $L_\Theta$,
  see~\cite{Alcock-Zeilinger:2016sxc, Littlewood:1950}), the Hermitian
  Young projectors $P_\Theta$ are automatically orthogonal with
  respect to the scalar product on $\API{\SUN,\Pow{m}}$
  \begin{equation}
    \label{eq:PTorth}
    \langle P_\Theta , P_\Phi \rangle = \Tr{P^\dagger_\Theta P_\Phi}
    \xlongequal{\eqref{eq:PTherm}}
    \Tr{P_\Theta P_\Phi}
    \xlongequal{\eqref{eq:PTproj}}
    0 \hspace{1cm}\text{ for $\Theta\neq\Phi$}
  \end{equation}
\end{enumerate}
\end{subequations}

Both of these properties hinge crucially on Hermiticity -- standard
(Littlewood-) Young projectors do not share them.

The most direct way to achieve this is in terms of Clebsch-Gordan
operators, and this will immediately ensure that operators
satisfying~\eqref{eq:PTmult} exist. This method remains
computationally expensive and keeping $N$ a parameter appears a
hopeless task, sec.~\ref{sec:ClebschBasis}. In
section~\ref{sec:HermitianYoungProjectorsSection} we give an effective
construction of transition operators for Hermitian Young projection
operators.  \ytableausetup {mathmode, boxsize=1em}

\subsection{A full orthogonal basis for
  \texorpdfstring{$\API{\SUN,\Pow{m}}$}{API(SU(N)Vm)} via
  Clebsch-Gordan operators}\label{sec:ClebschBasis}

To make this explicit, consider a general Clebsch-Gordan operator
$C_{\lambda\kappa;j_1 m_1\ldots j_n m_n}$ that implements the
projection and basis change from a product of irreducible
representations labelled by $j_1, \ldots, j_n$ (with states labeled by
$m_1,\ldots,m_n$) into an irreducible representation labelled by
$\lambda$ (where, of course, $\lambda$ stands in for a tableau $\Theta$
and with states labeled by $\kappa$)~\cite{Tung:1985na}
\begin{equation}
  \label{eq:Clebsch1}
  C_{\lambda\kappa;j_1 m_1\ldots j_n m_n} 
  = 
  \ket{\lambda,\kappa}
  \underbrace{\langle\lambda,\kappa\vert j_1, m_1\rangle
    \ket{j_2, m_2}\ldots\ket{j_n,m_n}}_{    \mathcal{C}_{\lambda\kappa;j_1 m_1\ldots j_n m_n} }  \bra{j_1, m_1}\bra{j_2, m_2}\ldots\bra{j_n, m_n} 
  \; =: \; 
  \scalebox{0.75}{\FPic{mqnqbClebschKappaRepLabels}
    \FPic{nqClebschLambda}
    \FPic{nqClebschjmRightLabels}}
\ ,
\end{equation}
the part marked as $\mathcal{C}_{\lambda\kappa;j_1 m_1\ldots j_n m_n}$
is the usual Clebsch-Gordan coefficient and the diagram on the right is
the birdtrack representation of this operator
(c.f.~\cite{Cvitanovic:2008zz}). 
Since we are interested only in
products of the fundamental representation acting on $\Pow{n}$ (so
that the $j_i$ all refer to this one representation) we can suppress
the corresponding label, but we must retain $\lambda$ to reference a
specific irreducible representation contained in this
product. Accordingly, we simplify notation in the birdtrack spirit by
removing the redundant indices according to
\begin{equation}
  \label{eq:clebsch-simplify}
  C_{\lambda\kappa;j_1 m_1\ldots j_n m_n} 
  = 
  \scalebox{0.75}{\FPic{mqnqbClebschKappaRepLabels}
    \FPic{nqClebschLambda}
    \FPic{nqClebschjmRightLabels}
  } 
  \to
  C_{\lambda,n} :=\scalebox{0.75}{    \FPic{nqClebschLambda}
     }
\ .
\end{equation}
By its very nature, the Clebsch-Gordan operator translates a product representation into its irreducib le sub-blocks labeled by $\lambda$, i.e.
\begin{align}
  \label{eq:transl-to-lambda}
   \scalebox{0.75}{
    \FPic{nqClebschLambda}\FPic{nqClebschUDagLabels}
  } 
  =
   U_{(\lambda)}^\dagger\scalebox{0.75}{
    \FPic{nqClebschLambda}
  } 
\end{align}
for all $U\in\SUN$ and $U_{(\lambda)}$ in the $\lambda$ representation of $\SUN$.
Orthonormality of the new states,
\begin{equation}
  \label{eq:ClebschOn}
  \scalebox{0.75}{\FPic{mqnqbClebschKappaRepLabels}\FPic{nqClebschLambdaLambdapOn}\FPic{mqnqbClebschKappapRepLabels}} = \delta_{\lambda,\lambda'} \delta_{\kappa,\kappa'}
\ ,
\end{equation}
allows us to cast projection operators in the form
\begin{equation}
  \label{eq:Clebsch2}
P_\lambda := \sum\limits_\kappa \ket{\lambda,\kappa}\bra{\lambda,\kappa}
= 
C_{\lambda,n}^{\dagger} \cdot C_{\lambda,n} 
\quad = \quad 
\scalebox{0.75}{\FPic{nqClebschProjOpsLambda}}
\end{equation}
which are clearly mutually orthogonal,
\begin{equation}
  \label{eq:Clebsch-Proj-Ops-ON}
P_\lambda  P_{\lambda'}
= \lambda_{\lambda,\lambda'} P_\lambda
\ .
\end{equation}
Equation~\eqref{eq:Clebsch2} also introduces the birdtrack notation of
$C_{\lambda,n}^{\dagger}$, the Hermitian conjugate of $C_{\lambda,n}$
in Eq.~\eqref{eq:Clebsch1}.  The $P_\lambda$ are mutually orthogonal
elements of the algebra of primitive invariants $\API{\SUN,\Pow{n}}$
due to eq.~\eqref{eq:transl-to-lambda}:
\begin{align}
  \label{eq:proj-PI-via-Clebsch-inv} 
  \scalebox{0.75}{
    \FPic{nqClebschULabels}
    \FPic{nqClebschProjOpsLambda}
    \FPic{nqClebschUDagLabels}
}
=
  \scalebox{0.75}{$
    \FPic{nqClebschLambdaDag} U_{(\lambda)} U_{(\lambda)}^\dagger\FPic{nqClebschLambda}$
  }
  =
  \scalebox{0.75}{
\FPic{nqClebschProjOpsLambda}
}
\end{align}
and general theory assures us that these yield projectors on
all irreducible subspaces contained in $\Pow{n}$~\cite{Tung:1985na}. 

From the perspective of Clebsch-Gordan operators there are obvious
candidates for transition operators: When two equivalent
representations $\lambda$ and $\lambda'$ are isomorphic, one can
choose the states $\ket{\lambda,\kappa}$ and $\ket{\lambda',\kappa}$
such that the representation matrices are identical, $U_{(\lambda')} =
U_{(\lambda )}$. This allows us to identify the transition operators
\begin{align}
  \label{eq:trans-op-def}
  T_{\lambda'\lambda} 
  := 
  \sum\limits_\kappa \ket{\lambda',\kappa}\bra{\lambda,\kappa}
  = 
  C_{\lambda',n}^{\dagger} \cdot C_{\lambda,n} 
\quad = \quad 
\scalebox{0.75}{\FPic{nqClebschProjOpsLambdaLambdap}}
\end{align}
as additional algebra elements, since, with this particular basis
choice, they also are invariant:
\begin{align}
  \label{eq:trans-op-inv} 
  \scalebox{0.75}{
    \FPic{nqClebschULabels}
    \FPic{nqClebschProjOpsLambdaLambdap}
    \FPic{nqClebschUDagLabels}
}
=
  \scalebox{0.75}{$
    \FPic{nqClebschLambdapDag} U_{(\lambda')} U_{(\lambda)}^\dagger\FPic{nqClebschLambda}$
  }
  \xlongequal{U_{(\lambda')} = U_{(\lambda)}}
  \scalebox{0.75}{
\FPic{nqClebschProjOpsLambdaLambdap}
}
\ .
\end{align}
Unlike the projection operators $P_\lambda$ the transition operators
$T_{\lambda'\lambda}$ are clearly not Hermitian. From their definition
in terms of states~\eqref{eq:trans-op-def} it however follows that
they are unitary (on the subspaces corresponding to $\lambda$ and $\lambda'$),
\begin{align}
  \label{eq:T-not-herm}
  (T_{\lambda'\lambda})^\dagger = T_{\lambda\lambda'}
  \ .
\end{align}
These operators in fact define the isomorphisms that in the standard
perspective allow us to claim equivalence between the two
representations in the first place. Our point here is that these
isomorphisms are elements of $\API{\SUN,\Pow{n}}$.

The totality of all projection and transition
operators~\eqref{eq:Clebsch2} and~\eqref{eq:trans-op-def} obviously
exhausts the algebra of invariants due to the completeness of
Clebsch-Gordan operators; the transition
operators~\eqref{eq:trans-op-def} provide \emph{all} the missing basis
elements, which fixes their total number. This matches with the
counting arguments of section~\ref{sec:YoungBasis} and seamlessly fits
into the multiplication pattern of closed subalgebras discussed in
sec.~\ref{sec:MultiplicationTable}.

Note that this procedure leads to a particular realization of
projectors and associated transition operators, any other equivalent
construction may produce results that differ by a similarity
transformation as discussed in
section~\ref{sec:MultiplicationTable}.\footnote{We will see that this
  similarity transform leaves the block-structure of the associated
  matrix $\mathfrak{M}$ invariant, sec.~\ref{sec:MultiplicationTable} and
refined below in eq.~\eqref{eq:Pab-transf-conj-constr}.} As a means to
find a basis in a practical calculation this procedure is exceedingly
inefficient: It relies on finding a total of $N^n$ normalized states
in $\Pow{n}$ as a stepping stone to produce $n!$ basis states for
$\API{\SUN,\Pow{n}}$ while keeping $N\ge n$ to avoid dimensional
zeroes.\footnote{This forces us into the domain where $N^n \ge n^n >
  n!$. There will always be more states than multiplets.} It clearly is not the most efficient option to achieve this
goal, in particular if one aims to keep $N$ as a parameter. Therefore
we use the Clebsch-Gordan method as a proof of concept and abstract
the main features of the resulting basis as the goalposts for a more
efficient construction to be presented in
sec.~\ref{sec:HermitianYoungProjectorsSection}.

We observe:
\begin{enumerate}
\item  $T_{\Theta \Phi}$, as a map from $\Pow{n}$ to $\Pow{n}$, projects
  onto the image of $P_{\Phi}$ and maps that surjectively onto the
  image of $P_{\Theta}$,
    \begin{equation}
      T_{\Theta \Phi}P_{\Phi} = T_{\Theta
      \Phi} = P_{\Theta}T_{\Theta \Phi},
    \end{equation}
    It thus can be considered a map from the image of $P_{\Phi}$,
    $P_{\Phi}\left(\Pow{n}\right)$, to the image of $P_{\Theta}$,
    $P_{\Theta}\left(\Pow{n}\right)$.
  \item \label{itmTransitionElementDef2}
    $T_{\Theta\Phi}^{\dagger}$
    is the right inverse of $T_{\Theta\Phi}$ on $P_{\Theta}\left(\Pow{n}\right)$,
    \begin{equation}
      T_{\Theta \Phi}T_{\Theta
      \Phi}^{\dagger} = P_{\Theta}.
    \end{equation}
  \item \label{itmTransitionElementDef3}
    $T_{\Theta\Phi}^{\dagger}$
    is the left inverse of $T_{\Theta\Phi}$ on $P_{\Phi}\left(\Pow{n}\right)$,
    \begin{equation}
      T_{\Theta \Phi}^{\dagger}T_{\Theta
      \Phi} = P_{\Phi}.
    \end{equation}
  \end{enumerate}

  We see that $T_{\Theta\Phi}$ maps the image $P_{\Phi}(\Pow{n})$
  bijectlively onto $P_{\Theta}(\Pow{n})$ -- the inverse is
  $T_{\Phi\Theta}=T_{\Theta\Phi}^\dagger$. The $T_{\Theta\Phi}$ are
  \emph{unitary transition operators} between the irreducible
  representations.

  These properties are sufficient to uniquely characterize the
  $T_{\Theta\Phi}$. The argument for uniqueness follows a similar
  pattern as the uniqueness proof for inverses in a group.

  Alternatively one may cast the
  statements~\ref{itmTransitionElementDef2}
  and~\ref{itmTransitionElementDef3} as
  \begin{enumerate}
 \item[2'.]   $T_{\Theta\Phi}^\dagger = T_{\Phi\Theta}$ 
 \item[3'.]   $T_{\Theta\Phi} T_{\Phi\Theta} = P_\Theta$
  \end{enumerate}
This is the form we use in the definition:
\begin{definition}[unitary transition operators]\label{thm:TransitionElementDef}
  Let $\Theta, \Phi \in \mathcal{Y}_n$ be two Young tableaux with
  the same underlying Young diagram, and let $P_{\Theta}$ and
  $P_{\Phi}$ be their respective Hermitian Young projection
  operators. Then, the operator $T_{\Theta \Phi}$ satisfying the
  following three properties is called the \emph{transition operator}
  between $P_{\Theta}$ and $P_{\Phi}$.
  \begin{subequations}
    \label{eq:unitary-trans-LR}
    \begin{align}
      \label{eq:TP}
      & 
      T_{\Theta \Phi}P_{\Phi} 
      = T_{\Theta\Phi} = P_{\Theta}T_{\Theta \Phi} 
      \\
      \label{eq:Tunitary}
      & T_{\Theta\Phi}^\dagger = T_{\Phi\Theta} \\
      \label{eq:TTinv}
     &  T_{\Theta\Phi} T_{\Phi\Theta} = P_\Theta
    \end{align}
  \end{subequations}
\end{definition}

\subsection{The multiplication table of the algebra of
  invariants}\label{sec:MultiplicationTable}

If we look at a given Young diagram $\mathbf{Y}_i$, then the set of
all projection and transition operators corresponding to tableaux with
shape $\mathbf{Y}_i$, $\mathfrak{S}_{\mathbf{Y}_i}$,
forms a closed subalgebra of $\API{\SUN,\Pow{m}}$. Its multiplication
table is given by eqns.~\eqref{eq:Clebsch-Proj-Ops-ON} and~\eqref{eq:unitary-trans-LR} and
evidently decouples from the rest of the algebra. A simple relabelling
allows to condense these equations into a single one (see
eq.~\eqref{eq:MultipletMultTable} below). To do so, form a matrix
pattern in which the projection operators are placed on the diagonals,
such that
\begin{subequations}
\label{eq:Mdef}
\begin{equation}
  \label{eq:Mdiag}
  \mathfrak{m}_{i i} = P_{\Theta_i} \hspace{1cm}\text{ for all
    $\Theta_i\in\mathcal{Y}_m$ with underlying diagram $\mathbf{Y}_i$}
\end{equation}
and poplulate the off-diagonal sites with the transition operators, such that
\begin{equation}
  \label{eq:Moffdiag}
  \mathfrak{m}_{i j} = T_{\Theta_i \Theta_j}  \hspace{1cm}\text{ for
    all $\Theta_{i},\Theta_{j}\in\mathcal{Y}_m$  with underlying
    diagram $\mathbf{Y}_i$}
\ .
\end{equation}
\end{subequations}
Calling the matrix of elements for this subalgebra
$\mathfrak{M}_{\mathbf{Y}_i}$ we can then assemble all such blocks
into a matrix pattern of indepdendent closed subalgebras by placing
the blocks along the diagonal while filling the remainder with
zeroes (this makes sense since the ``transition operators'' between
projectors belonging to different blocks should be zero). Schematically we obtain
\begin{subequations}
  \label{eq:blockbasis}
\begin{equation}
  \label{eq:Mblocks}
  \mathfrak{M} 
 = 
  \begin{tikzpicture}[baseline=(current bounding box.east),every node/.style={inner sep=0,outer sep=0}]
  \matrix(mm)[matrix of nodes, nodes in empty cells,row sep=0,column sep=0,inner sep=0mm, outer sep=0,left delimiter  = (,right delimiter =)] {
     {\fblockmatrix[0.8,1.0,0.8]{1.1cm}{1.1cm}{}} & & & &\\
    & {\fblockmatrix[0.8,1.0,0.8]{.7cm}{.7cm}{}} & & & &\\
    & &  {$\phantom{\ddots}$} & & &\\
    & & &  {\fblockmatrix[0.8,1.0,0.8]{1cm}{1cm}{}} & & \\
    & & & &  {$\phantom{\ddots}$} & \\
    & & & & &   {\fblockmatrix[0.8,1.0,0.8]{.9cm}{.9cm}{}}\\
    };
\draw[loosely dotted, thick ] (mm-2-2.south east) -- (mm-4-4.north west); 
\draw[loosely dotted, thick ] (mm-4-4.south east) -- (mm-6-6.north west); 

\node at (mm-1-1) {\resizebox{.6cm}{!}{$\mathfrak{M}_{\mathbf{Y}_1}$}};
\node at (mm-2-2) {\resizebox{.6cm}{!}{$\mathfrak{M}_{\mathbf{Y}_2}$}};
\node at (mm-4-4) {\resizebox{.6cm}{!}{$\mathfrak{M}_{\mathbf{Y}_i}$}};
\node at (mm-6-6) {\resizebox{.6cm}{!}{$\mathfrak{M}_{\mathbf{Y}_k}$}};
\end{tikzpicture}
 \ .
\end{equation}
This matrix once again illustrates the fact that the sum of all
projection operators and transition operators of $\SUN$ must be a sum
of squares (\emph{c.f.} eq.~\eqref{eq:FullAlgebra0}), as $\mathfrak{M}$ is block diagonal.  It is clear that
the matrix elements $\mathfrak{m}_{i j}$ of $\mathfrak{M}$
in~\eqref{eq:Mblocks} satisfy the property
\begin{equation}
  \label{eq:MultipletMultTable}
   \mathfrak{m}_{i j} \mathfrak{m}_{kl} = \delta_{jk} \mathfrak{m}_{i l}
   \ ,
\end{equation}
Eq.~\eqref{eq:MultipletMultTable} (together with the statement which
of the $\mathfrak{m}_{i j}$ are zero) is probably the most compact form
to encode both eqns. ~\eqref{eq:Clebsch-Proj-Ops-ON}
and~\eqref{eq:unitary-trans-LR} simultaneously.  In the new notation,
we have
\begin{equation}
  \label{eq:SM-span-ij}
  \API{\SUN,\Pow{m}} 
  = 
  \Bigl\{ \alpha_{i j} \mathfrak{m}_{i j} |
  \alpha_{i j}\in\mathbb R, \mathfrak{m}_{i j}\in\mathfrak{S}_m 
  \Bigr\}
\end{equation}
where the sum is over all $i, j \in \{1,\ldots,m!\}$ and
$\mathfrak{S}_m$ once again denotes the set of all projection and
transition operators corresponding to the irreducible representations
of $\SUN$ over $\Pow{m}$.  This basis has the advantage that
dimensional zeroes manifest themselves directly as zeroes of the basis
elements: if a dimensional zero appears at $N < m$ it affects whole
equivalence blocks. All operators in the block turn into null-operators
simultaneously, \emph{c.f.}  appendix~\ref{sec:VanishingReps} or
section~\ref{sec:Herm-Algebra-Example} for an explicit example. No
\emph{additional} dimensional zeroes can arise from linear
combinations of the remaining basis elements.
\end{subequations}

In particular, the product~\eqref{eq:MultipletMultTable} yields a
non-zero result only if either $i=j=k=l$ (squaring a projection
operator), or if we multiply a diagonal element with an off-diagonal
element of the same block in the correct order. In
fact,~\eqref{eq:MultipletMultTable} is the structure of the
multiplication table of the multiplet basis for $\API{\SUN,\Pow{m}}$,
and even for $\API{\SUN,\MixedPow{m}{m'}}$.

\begin{sloppypar}
  The multiplication table~\eqref{eq:MultipletMultTable} is clearly
  more structured than that of the primitive invariant basis of
  $\API{\SUN,\Pow{m}}$, which is directly the multiplication table of
  $S_m$:
\begin{equation}
  \label{eq:PrimitiveMultTable}
  \rho_i \rho_j = \tensor{A}{^k_{i j}} \rho_k
\ .
\end{equation}
This has consequences: The simpler structure
of~\eqref{eq:MultipletMultTable} also gives access to the uniqueness
of the operators $\mathfrak{m}_{i j}$ appearing in it. While the types
and equivalence patterns (the block structure) of irreducible
representations contained in $\API{\SUN,\Pow{m}}$, or
$\API{\SUN,\MixedPow{m}{m'}}$ are uniquely determined by $N$, $m$ and
$m'$ the operators themselves are not uniquely determined by the
multiplication table and decomposition of unity reqirements alone, if
the size of the block it falls into is greater than
one.
\end{sloppypar}

This can be seen as follows: Since the block structure is fixed, two
realizations $\mathfrak{M}$ and $\Tilde{\mathfrak{M}}$ of bases with
the same block structure must satisfy
\begin{align}
  \label{eq:mult-tables}
  \mathfrak{m}_{i j} \mathfrak{m}_{kl} = \delta_{j k} \mathfrak{m}_{i l}
  \ \text{ and } \
  \Tilde{\mathfrak{m}}_{i j} \Tilde{\mathfrak{m}}_{kl} 
  = 
  \delta_{j k} \Tilde{\mathfrak{m}}_{i l}
\end{align}
and be related by a general \emph{real} linear transformation as
\begin{align}
  \label{eq:genlinPab}
  \Tilde{\mathfrak{m}}_{i j} := a_{i \alpha} \mathfrak{m}_{\alpha\beta} b_{\beta j}
\end{align}
This implies that (note that the $\mathfrak{m}$ are \emph{operators}, while the $a_{i j}$ and $b_{i j}$ are real coefficients
 that commute with the $\mathfrak{m}$)
\begin{align}
  \label{eq:Pab-transf-table-constr}
  \Tilde{\mathfrak{m}}_{i j}\Tilde{\mathfrak{m}}_{k l} 
  =  
  a_{i \alpha} \mathfrak{m}_{\alpha\beta} b_{\beta j} a_{k \gamma} \mathfrak{m}_{\gamma\delta} b_{\delta l}
  \xlongequal{\mathfrak{m}_{\alpha\beta}\mathfrak{m}_{\gamma\delta} 
  =\mathfrak{m}_{\alpha\delta} \delta_{\beta\gamma}
   }
  a_{i \alpha}\mathfrak{m}_{\alpha\delta} b_{\delta l} \underbrace{(a_{k\beta} b_{\beta j})}_{\overset{!}{=} \delta_{k j}}
  \overset{!}{=} \delta_{k j}\Tilde{\mathfrak{m}}_{i l}
\ ,
\qquad\text{i.e. $b= a^{-1}$}
\ .
\end{align}
With this constraint $a$ must have the same block structure as both
$\mathfrak{M}$ and $\Tilde{\mathfrak{M}}$.

Everything we said up to this point also holds for a basis of Young
projection and transition operators over $\API{\SUN,\Pow{m}}$ provided
$m\leq4$.\footnote{We have not provided transition operators for the
  Littlewood-Young operators, so that at this point we need to switch
  to Hermitian operators as soon as $m \ge 5$.} The main disadvantage
(besides being restricted to small $m$) of this basis remains that it
is not orthogonal under the standard scalar product on
$\API{\SUN,\Pow{m}}$ provided by $\langle A,B\rangle = \Tr{A^\dagger
  B}$, as is exemplified in~\eqref{eq:not-orth}.

If we populate the subalgebra listing $\mathfrak{M}$ with 
Hermitian projectors and their unitary transition operators results in
an orthogonal basis:
\begin{align}
  \label{eq:herm-unit-orth}
  \langle\mathfrak m_{i j}, \mathfrak m_{k l}\rangle
  =
  \Tr{\mathfrak m_{i j}^\dagger \mathfrak m_{k l}}
  \xlongequal[\eqref{eq:PTherm}]{\eqref{eq:Tunitary}}
  \Tr{\mathfrak m_{j i} \mathfrak m_{k l}}
  \xlongequal{\eqref{eq:MultipletMultTable}}
  \delta_{i k} \Tr{\mathfrak m_{j l}}
  =
  \delta_{i k} \delta_{j l} \text{dim}(\Theta_j)
\ ,
\end{align}
where $\Theta_j$ labels the representation corresponding to the
projection operator $\mathfrak{m}_{j j}$.
Note that this is a general statement based purely on the
multiplication table, Hermiticity and unitarity without any reference
to a specific realization of the basis elements and thus automatically
also applies to the basis we construct in
sec.~\ref{sec:HermitianYoungProjectorsSection}.

Hermiticity and unitarity also restrict the freedom to change the $\mathfrak m_{i j}$ beyond what we had seen in eq.~\eqref{eq:Pab-transf-table-constr}:
Using the Hermiticity properties ($\mathfrak{m}^\dagger_{i j} =
\mathfrak{m}_{j i}$, $\Tilde{\mathfrak{m}}^\dagger_{i j} =
\Tilde{\mathfrak{m}}_{j i}$ and $a_{i j}\in\mathbb R$ since the
algebra is real) then lead to
\begin{align}
  \label{eq:Pab-transf-conj-constr}
   \Tilde{\mathfrak{m}}_{i j} ^\dagger = a_{i \alpha} \mathfrak{m}_{\alpha\beta}^\dagger a^{-1}_{\beta j} 
   = a_{i \alpha} \mathfrak{m}_{\beta\alpha} a^{-1}_{\beta j}
   =  (a^{-1})^t _{j\beta} \mathfrak{m}_{\beta\alpha}  a^t_{\alpha i} \overset{!}{=} \Tilde{\mathfrak{m}}_{j i}
   \hspace{1cm}\text{i.e.}\hspace{1cm}
   a^{-1} = a^t
\ ,
\end{align}
the freedom is restricted to (blockwise!) orthogonal transformations
of the $\mathfrak m_{i j}$.

\section{Unitary transition operators for Hermitian Young projectors}
\label{sec:HermitianYoungProjectorsSection}

Like the (non-unitary) transition operators for Young projectors over
$\Pow{m}$ ($m\leq4$) introduced in sec.~\ref{sec:YoungBasis}, the
unitary transition operators associated with Hermitian Young
projectors are based on the projectors themselves. In the present
case, the building blocks will be a set
$\{P_\Theta|\Theta\in\mathcal Y_m\}$ (with the full list of properties
listed in sec.~\ref{sec:Herm-Proj-Ops}), where we need not put a
restriction on $m$. We first recapitulate their ingredients in
sec.~\ref{sec:Three-Hermitian-Ops} before we use them and the tableau
permutations of Definition~\ref{thm:TableauPermutation-birdtrack} to
construct the transition operators in sec.~\ref{sec:TransitionOps}.

\subsection{Construction methods of Hermitian Young
  projection operators}\label{sec:Three-Hermitian-Ops}

At the present time, there exist three ways of constructing
(completely equivalent) Hermitian
Young projection operators. The first is an iterative method that goes back to Keppeler and
Sj{\"o}dahl (KS) and is discussed in~\cite{Keppeler:2013yla}. The
second method is based on the KS-algorithm but produces substantially
shorter operators~\cite{Alcock-Zeilinger:2016sxc}. The
third method exploits the structure of Young tableaux and their
``lexical
ordering''~\cite{Alcock-Zeilinger:2016sxc}. Since the
second and third method are used in this paper, we summarize these two
construction algorithms in the present section.

In the first method, Hermitian Young projection operators
are constructed by forming a product of consecutively ``older'' Young
projection operators:

\begin{theorem}[staircase form of Hermitian Young
  projectors~\cite{Alcock-Zeilinger:2016sxc}]
\label{thm:ShortKSProjectors}
  Let $\Theta\in\mathcal{Y}_n$ be a Young tableau. Then, the 
  corresponding Hermitian Young projection operator $P_{\Theta}$ is
  given by
  \begin{equation}
    P_{\Theta} = \; Y_{\Theta_{(n-2)}}
    Y_{\Theta_{(n-3)}} Y_{\Theta_{(n-4)}} \ldots
    Y_{\Theta_{(2)}} Y_{\Theta_{(1)}} Y_{\Theta} \;
    Y_{\Theta_{(1)}} Y_{\Theta_{(2)}} \ldots
    Y_{\Theta_{(n-4)}} Y_{\Theta_{(n-3)}} Y_{\Theta_{(n-2)}}.
  \end{equation}
\end{theorem}

In the second method, one takes into account the \emph{lexical
  ordering} of the Young tableau. In order to accomplish this we
require a few more definitions.

\begin{definition}[column- \& row-words and lexical ordering]
  \label{TableauxWords}
  Let $\Theta \in \mathcal{Y}_n$ be a Young tableau. We define the
  \emph{column-word} of $\Theta$, $\mathfrak{C}_{\Theta}$, to be the
  column vector whose entries are the entries of $\Theta$ as read
  column-wise from left to right. Similarly, the \emph{row-word} of
  $\Theta$, $\mathfrak{R}_{\Theta}$, is defined to be the row vector
  whose entries are those of $\Theta$ read row-wise from top to
  bottom.

We say that a tableau $\Theta$ is \emph{(lexically) ordered} if either its
row-word or its column-word (or both) is in lexical order. If we want to
be more specific, we might call $\Theta$ row-ordered resp. column-ordered.
\end{definition}
\ytableausetup{mathmode, boxsize=normal}

 For example, the tableau
\begin{equation}
\label{eq:ExWords1}
  \Phi :=
  \begin{ytableau}
    1 & 5 & 7 & 9 \\
    2 & 6 & 8 \\
    3 \\
    4
  \end{ytableau}
\end{equation}
has a column-word
\begin{equation}
 \mathfrak{C}_{\Phi} = (1,2,3,4,5,6,7,8,9)^t,
\end{equation}
and a row-word
\begin{equation}
  \mathfrak{R}_{\Phi} = (1,5,7,9,2,6,8,3,4).
\end{equation}
Since $\mathfrak{C}_{\Phi}$ is lexically ordered, we say that $\Phi$
is a (column-) ordered tableau.

It should be noted that the above definition of the row-word is
\emph{different} to the definition given in the standard literature
such as~\cite{Sagan:2000,Fulton:1997} (there, the row word is read
from bottom to top rather than from top to bottom). However, for the
purposes of this paper, the above given definition is more useful than
the standard definition.

\begin{definition}[measure of lexical disorder (MOLD)]\label{MOLDDef}
  Let $\Theta \in \mathcal{Y}_n$ be a Young tableau. We define its
  \emph{Measure Of Lexical Disorder} (MOLD) to be the smallest natural number
  $\mathcal{M}(\Theta) \in \mathbb{N}$ such that
  \begin{equation}
    \Theta_{\left(\mathcal{M}(\Theta)\right)} = \pi^{\mathcal{M}(\Theta)} \left(\Theta\right)
  \end{equation}
is a lexically ordered tableau. (Recall from Definition~\ref{ParentMap} that
$\pi^{\mathcal{M}(\Theta)}$ refers to $\mathcal{M}(\Theta)$ consecutive
applications of the parent map $\pi$ to the tableau $\Theta$.)
\end{definition}
We note that the MOLD of a Young tableau is a well-defined quantity, since one
will always eventually arrive at a lexically ordered tableau, as, for
example, all tableaux in $\mathcal{Y}_3$ are lexically ordered. This
then implies that the MOLD of a tableau $\Theta \in
\mathcal{Y}_n$ has an upper bound,
\begin{equation}
\label{eq:MOLDUpperBound}
\mathcal{M}(\Theta) \leq n-3,
\end{equation}
making it a well-defined quantity. As an example, consider the tableau
\begin{equation}
\Phi :=
  \begin{ytableau}
    1 & 2 & 4 \\
    3 & 5
  \end{ytableau}
\ .
\end{equation}
The MOLD of the above tableau is $\mathcal{M}(\Theta)=2$, one needs to
apply $\pi$ twice to arrive at the first lexically ordered ancestor,
which in this case is row ordered:
\begin{equation}
  \underbrace{\begin{ytableau}
    1 & 2 & *(cyan!20) 4 \\
    3 & *(cyan!40) 5
  \end{ytableau}}_{\Phi} \quad \xlongrightarrow[]{\pi} \quad 
  \underbrace{\begin{ytableau}
    1 & 2 & *(cyan!20) 4 \\
    3 
  \end{ytableau}}_{\Phi_{(1)}} \quad \xlongrightarrow[]{\pi} \quad 
  \underbrace{\begin{ytableau}
    1 & 2 \\
    3 
  \end{ytableau}}_{\Phi_{(2)}}
\ .
\end{equation}
The following construction algorithm of Hermitian Young projection
operators uses the MOLD of the corresponding Young tableau.

\begin{theorem}[MOLD operators~\cite{Alcock-Zeilinger:2016sxc}]
\label{thm:MOLDConstruction}
  Consider a Young tableau $\Theta  \in
  \mathcal{Y}_n$ with MOLD $\mathcal{M}(\Theta)=m$. Furthermore, suppose that $\Theta_{(m)}$
  has a lexically ordered \emph{row-word}. Then, the Hermitian Young
  projection operator corresponding to $\Theta$, $P_{\Theta}$, is,
  \emph{for $m$ even},
  \begin{subequations}
    \begin{equation}\label{eq:MOLDHermitianOperatorConstruction1}
    P_{\Theta} \; = \beta_{\Theta} \cdot\mathbf{S}_{\Theta_{(m)}}
     \; \mathbf{A}_{\Theta_{(m-1)}}  \; \mathbf{S}_{\Theta_{(m-2)}}  \; \ldots
     \; \mathbf{S}_{\Theta_{(2)}}  \; \mathbf{A}_{\Theta_{(1)}}
     \; \colorbox{red!20}{$\bar{Y}_{\Theta} \bar{Y}_{\Theta}^\dagger$}  \; \mathbf{A}_{\Theta_{(1)}}  \; \mathbf{S}_{\Theta_{(2)}}  \; \ldots
     \; \mathbf{S}_{\Theta_{(m-2)}}  \; \mathbf{A}_{\Theta_{(m-1)}}
     \; \mathbf{S}_{\Theta_{(m)}},
\end{equation}
and, \emph{for $m$ odd},
\begin{equation}\label{eq:MOLDHermitianOperatorConstruction2}
P_{\Theta} \; = \beta_{\Theta} \cdot\mathbf{S}_{\Theta_{(m)}}
     \; \mathbf{A}_{\Theta_{(m-1)}}  \; \mathbf{S}_{\Theta_{(m-2)}}  \; \ldots
     \; \mathbf{A}_{\Theta_{(2)}}  \; \mathbf{S}_{\Theta_{(1)}}  \; 
    \colorbox{red!20}{$\bar{Y}_{\Theta}^\dagger \bar{Y}_{\Theta}$}
     \; \mathbf{S}_{\Theta_{(1)}}  \; \mathbf{A}_{\Theta_{(2)}}  \;
     \ldots \; 
     \; \mathbf{S}_{\Theta_{(m-2)}} \; \mathbf{A}_{\Theta_{(m-1)}}
     \; \mathbf{S}_{\Theta_{(m)}}.
  \end{equation}
Similarly, if $\Theta_{(m)}$ has a lexically ordered \emph{column-word},
$P_{\Theta}$ is given by, \emph{for $m$ even},
 \begin{equation}\label{eq:MOLDHermitianOperatorConstruction3}
    P_{\Theta} \; = \beta_{\Theta} \cdot\mathbf{A}_{\Theta_{(m)}}
     \; \mathbf{S}_{\Theta_{(m-1)}}  \; \mathbf{A}_{\Theta_{(m-2)}}  \; \ldots
     \; \mathbf{A}_{\Theta_{(2)}}  \; \mathbf{S}_{\Theta_{(1)}}
     \; \colorbox{red!20}{$\bar{Y}_{\Theta}^{\dagger} \bar{Y}_{\Theta}$}  \; \mathbf{S}_{\Theta_{(1)}}  \; \mathbf{A}_{\Theta_{(2)}}  \; \ldots
     \; \mathbf{A}_{\Theta_{(m-2)}}  \; \mathbf{S}_{\Theta_{(m-1)}}
     \; \mathbf{A}_{\Theta_{(m)}},
\end{equation}
and, \emph{for $m$ odd},
\begin{equation}\label{eq:MOLDHermitianOperatorConstruction4}
P_{\Theta} \; = \beta_{\Theta} \cdot\mathbf{A}_{\Theta_{(m)}} \; 
    \mathbf{S}_{\Theta_{(m-1)}}  \; \mathbf{A}_{\Theta_{(m-2)}}  \; \ldots
     \; \mathbf{S}_{\Theta_{(2)}}  \; \mathbf{A}_{\Theta_{(1)}} \; 
    \colorbox{red!20}{$\bar{Y}_{\Theta} \bar{Y}_{\Theta}^\dagger$}
     \; \mathbf{A}_{\Theta_{(1)}} \; \mathbf{S}_{\Theta_{(2)}}  \; \ldots \; 
     \; \mathbf{A}_{\Theta_{(m-2)}}  \; \mathbf{S}_{\Theta_{(m-1)}}
     \; \mathbf{A}_{\Theta_{(m)}}.
  \end{equation}
\end{subequations}
In the above, all symmetrizers and antisymmetrizers are understood to
be canonically embedded into the algebra over $V^{\otimes n}$;
$\beta_{\Theta}$ is a \emph{non-zero constant} chosen such that
$P_{\Theta}$ is idempotent.
\end{theorem}

This construction seems complicated at first glance as four cases need
to be considered. In~\cite{Alcock-Zeilinger:2016sxc} we
discuss why this is necessary and how the structure of
Theorem~\ref{thm:MOLDConstruction} can be understood.

\subsection{Unitary transition operators for Hermitian Young projectors}\label{sec:TransitionOps}

For the standard Young projection operators, the tableau permutation
$\rho_{\Theta\Phi}$, viewed as an element of $\Lin{\Pow{m}}$,
directly relates any associated Young projectors:
\begin{equation}
  Y_{\Theta} = \rho_{\Theta\Phi} Y_{\Phi}
  \rho_{\Theta\Phi}^{-1} 
\ ,
\end{equation}
\emph{c.f.} eq.~\eqref{eq:permute-tableaux2}.  For Hermitian Young
projection operators, this is no longer true in general: There exist
tableaux $\Theta$ and $\Phi$ such that
\begin{equation}
  \label{eq:permute-tableaux3}
  P_{\Theta} \neq \rho_{\Theta\Phi} P_{\Phi}
  \rho_{\Theta\Phi}^{-1} 
\ .
\end{equation}
The simplest example for such a mismatch is probably the equivalence
pair corresponding to the Young tableaux from eq.~\eqref{eq:Equiv3quarks}
\begin{equation}
  \Theta :=
  \begin{ytableau}
    1 & 2 \\
    3
  \end{ytableau}
\qquad \text{and} \qquad \Phi :=
\begin{ytableau}
  1 & 3 \\
  2
\end{ytableau}
\end{equation}
with
\begin{equation}
  \label{eq:young3-mixed}
  Y_{\Theta} = \sfrac{4}{3}\cdot\FPic{3Sym12ASym13} \qquad
  \text{and} \qquad Y_{\Phi} = \sfrac{4}{3}\cdot\FPic{3Sym13ASym12}
\end{equation}
and
\begin{equation}
  \label{eq:3-mixed-proj} 
  P_{\Theta} = \sfrac{4}{3}\cdot\FPic{3Sym12ASym23Sym12} \qquad
  \text{and} \qquad P_{\Phi} = \sfrac{4}{3}\cdot\FPic{3ASym12Sym23ASym12}
\end{equation}
respectively. We recall the associated tableau permutation from
eq.~\eqref{eq:3-Rho-ThetaThetap}: $\rho_{\Theta \Phi}
=\FPic{3s23SN}$. Evidently
\begin{align}
  \label{eq:Pnotequiv}
  \sfrac{4}{3}\cdot\FPic{3Sym12ASym13}  = \sfrac{4}{3}\cdot\FPic{3s23N}\FPic{3Sym13ASym12}\FPic{3s23N}
\quad \text{ while } \quad
  \sfrac{4}{3}\cdot\FPic{3Sym12ASym23Sym12} \neq \sfrac{4}{3}\cdot\FPic{3s23N}\FPic{3ASym12Sym23ASym12}\FPic{3s23N}
\end{align}
as claimed.

However, what remains
true is that
\begin{equation}
  \label{eq:permute-tableaux4}
  P_{\Theta} \cdot \rho_{\Theta\Phi} P_{\Phi}
  \rho_{\Theta\Phi}^{-1} \neq 0
\ ,
\end{equation}
since all symmetrizers and anti-symmetrizers in
$\rho_{\Theta\Phi}P_{\Phi}\rho_{\Theta\Phi}^{-1}$ can be absorbed into
$\mathbf{S}_{\Theta}$ and $\mathbf{A}_{\Theta}$ respectively, see
part~\ref{itm:nonzero-O-2} (eq.~\eqref{eq:CancelWedgedParent2}) in
section~\ref{sec:CancellationRules}. The fact that
$P_{\Theta} \cdot \rho_{\Theta\Phi} P_{\Phi}
\rho_{\Theta\Phi}^{-1}\neq 0$
in eq.~\eqref{eq:permute-tableaux4} is the main ingredient that
guarantees that the transition operators constructed below fulfill all
necessary criteria.  A more involved example illustrating the action
of $\rho_{\Theta\Phi}$ on $P_{\Phi}$ is given in
appendix~\ref{sec:Illustration-Rho}.

Here is a first version of the construction algorithm for transition operators:

\begin{theorem}[unitary transition operators]\label{thm:TransitionElement}
  Let $\Theta, \Phi \in \mathcal{Y}_n$ be two Young tableaux with
  the same underlying Young diagram, and let $P_{\Theta}$ and
  $P_{\Phi}$ be their respective Hermitian Young projection
  operators and $T_{\Theta\Phi}$ the transition operator between
  them. Then, $T_{\Theta\Phi}$ is given by
  \begin{equation}
    \label{eq:TransitionElement}
    T_{\Theta\Phi} = \tau \cdot P_{\Theta} \rho_{\Theta\Phi} P_{\Phi},
  \end{equation}
  where $\tau$ is a non-zero constant and $\rho_{\Theta\Phi} \in
  S_n$ is the permutation constructed according to
  Definition~\ref{thm:TableauPermutation-birdtrack}. The constant $\tau$ is
  constrained by~\eqref{eq:TTinv} and can be determined through
  explicit calculation (c.f eq.~\eqref{eq:tau-expr}).
\end{theorem}

That the operator~\eqref{eq:TransitionElement} defined in Theorem~\ref{thm:TransitionElement} satisfies all
conditions~\eqref{eq:unitary-trans-LR} is readily seen:

\paragraph{\emph{Property~\eqref{eq:TP}}, $T_{\Theta
  \Phi}P_{\Phi} = T_{\Theta \Phi} = P_{\Theta}T_{\Theta
  \Phi}$, is easily shown:} Let
\begin{equation}\label{eq:TransElProof1}
  T_{\Theta\Phi} := \tau \cdot P_{\Theta} \rho_{\Theta\Phi} P_{\Phi}
\qquad \text{with $\tau \in \mathbb{R}\setminus\lbrace 0\rbrace$}
\ .
\end{equation}
Then, 
\begin{equation}
  T_{\Theta\Phi} \cdot P_{\Phi} := \tau \cdot P_{\Theta}
  \rho_{\Theta\Phi} \underbrace{P_{\Phi} \cdot
    P_{\Phi}}_{=P_{\Phi}} = \tau \cdot P_{\Theta} \rho_{\Theta\Phi} P_{\Phi}
\end{equation}
since $P_{\Phi}$ is a projection operator. Similarly
\begin{equation}
  P_{\Theta} \cdot T_{\Theta\Phi} := \tau \cdot \underbrace{P_{\Theta} \cdot  P_{\Theta}}_{=P_{\Theta}}
  \rho_{\Theta\Phi} P_{\Phi} = \tau \cdot P_{\Theta} \rho_{\Theta\Phi}
  P_{\Phi}
\ .
\end{equation}

\paragraph{\emph{Property~\eqref{eq:Tunitary}},
$T_{\Theta\Phi}^\dagger = T_{\Phi\Theta}$:}
\begin{equation}
  \label{eq:conjugateT}
  T_{\Theta\Phi}^\dagger 
  = 
  \Bigl(
  P_\Theta \rho_{\Theta\Phi} P_{\Phi} 
  \Bigr)^\dagger
  =
   P_{\Phi} \rho_{\Theta\Phi}^\dagger P_{\Theta} = T_{\Phi\Theta}
\end{equation}
where the last equality holds since $\rho_{\Theta\Phi}^\dagger =
\rho_{\Phi\Theta}$ is the inverse permutation of
$\rho_{\Theta\Phi}$, \emph{c.f.} Definition~\ref{thm:TableauPermutation-birdtrack}.

\paragraph{\emph{Property~\eqref{eq:TTinv}},
$T_{\Theta\Phi} T_{\Phi\Theta} = P_\Theta$:}
We unpack
\begin{equation}
  T_{\Theta\Phi}T_{\Phi\Theta} = \tau^2 \cdot 
  P_{\Theta} \rho_{\Theta\Phi} \underbrace{P_{\Phi} \cdot
    P_{\Phi}}_{=P_{\Phi}} \rho_{\Theta\Phi}^{\dagger}
  P_{\Theta} = \tau^2 \cdot P_{\Theta} \rho_{\Theta\Phi} P_{\Phi}
  \rho_{\Theta\Phi}^{\dagger} P_{\Theta}
  \ ,
\end{equation}
writing $\rho_{\Phi\Theta}$ as $\rho_{\Theta\Phi}^{\dagger}$ for
clarity in the steps to follow.  Of the equivalent ways to
express the projectors $P_\Theta$ and $P_{\Phi}$~\cite{Keppeler:2013yla,Alcock-Zeilinger:2016sxc}, we choose
$P_{\Theta}$ and $P_{\Phi}$ to be constructed according to 
Theorem~\ref{thm:ShortKSProjectors} (sec.~\ref{sec:Three-Hermitian-Ops}):
\begin{equation}
  \label{eq:TransElProof2}
  \frac{T_{\Theta\Phi}T_{\Phi\Theta}}{\tau^2} =
\underbrace{Y_{\Theta_{(n-2)}}
  \cdots Y_{\Theta} \cdots Y_{\Theta_{(n-2)}}}_{P_{\Theta}} \rho_{\Theta\Phi} \underbrace{Y_{\Phi_{(n-2)}}
  \cdots Y_{\Phi} \cdots Y_{\Phi_{(n-2)}}}_{P_{\Phi}}
  \rho_{\Theta\Phi}^{\dagger} 
\underbrace{Y_{\Theta_{(n-2)}}
  \cdots Y_{\Theta} \cdots Y_{\Theta_{(n-2)}}}_{P_{\Theta}}.
\end{equation}
Writing each Young projection operator as a product of symmetrizers
and antisymmetrizers,
$Y_\Xi=\alpha_\Xi\mathbf{S}_{\Xi}\mathbf{A}_{\Xi}$,
eq.~\eqref{eq:TransElProof2} becomes

\begin{align}
\label{eq:TTDag-Green-Box}
&\frac{T_{\Theta\Phi}T_{\Phi\Theta}}{\tau^2 \beta_{\Theta}^2
        \beta_{\Phi}} 
=  \\ \notag
  & \begin{tikzpicture}[baseline=(current bounding box.west),
  every node/.style={inner sep=.5pt,outer sep=-1pt}        ]
    \matrix(ID)[
    matrix of math nodes,
    ampersand replacement=\&,
    row sep =0mm,
    column sep =0mm
    ]
    {\mathbf{S}_{\Theta_{(n-2)}}
            \& {\cdot}{\cdot}{\cdot}      \& \mathbf{S}_{\Theta}
      \& \mathbf{A}_{\Theta}
      \& \mathbf{S}_{\Theta_{(1)}}
      \& {\cdot}{\cdot}{\cdot}            \& \mathbf{A}_{\Theta_{(n-2)}}
      \& \rho_{\Theta\Phi}
      \& \mathbf{S}_{\Phi_{(n-2)}}
            \& {\cdot}{\cdot}{\cdot}      \& \mathbf{S}_{\Phi}
      \& \mathbf{A}_{\Phi}
      \& {\cdot}{\cdot}{\cdot}            \& \mathbf{A}_{\Phi_{(n-2)}}
      \& \rho_{\Phi\Theta}
      \& \mathbf{S}_{\Theta_{(n-2)}}
            \& {\cdot}{\cdot}{\cdot}      \& \mathbf{A}_{\Theta_{(1)}}
      \& \mathbf{S}_{\Theta}
      \& \mathbf{A}_{\Theta}
      \& {\cdot}{\cdot}{\cdot}            \& \mathbf{A}_{\Theta_{(n-2)}}
      \& ,
      \\
};
\draw[color=green,thick] ($(ID-1-3.north west) + (-1pt,6pt)$) rectangle
($(ID-1-20.south east) + (1pt,-6pt)$);
\draw[decorate,decoration={brace,amplitude=6pt},thick] ($(ID-1-4.north west) +(0,5.4pt)$) --
($(ID-1-7.north east)+(0,5.4pt)$) node[pos=.5,anchor=south,yshift=3mm]
{\scriptsize$M^{(1)}$};
\draw[decorate,decoration={brace,amplitude=6pt},thick] ($(ID-1-8.north west) +(0,8pt)$) --
($(ID-1-15.north east)+(0,8pt)$) node[pos=.5,anchor=south,yshift=3mm]
{\scriptsize$M^{(2)}$};
\draw[decorate,decoration={brace,amplitude=6pt},thick] ($(ID-1-16.north west) +(0,5.4pt)$) --
($(ID-1-19.north east)+(0,5.4pt)$) node[pos=.5,anchor=south,yshift=3mm]
{\scriptsize$M^{(1)}$};
\draw[decorate,decoration={brace,amplitude=6pt},thick] ($(ID-1-7.south east)+(0,-3pt)$) --
($(ID-1-1.south west) +(0,-3pt)$) node[pos=.5,anchor=north,yshift=-3mm]
{\scriptsize$\bar{P}_{\Theta}$};
\draw[decorate,decoration={brace,amplitude=6pt},thick] ($(ID-1-14.south east)+(0,-3pt)$) --
($(ID-1-9.south west) +(0,-3pt)$) node[pos=.5,anchor=north,yshift=-3mm]
{\scriptsize$\bar{P}_{\Phi}$};
\draw[decorate,decoration={brace,amplitude=6pt},thick] ($(ID-1-22.south east)+(0,-3pt)$) --
($(ID-1-16.south west) +(0,-3pt)$) node[pos=.5,anchor=north,yshift=-3mm]
{\scriptsize$\bar{P}_{\Theta}$};
\end{tikzpicture}
\end{align}
where the constants $\beta_{\Theta}$ and $\beta_{\Phi}$ lump
together all the constants $\alpha_\Xi$ appearing in $P_{\Theta}$ and
$P_{\Phi}$ respectively. Let us now take a closer look
the part of $T_{\Theta\Phi}T_{\Phi\Theta}$ that is enclosed in a
green box in~\eqref{eq:TTDag-Green-Box}: We notice that this part is
of the form
\begin{equation}
\label{eq:TTDag-O}
  O := \mathbf{S}_{\Theta} \; M^{(1)} \; M^{(2)} \; M^{(1)} \;
  \mathbf{A}_{\Theta}
\ ,
\end{equation}
where the $M^{(i)}$ are defined
in~\eqref{eq:TTDag-Green-Box}. According to the
Cancellation-Theorem~\ref{thm:CancelMultipleSets}, there exists a
constant $\lambda$ such that
\begin{equation}
\label{eq:TTDag-O-Y}
  O = \lambda Y_{\Theta}
\ . 
\end{equation}
Furthermore, we know that $\lambda\neq 0$, if the operator $O$ itself
is non-zero. In section~\ref{sec:CancellationRules}, we gave two
conditions under which $O$ is guaranteed to be non-zero. From the
definition of the $M^{(i)}$~\eqref{eq:TTDag-Green-Box}, it is clear
that $M^{(1)}$ satisfies the first such condition
(condition~\ref{itm:nonzero-O-1}), while $M^{(2)}$ satisfies the
second condition (condition~\ref{itm:nonzero-O-2}). Thus, a
combination of the two conditions hold and $O$ is non-zero
(condition~\ref{itm:nonzero-O-3}). This implies
that~\eqref{eq:TTDag-O-Y} holds for a non-zero constant $\lambda$. We
may therefore simplify~\eqref{eq:TTDag-Green-Box} as
\begin{equation}
\begin{tikzpicture}[baseline=(current bounding box.west),
  every node/.style={inner sep=1pt,outer sep=-1pt}        ]
    \matrix(ID)[
    matrix of math nodes,
    ampersand replacement=\&,
    row sep =0mm,
    column sep =0mm
    ]
    {  \frac{T_{\Theta\Phi}T_{\Phi\Theta}}{\tau^2 \beta_{\Theta}^2
        \beta_{\Phi}} \; = \; \lambda \cdot 
      \& \mathbf{S}_{\Theta_{(n-2)}}
      \& \cdots
      \& \mathbf{A}_{\Theta_{(1)}}
      \& \mathbf{S}_{\Theta}
      \& \mathbf{A}_{\Theta}
      \& \; \mathbf{S}_{\Theta_{(1)}}
      \& \cdots
      \& \mathbf{A}_{\Theta_{(n-2)}}
      \& .
      \\
};
\draw[color=green,thick] ($(ID-1-5.north west) + (-1pt,6pt)$) rectangle
($(ID-1-6.south east) + (1pt,-4pt)$);
\end{tikzpicture}
\end{equation}
Once again writing the sets of symmetrizers and antisymmetrizers as
Young projection operators,
$Y_\Xi=\alpha_\Xi\mathbf{S}_{\Xi}\mathbf{A}_{\Xi}$ (where the
$\alpha_{\Xi}$ are encoded in the constants $\beta$), the product
$T_{\Theta\Phi}T_{\Phi\Theta}$ becomes
\begin{equation}
\label{eq:TTDag1}
    T_{\Theta\Phi}T_{\Phi\Theta} = \;
\left(\tau^2 \beta_{\Theta} \beta_{\Phi} \lambda \right) 
\cdot
\underbrace{
Y_{\Theta_{(n-2)}}
  \cdots
Y_{\Theta}
\cdots
Y_{\Theta_{(n-2)}}}_{P_{\Theta}}
\ . 
\end{equation}
Thus, for
\begin{equation}
  \label{eq:tau-expr}
  \tau=\frac{1}{\sqrt{\beta_{\Theta} \beta_{\Phi}
    \lambda}}
\ ,
\end{equation}
 the transition operator $T_{\Theta\Phi}$ also
satisfies Property~\ref{itmTransitionElementDef3} of
  Definition~\ref{thm:TransitionElementDef}. 

Since $T_{\Theta\Phi}$ does indeed satisfy all properties laid out in
eqs.~\eqref{eq:unitary-trans-LR}, we conclude that it is the transition
operator between the Hermitian Young projection operators $P_{\Theta}$
and $P_{\Phi}$. \qed

Due to the length of the operator expressions Theorem~\ref{thm:TransitionElement}
becomes inefficient very easily.  We will build on this result to
refine our methods in Theorem~\ref{thm:TransitionCompact}, which
provides a more efficient way of constructing the transition
operators.

Returning to our example of eq.~\eqref{eq:3-mixed-proj} we obtain
\begin{equation}
  T_{\Theta \Phi} = \tau \cdot 
  \underbrace{\FPic{3Sym12ASym23Sym12}}_{P_\Theta}
  \underbrace{\FPic{3s23SN}}_{\rho_{\Theta\Phi}}
  \underbrace{\FPic{3ASym12Sym23ASym12}}_{P_{\Phi}}
\ .
\end{equation}
Using Theorem~\ref{thm:CancelMultipleSets}, this can be simplified to
\begin{equation}
  T_{\Theta \Phi} = \sqrt{\sfrac{4}{3}} \cdot \FPic{3Sym12s23ASym12}
  \ .
\end{equation}
The constant $\sqrt{\frac{4}{3}}$ is determined via implementing
eq.~\eqref{eq:TTinv}. In fact, one can incorporate this simplification
step directly in the construction, arriving at a general efficient
algorithm.

Our description of the algorithm is based on a specific graphical
convention for the birdtracks used to represent the projection
operators: For any birdtrack operator we will align all sets of symmetrizers and antisymmetrizer at
the top.  If a particular set of symmetrizers
$\mathbf{S}_{\Theta}$ contains several symmetrizers such that each
$\bm{S}_i\in\mathbf{S}_{\Theta}$ corresponds to the $i^{th}$ row of
$\Theta$, then we draw $\bm{S}_i$ above $\bm{S}_j$ if $i<j$. A similar
convention is used for antisymmetrizers corresponding to the columns
of $\Theta$.

For birdtrack operators containing $3$ index lines, we have neglected
to follow this convention in two cases for consistency with the literature, for
example~\cite{Cvitanovic:2008zz}. This is remedied using the two identities
\begin{equation}
  \FPic{3Sym12ASym23Sym12}
 = \FPic{3Sym12N}\FPic{3s23N}\FPic{3ASym12N}\FPic{3s23N}\FPic{3Sym12N} \qquad \text{and} \qquad   \FPic{3ASym12Sym23ASym12}
 = \FPic{3ASym12N}\FPic{3s23N}\FPic{3Sym12N}\FPic{3s23N}\FPic{3ASym12N} \ .
\end{equation}

\begin{theorem}[compact transition operators]
\label{thm:TransitionCompact}
  Let $\Theta$ and $\Phi$ be two Young tableaux of equivalent
  representations of $\SUN$. They therefore have the same shape and
  sets of antisymmetrizers $\mathbf{A}_{\Theta}$ and
  $\mathbf{A}_{\Phi}$ are in one to one correspondence: For each
  element of $\mathbf{A}_{\Theta}$ there exists a counterpart in
  $\mathbf{A}_{\Phi}$ with the same length (this is important for
  the graphical matching described below). Let $\bar{P}_{\Theta}$ and
  $\bar{P}_{\Phi}$ be the birdtracks of two Hermitian Young
  projection operators constructed according to the MOLD-Theorem~\ref{thm:MOLDConstruction},
  drawn using the conventions listed in the previous paragraph. Then
  $\bar{P}_{\Theta}$ and $\bar{P}_{\Phi}$ contain
  $\mathbf{A}_{\Theta}$ and $\mathbf{A}_{\Phi}$ at least once, but
  at most twice. This determines how to proceed:
  \begin{enumerate}
  \item\label{itm:CompactTrans1} If both $\bar{P}_{\Theta}$ and
    $\bar{P}_{\Phi}$ each contain exactly one set of
    $\mathbf{A}_{\Theta}$ respectively $\mathbf{A}_{\Phi}$, then
    pick this set in each operator.
  \item\label{itm:CompactTrans2} If one of $\bar{P}_{\Theta}$ and
    $\bar{P}_{\Phi}$ contains one copy of $\mathbf{A}_{\Theta}$
    respectively $\mathbf{A}_{\Phi}$, the other contains two, then
    pick the left-most set $\mathbf{A}_{\Theta}$ in $\bar{P}_{\Theta}$
    and the right-most set $\mathbf{A}_{\Phi}$ in
    $\bar{P}_{\Phi}$.
  \item\label{itm:CompactTrans3} If both $\bar{P}_{\Theta}$ and
    $\bar{P}_{\Phi}$ each contain two sets of $\mathbf{A}_{\Theta}$
    respectively $\mathbf{A}_{\Phi}$, then pick either the
    left-most set or the right-most set in \emph{both}
    operators. (It does not matter which one, but it needs to be
      the \emph{same in both} operators.)
  \end{enumerate}
  Now split $\bar{P}_{\Theta}$ and $\bar{P}_{\Phi}$ by
  vertically cutting through the tower of antisymmetrizers chosen
  according to these rules. The next step discards everything to the
  right of the cut in $\bar{P}_{\Theta}$ and everything to the left of
  the cut in $\bar{P}_{\Phi}$ and glues the remaining pieces
  together at the cut.  The resulting birdtrack is $\bar{T}_{\Theta
    \Phi}$\footnote{It should be noted that this gluing can always
    be done, since the two Young tableaux $\Theta$ and $\Phi$ have
    the same shape, thus do their sets of antisymmetrizers
    $\mathbf{A}_{\Theta}$ and $\mathbf{A}_{\Phi}$, and the two sets
    are top-aligned.}.
\end{theorem}

\noindent The proof of this Theorem is rather lengthy and thus
deferred to Appendix~\ref{sec:ProofsCompactTransition}. This proof will also shed light
on the three distinctions~\ref{itm:CompactTrans1},
\ref{itm:CompactTrans2} and~\ref{itm:CompactTrans3} we had to make in
the Theorem.

To forestall any misunderstanding about the cut, discard and glue
procedures (the significance of which is discussed in appendix~\ref{sec:Cut-Glue-Procedure}), we will now clarify them with an example: Consider the two
Hermitian Young projection operators
\begin{equation}
  P_{\Theta} = \; \sfrac{3}{2} \cdot 
\FPic{4ASym123N}\FPic{4s243N}\FPic{4Sym12N}\FPic{4s234N}\FPic{4ASym123N}
\qquad \text{and} \qquad P_{\Phi} = \; 2 \cdot 
\FPic{4ASym12N}\FPic{4s23N}\FPic{4Sym12N}\FPic{4s234N}\FPic{4ASym123N}\FPic{4s243N}\FPic{4Sym12N}\FPic{4s23N}\FPic{4ASym12N}
\end{equation}
corresponding to the Young tableaux
\begin{equation}
  \Theta =
  \begin{ytableau}
    1 & 4 \\
    2 \\
    3
  \end{ytableau} \qquad \text{and} \qquad \Phi =
  \begin{ytableau}
    1 & 3 \\
    2 \\
    4
  \end{ytableau}
\end{equation}
respectively. We construct $\bar{T}_{\Theta \Phi}$
according to the compact Theorem~\ref{thm:TransitionCompact}: we 
first split the left-most antisymmetrizer $\bm{A}_{123}$ of
$\bar{P}_{\Theta}$ and discard everything to the right of it,
\begin{equation}
  \label{eq:SplitEx1}
  \bar{P}_{\Theta} = \; \FPic{4ASym123SplitN}\FPic{4s243N}\FPic{4Sym12N}\FPic{4s234N}\FPic{4ASym123N}
\quad \mapsto \quad
\FPic{4ASym123SplitLeftN}\;\cancel{\FPic{4ASym123SplitRightN}\FPic{4s243N}\FPic{4Sym12N}\FPic{4s234N}\FPic{4ASym123N}}
\; = \; \FPic{4ASym123SplitLeftN}
\ .
\end{equation}
Similarly,
\begin{equation}
  \label{eq:SplitEx2}
  \bar{P}_{\Phi} = \; 
\FPic{4ASym12N}\FPic{4s23N}\FPic{4Sym12N}\FPic{4s234N}\FPic{4ASym123SplitN}\FPic{4s243N}\FPic{4Sym12N}\FPic{4s23N}\FPic{4ASym12N}
\quad \mapsto \quad 
\cancel{\FPic{4ASym12N}\FPic{4s23N}\FPic{4Sym12N}\FPic{4s234N}\FPic{4ASym123SplitLeftN}}\;\FPic{4ASym123SplitRightN}\FPic{4s243N}\FPic{4Sym12N}\FPic{4s23N}\FPic{4ASym12N}
\; = \;
\FPic{4ASym123SplitRightN}\FPic{4s243N}\FPic{4Sym12N}\FPic{4s23N}\FPic{4ASym12N}\
.
\end{equation}
Gluing the remaining pieces together at the cut then yields 
\begin{equation}
  \label{eq:2}
  \bar{T}_{\Theta \Phi} = \;
  \FPic{4ASym123SplitN}\FPic{4s243N}\FPic{4Sym12N}\FPic{4s23N}\FPic{4ASym12N}
\ ;
\end{equation}
and indeed, the transition operator
$T_{\Theta\Phi}=\sqrt{2}\bar{T}_{\Theta\Phi}$, as
can be easily checked by direct calculation.

Readers should note that one can replace antisymmetrizer sets
($\mathbf{A}_{\Theta}$ respectively $\mathbf{A}_{\Phi}$) by
symmetrizer set ($\mathbf{S}_{\Theta}$ respectively
$\mathbf{S}_{\Phi}$) in \emph{all} the steps outlined in
Theorem~\ref{thm:TransitionCompact}. This leads to the same birdtrack
$\bar{T}_{\Theta\Theta}$ as becomes evident in the proof.  Basing the
procedure on antisymmetrizers however makes the discussion on
``vanishing representations'' in appendix~\ref{sec:VanishingReps}
clearer.

To obtain $T_{\Theta\Phi} = \tau \bar{T}_{\Theta\Phi}$ one still
needs to find the normalization constant $\tau$ from direct
calculation by requiring eq.~\eqref{eq:TTinv} to hold. The relatively compact expression are well suited for
automated treatment.

\section{Examples}\label{sec:Herm-Algebra-Example}

\subsection{\texorpdfstring{$\API{\SUN,\Pow{3}}$}{API(SU(N)V3)} --
  the full algebra of \texorpdfstring{$3$}{3}
  quarks}\label{sec:3qAlgebraSectionHerm}

 Revisiting the Young tableaux in $\mathcal{Y}_3$ (eq.~\eqref{S3YTableaux}),
\ytableausetup
{mathmode, boxsize=1.2em}
\begin{equation}\label{S3Tableaux}
  \raisebox{-0.5\height}{\begin{tikzpicture}[level distance=1.4cm,
level 1/.style={sibling distance=1.2cm}]
\tikzstyle{every node}=[rectangle, inner sep = 4pt]
\node{$\ydiagram{3}$}
        child{node{$\begin{ytableau}
            \scriptstyle 1 & \scriptstyle 2 & \scriptstyle 3
          \end{ytableau}$}};
\end{tikzpicture}} \qquad
\raisebox{-0.5\height}{\begin{tikzpicture}[level distance=1.4cm,
level 1/.style={sibling distance=1.2cm}]
\tikzstyle{every node}=[rectangle, inner sep = 4pt]
 \node{$\ydiagram{2,1}$}
        child{node{$\begin{ytableau}
            \scriptstyle 1 & \scriptstyle 2 \\
            \scriptstyle 3
          \end{ytableau}$}}
        child{node{$\begin{ytableau}
            \scriptstyle 1 & \scriptstyle 3\\
            \scriptstyle 2
          \end{ytableau}$}};
\end{tikzpicture}} \qquad
\raisebox{-0.5\height}{\begin{tikzpicture}[level distance=1.4cm,
level 1/.style={sibling distance=1.2cm}]
\tikzstyle{every node}=[rectangle, inner sep = 4pt]
\node{$\ydiagram{1,1,1}$}
        child{node{$\begin{ytableau}
            \scriptstyle 1 \\
            \scriptstyle 2 \\
            \scriptstyle 3 
          \end{ytableau}$}};
\end{tikzpicture}}.
\end{equation}
Denote the Hermitian projection operator corresponding to the $i^{th}$
tableau in~\eqref{S3Tableaux} (read from left to right) by $P_i$.
The Hermitian projection operators corresponding to the first and last tableau
in~\eqref{S3Tableaux} are equal to the Young projection operators~\eqref{eq:3Herm-Youngs}
\begin{equation}
  P_1 = \FPic{3Sym123} = Y_1 
\quad \text{and} \quad 
P_4 = \FPic{3ASym123} = Y_4 
\ ,
\end{equation}
since the Young projectors are Hermitian to begin with.
The Hermitian projection operators corresponding to the central two
tableaux are different from their Young counterparts~\eqref{eq:3nonHerm-Youngs}
\begin{equation}
  P_2 = \sfrac{4}{3} \; \FPic{3Sym12ASym23Sym12} \neq Y_2
\quad \text{and} \quad 
P_3 = \sfrac{4}{3} \; \FPic{3ASym12Sym23ASym12} \neq Y_3
\ ,
\end{equation}
and similarly for their transition operators $T_{i j}$ between $P_i$ and $P_j$,
\begin{equation}
  T_{23} = \sqrt{\sfrac{4}{3}} \; \FPic{3Sym12s23ASym12} \quad
  \text{and} \quad T_{32} = \sqrt{\sfrac{4}{3}} \; \FPic{3ASym12s23Sym12}
\ .
\end{equation}
The birdtracks of the $\bar{T}_{i j}$ were constructed using
Theorem~\ref{thm:TransitionCompact}, and the constants were determined
to match eq~\eqref{eq:TTinv}. Arranging all projection operators and
transition operators in a matrix $\mathfrak{M}$ as
in~\eqref{eq:Mblocks}, one obtains
\begin{equation}
\label{eq:PMatrixRepAlgS3}
\mathfrak{M}= \;
\begin{pmatrix}
\colorbox{blue!25}{$\FPic{3Sym123SN}$} & 0 & 0 & 0 \\%\\
0 & \colorbox{blue!25}{$\frac{4}{3} \; \FPic{3Sym12ASym23Sym12}$} & \sqrt{\frac{4}{3}} \; \FPic{3Sym12s23ASym12} & 0 \\%\\
0 & \sqrt{\frac{4}{3}} \; \FPic{3ASym12s23Sym12} & \colorbox{blue!25}{$\frac{4}{3} \; \FPic{3ASym12Sym23ASym12}$} & 0 \\%\\
0 & 0 & 0 & \colorbox{blue!25}{$\FPic{3ASym123SN}$}
\end{pmatrix}
\ ,
\end{equation}
where all projection operators are highlighted in blue. The Hermitian
Young projection operators in~\eqref{eq:PMatrixRepAlgS3} were already
known~\cite{Cvitanovic:2008zz}, the transition operators are a new
result.

As is the case for the Young projector
matrix~\eqref{eq:YMatrixRepAlgS3}, the operator in the bottom right
corner in~\eqref{eq:PMatrixRepAlgS3} becomes a null-operator for $N\leq 2$, and so does the $2\times
2$-block for $N\leq 1$.

\subsection{\texorpdfstring{$\API{\SUN,\Pow{4}}$}{API(SU(N)V4)} -- the full algebra of \texorpdfstring{$4$}{4} quarks}\label{sec:4qAlgebra}

The general pattern analyzed in sec.~\ref{sec:MultiplicationTable} and
once again observed in sec.~\ref{sec:3qAlgebraSectionHerm} must reappear
if $m$ increases. The case $m=4$ provides additional illustration.

All Young tableaux of $\mathcal{Y}_4$ and the Young diagrams from
which they originate are \ytableausetup {mathmode, boxsize=1.2em}
\begin{equation}
  \label{eq:Y4Tableaux}
  \raisebox{-0.5\height}{\begin{tikzpicture}[level distance=1.7cm,
level 1/.style={sibling distance=1.2cm}]
\tikzstyle{every node}=[rectangle, inner sep = 4pt]
\node{$\ydiagram{4}$}
        child{node{$\begin{ytableau}
            \scriptstyle 1 & \scriptstyle 2 & \scriptstyle 3 & \scriptstyle 4
          \end{ytableau}$}};
\end{tikzpicture}} \hspace{0.5cm}
\raisebox{-0.5\height}{\begin{tikzpicture}[level distance=1.7cm,
level 1/.style={sibling distance=1.5cm}]
\tikzstyle{every node}=[rectangle, inner sep = 4pt]
 \node{$\ydiagram{3,1}$}
        child{node{$\begin{ytableau}
            \scriptstyle 1 & \scriptstyle 2 & \scriptstyle 3 \\
            \scriptstyle 4
          \end{ytableau}$}}
        child{node{$\begin{ytableau}
            \scriptstyle 1 & \scriptstyle 2 & \scriptstyle 4 \\
            \scriptstyle 3
          \end{ytableau}$}}
        child{node{$\begin{ytableau}
            \scriptstyle 1 & \scriptstyle 3 & \scriptstyle 4 \\
            \scriptstyle 2
          \end{ytableau}$}};
\end{tikzpicture}} \hspace{0.5cm}
\raisebox{-0.5\height}{\begin{tikzpicture}[level distance=1.7cm,
level 1/.style={sibling distance=1.2cm}]
\tikzstyle{every node}=[rectangle, inner sep = 4pt]
 \node{$\ydiagram{2,2}$}
        child{node{$\begin{ytableau}
            \scriptstyle 1 & \scriptstyle 2 \\
            \scriptstyle 3 & \scriptstyle 4
          \end{ytableau}$}}
        child{node{$\begin{ytableau}
            \scriptstyle 1 & \scriptstyle 3 \\
            \scriptstyle 2 & \scriptstyle 4
          \end{ytableau}$}};
\end{tikzpicture}} \hspace{0.5cm}
\raisebox{-0.5\height}{\begin{tikzpicture}[level distance=1.7cm,
level 1/.style={sibling distance=1.2cm}]
\tikzstyle{every node}=[rectangle, inner sep = 4pt]
 \node{$\ydiagram{2,1,1}$}
        child{node{$\begin{ytableau}
            \scriptstyle 1 & \scriptstyle 2 \\ 
            \scriptstyle 3 \\
            \scriptstyle 4
          \end{ytableau}$}}
        child{node{$\begin{ytableau}
            \scriptstyle 1 & \scriptstyle 3 \\ 
            \scriptstyle 2 \\
            \scriptstyle 4
          \end{ytableau}$}}
        child{node{$\begin{ytableau}
            \scriptstyle 1 & \scriptstyle 4 \\ 
            \scriptstyle 2 \\
            \scriptstyle 3
          \end{ytableau}$}};
\end{tikzpicture}} \hspace{0.5cm}
\raisebox{-0.5\height}{\begin{tikzpicture}[level distance=1.7cm,
level 1/.style={sibling distance=1.2cm}]
\tikzstyle{every node}=[rectangle, inner sep = 4pt]
\node{$\ydiagram{1,1,1,1}$}
        child{node{$\begin{ytableau}
            \scriptstyle 1 \\
            \scriptstyle 2 \\
            \scriptstyle 3 \\
            \scriptstyle 4
          \end{ytableau}$}};
\end{tikzpicture}}
\ .
\end{equation}\ytableausetup{mathmode, boxsize=normal}The first and last tableau each stem
  from a unique Young diagram and their corresponding representations
  thus are not equivalent to any other irreducible representation of
  $\SUN$. Tableaux $2$, $3$ and $4$ (as counted from the left) all have the same shape and
  therefore correspond to equivalent irreducible
  representations. Similarly for tableaux $5$ and $6$ and tableaux
  $7$, $8$ and $9$.

  If we arrange the Young projection operators corresponding to the
  tableaux in~\eqref{eq:Y4Tableaux}, as well as the transition
  operators in a block-diagonal matrix as was done
  in~\eqref{eq:Mblocks} (with projection operators on the diagonal and
  transition operators on the off-diagonal), the resulting block
  diagonal matrix will be of the form
\begin{equation}
\label{eq:S4-MBlock}
\mathfrak{M} = \; 
\begin{pmatrix}
    \FPic{4qBlockMatrix}
  \end{pmatrix},
\end{equation}
where the number in each block gives the size of the block. Indeed, we again find that
\begin{equation}
  4! = \vert S_4 \vert = \sum_{{\mathbf{Y}_i}} \left(\frac{4!}{\mathcal{H}_{{\mathbf{Y}_i}}}\right)^2 = 1^2 +
  3^2 + 2^2 + 3^2 + 1^2.
\end{equation}
Since the matrix~\eqref{eq:S4-MBlock} would be rather large, we will
now give each block separately. The first block, consisting only of
one Hermitian Young projection operator is \ytableausetup {mathmode,
  boxsize=1em}
\begin{equation}
  \scalebox{0.75}{\ydiagram{4}}: \qquad \scalebox{0.75}{$\begin{pmatrix}\colorbox{blue!25}{$\FPic{4Sym1234SN}$}\end{pmatrix}$},
\end{equation}
and corresponds to an irreducible representation of $\SUN$ with
dimension $\ud=\sfrac{N(N+1)(N+2)(N+3)}{24}$. The second block,
a $3 \times 3$-block, is
\begin{equation}
\scalebox{0.75}{\ydiagram{3,1}} : \qquad \scalebox{0.75}{$
  \begin{pmatrix}
\colorbox{blue!25}{$\frac{3}{2} \;
 \FPic{4Sym123N}\FPic{4s243N}\FPic{4ASym12N}\FPic{4s234N}\FPic{4Sym123N}$} &
 \sqrt{2} \; \FPic{4Sym123N}\FPic{4s243N}\FPic{4ASym12N}\FPic{4s23N}\FPic{4Sym12N} &
\sqrt{\frac{3}{2}} \; \FPic{4Sym123N}\FPic{4s243N}\FPic{4ASym12N} \\\\
\sqrt{2} \;
\FPic{4Sym12N}\FPic{4s23N}\FPic{4ASym12N}\FPic{4s234N}\FPic{4Sym123N}
&
\colorbox{blue!25}{$2 \;
\FPic{4Sym12N}\FPic{4s23N}\FPic{4ASym12N}\FPic{4s234N}\FPic{4Sym123N}\FPic{4s243N}\FPic{4ASym12N}\FPic{4s23N}\FPic{4Sym12N}$}
& 
\sqrt{3} \;
\FPic{4Sym12N}\FPic{4s23N}\FPic{4ASym12N}\FPic{4s234N}\FPic{4Sym123N}\FPic{4s243N}\FPic{4ASym12N}
\\\\
\sqrt{\frac{3}{2}} \;
 \FPic{4ASym12N}\FPic{4s234N}\FPic{4Sym123N} &
\sqrt{3} \;
\FPic{4ASym12N}\FPic{4s234N}\FPic{4Sym123N}\FPic{4s243N}\FPic{4ASym12N}\FPic{4s23N}\FPic{4Sym12N} &
\colorbox{blue!25}{$\frac{3}{2} \;
 \FPic{4ASym12N}\FPic{4s234N}\FPic{4Sym123N}\FPic{4s243N}\FPic{4ASym12N}$}
  \end{pmatrix}$}.
\end{equation}
All Hermitian Young projection operators on the diagonal of this block
correspond to equivalent irreducible representations of $\SUN$ with dimension
$\ud=\sfrac{N(N+2)(N^2-1)}{8}$. The following $2 \times 2$-block
has projection operators on its diagonal that correspond to equivalent
irreducible representations of dimension
$\ud=\sfrac{N^2(N^2-1)}{12}$,
\begin{equation}
\scalebox{0.75}{\ydiagram{2,2}} : \qquad \scalebox{0.75}{$
  \begin{pmatrix}
\colorbox{blue!25}{$\frac{4}{3} \;
\FPic{4Sym12Sym34N}\FPic{4s23N}\FPic{4ASym12ASym34N}\FPic{4s23N}\FPic{4Sym12Sym34N}$}
& \sqrt{\frac{4}{3}} \; \FPic{4Sym12Sym34N}\FPic{4s23N}\FPic{4ASym12ASym34N} \\\\
\sqrt{\frac{4}{3}} \;
\FPic{4ASym12ASym34N}\FPic{4s23N}\FPic{4Sym12Sym34N}
& \colorbox{blue!25}{$\frac{4}{3} \;
\FPic{4ASym12ASym34N}\FPic{4s23N}\FPic{4Sym12Sym34N}\FPic{4s23N}\FPic{4ASym12ASym34N}$}
  \end{pmatrix}$}.
\end{equation}
The next $3 \times 3$ block is given by
\begin{equation}
\scalebox{0.75}{\ydiagram{2,1,1}} : \qquad 
\scalebox{0.75}{$
  \begin{pmatrix}
\colorbox{blue!25}{$\frac{3}{2} \;
 \FPic{4Sym12N}\FPic{4s234N}\FPic{4ASym123N}\FPic{4s243N}\FPic{4Sym12N}$}
&
\sqrt{3} \;
\FPic{4Sym12N}\FPic{4s234N}\FPic{4ASym123N}\FPic{4s243N}\FPic{4Sym12N}\FPic{4s23N}\FPic{4ASym12N}
&
\sqrt{\frac{3}{2}} \;
 \FPic{4Sym12N}\FPic{4s234N}\FPic{4ASym123N}\\\\
\sqrt{3} \;
\FPic{4ASym12N}\FPic{4s23N}\FPic{4Sym12N}\FPic{4s234N}\FPic{4ASym123N}\FPic{4s243N}\FPic{4Sym12N}
&
\colorbox{blue!25}{$2 \;
\FPic{4ASym12N}\FPic{4s23N}\FPic{4Sym12N}\FPic{4s234N}\FPic{4ASym123N}\FPic{4s243N}\FPic{4Sym12N}\FPic{4s23N}\FPic{4ASym12N}$}
& 
\sqrt{2} \;
\FPic{4ASym12N}\FPic{4s23N}\FPic{4Sym12N}\FPic{4s234N}\FPic{4ASym123N}
\\\\
\sqrt{\frac{3}{2}} \; \FPic{4ASym123N}\FPic{4s243N}\FPic{4Sym12N} &
 \sqrt{2} \; \FPic{4ASym123N}\FPic{4s243N}\FPic{4Sym12N}\FPic{4s23N}\FPic{4ASym12N} &
\colorbox{blue!25}{$\frac{3}{2} \;
 \FPic{4ASym123N}\FPic{4s243N}\FPic{4Sym12N}\FPic{4s234N}\FPic{4ASym123N}$}
  \end{pmatrix}$};
\end{equation}
here the projection operators each correspond to irreducible
representations of dimension $\ud=\sfrac{N(N-2)(N^2-1)}{8}$. What
remains is a $1 \times 1$-block:
\begin{equation}
  \scalebox{0.75}{\ydiagram{1,1,1,1}} : \qquad 
\scalebox{0.75}{$\begin{pmatrix}\colorbox{blue!25}{$\FPic{4ASym1234SN}$}\end{pmatrix}$}
\ .
\end{equation}
This operator corresponds to an irreducible representation of
dimension $\ud=\sfrac{N(N-1)(N-2)(N-3)}{24}$.

All the projection operators given above have previously been
known~\cite{Cvitanovic:2008zz}. The transition operators again are a
new result.

Similarly to what we observed for the $3$-quark algebra, we find that
the blocks described above give null-operators from bottom right to
top left as we incrementally decrease $N$ below $4$: For $N=3$, only
the last $1 \times 1$-block turns into a null-operator. For $N=2$, the
last $1 \times 1$-block as well as the second-to-last $3 \times
3$-block consists of null-operators. All but the top-most $1 \times 1$
block give null-operators for $N=1$. The entire matrix will
(trivially) consist of null-operators if we decrease $N$ to $0$. In
fact, we can read off which operators will be null-operators from
their dimension formula, as $\ud=0$ for a null-operator.

\begin{sidewaysfigure}[p]
  \centering
  \includegraphics[width=\textwidth]{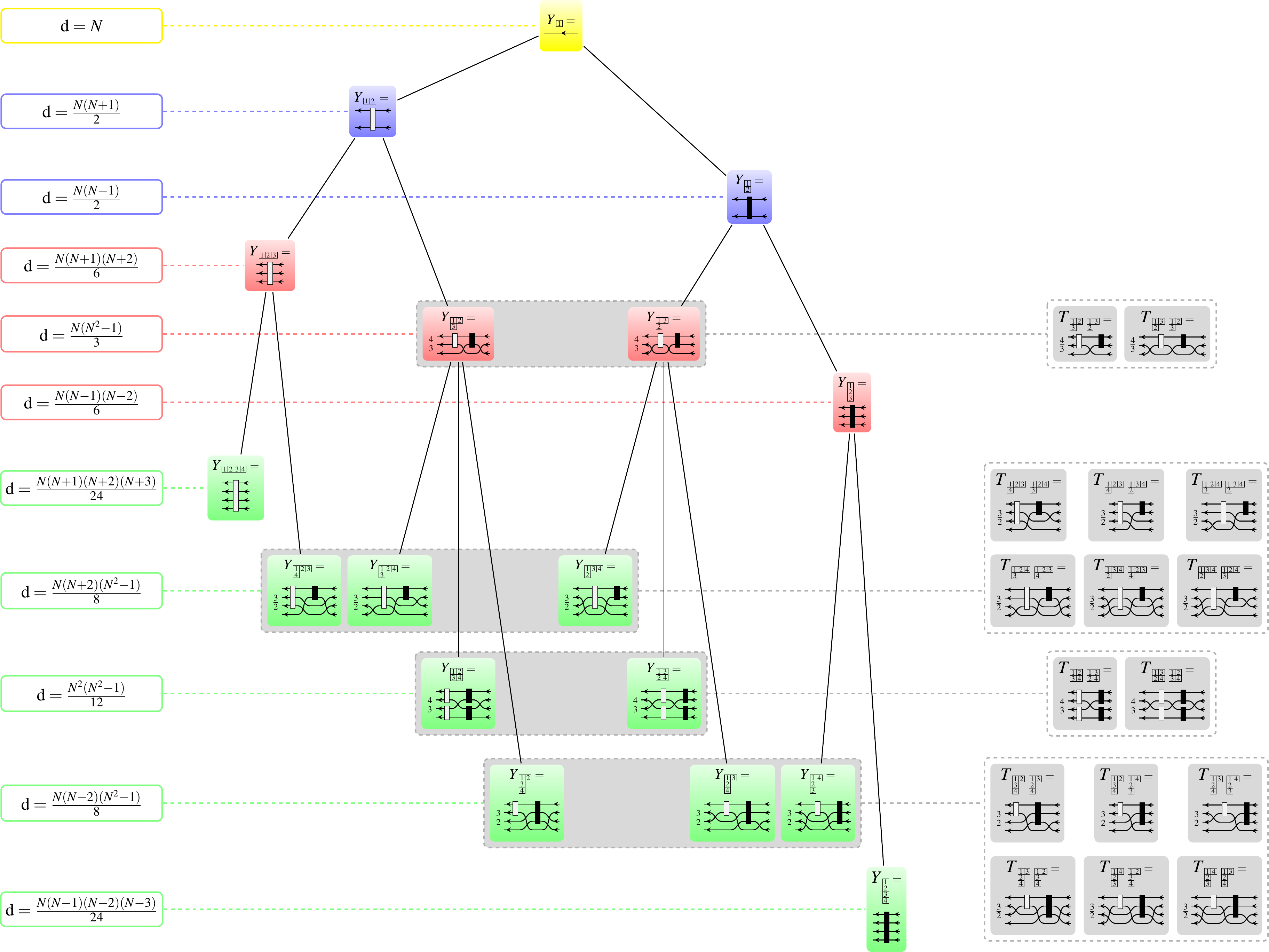}
  \caption{Hierarchy of Young tableaux and the associated (non-nested)
    Young projector decompositions over $\Pow{m}$ for $m=1,2,3,4$ (in
    the sense of embeddings into $\API{\SUN,\Pow{4}}$): The lines
    indicate ancestry. The associated transition operators for groups
    of equivalent representations are listed to the right. Note that
    this tree cannot be extended beyond $m=4$ due to the failure of
    the corresponding Young projectors to be orthogonal or complete.}
  \label{fig:YT-hierarchy-4}
\end{sidewaysfigure}

\begin{sidewaysfigure}[p]
  \centering
  \includegraphics[width=\textwidth]{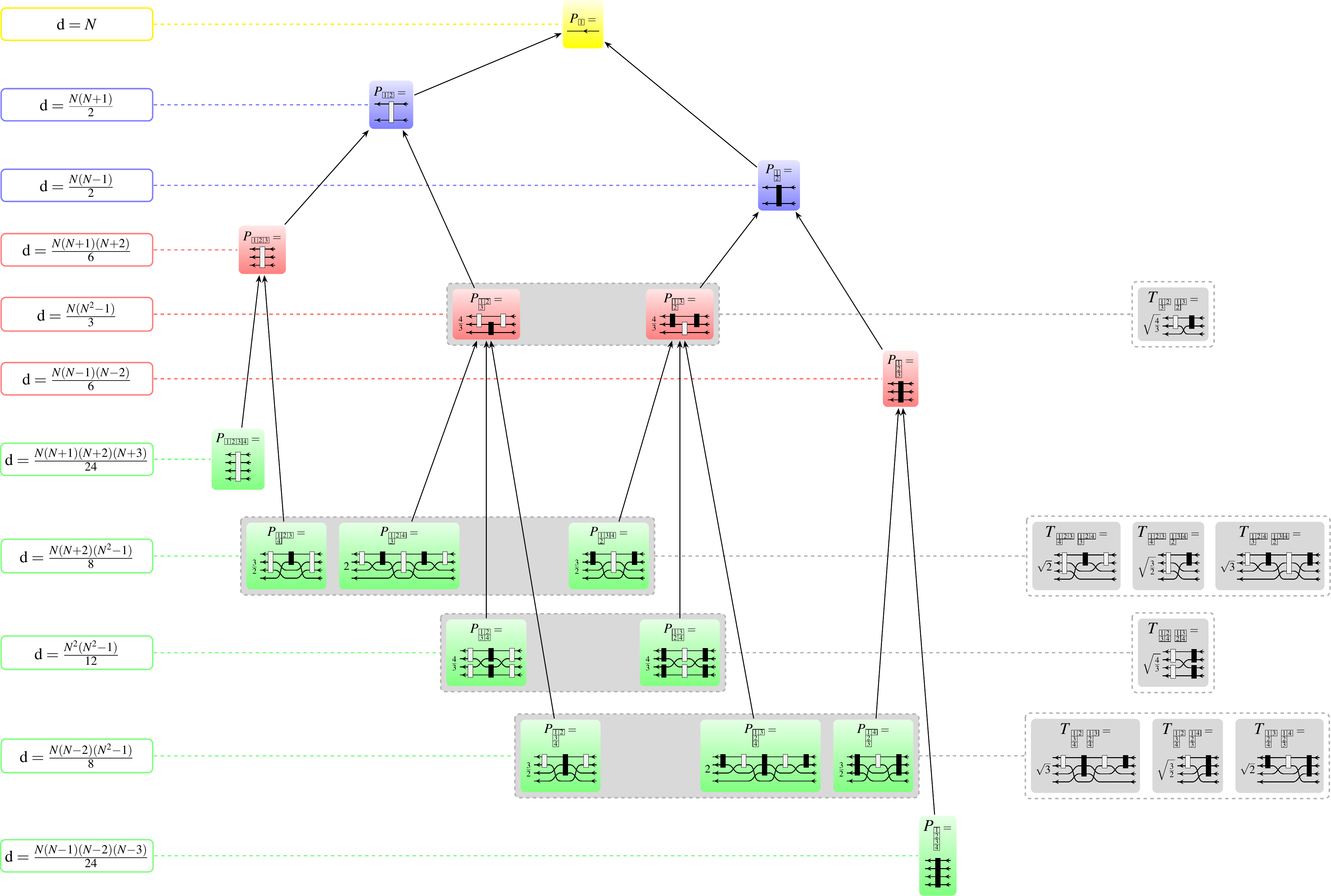}
  \caption{Hierarchy of Young tableaux and the associated nested
    Hermitian Young projector decompositions over $\Pow{m}$ for
    $m\leq4$ (in the sense of embeddings into $\API{\SUN,\Pow{4}}$):
    The arrows indicate which operators sum to which ancestors
    (see~\cite{Alcock-Zeilinger:2016sxc} -- this does \emph{not} apply
    to their standard Young counterparts shown in
    Fig.~\ref{fig:YT-hierarchy-4}). The associated transition
    operators for groups of equivalent representations are listed to
    the right (recall that these transition operators are unitary on
    the image of the projection operators,
    $T_{\Theta\Phi}=T_{\Phi\Theta}^{\dagger}$). This tree can be
    extended to arbitrary $m$ using the construction algorithm of
    Hermitian Young projectors~\cite{Alcock-Zeilinger:2016sxc} and
    that of transition operators given in this paper.}
  \label{fig:hierarchy-4}
\end{sidewaysfigure}

Figs.~\ref{fig:YT-hierarchy-4} and~\ref{fig:hierarchy-4} expand on
\cite[figs. 9.1 and 9.2]{Cvitanovic:2008zz}: The figures collect the
hierarchy of Young tableaux and the associated nested Hermitian
projector decompositions (in the sense of embeddings into
$\API{\SUN,\Pow{4}}$) and adds the transition operators we have
derived in this paper (recall that for $m\leq4$ the construction
algorithm for transition operators between Young projectors over
$\Pow{m}$ is well-defined). We would like to draw attention to the
fact that only the leftmost and rightmost branches in each tree,
consisting solely of a single symmetrizer or antisymmetrizer, are fully
unique as they are not connected by transition operators listed on the
right.

\ytableausetup{mathmode, boxsize=normal}

\section{Conclusion \& outlook}

The representation theory of $\SUN$ is an old theory with many
successful applications in physics. Yet some of the tools remain
awkward and only applicable in specific situations, like the general
theory of angular momentum or the construction of Young projection
operators that lack Hermiticity. Newer tools like the birdtrack
formalism remain only partially connected with these time honored
results. We have a very specific interest in applications to QCD in
the JIMWLK context, in jet physics, in energy loss and generalized
parton distributions, so we have aimed at creating a set of tools that
we know will aid in these applications and, in the process have
pointed out where the existing tools fall short of our needs.

Building on the previously found Hermitian Young projection
operators~\cite{Keppeler:2013yla,Alcock-Zeilinger:2016sxc}, the main
result of this paper is the inclusion of transition operators that
complement the set of multiplet projectors to a basis for the full
algebra of invariants $\API{\SUN,\Pow{m}}$; for Young projectors and
their associated transition operators this is only possible up to
$m=4$. Any subset of projectors encoding mutually equivalent
representations together with their transition operators form closed
subalgebras. Relabeling the set of basis operators as
$\mathfrak m_{i j}$ (with double indices according to~\eqref{eq:Mdef})
so that
\begin{equation}
  \API{\SUN,\Pow{m}} 
  = 
  \Bigl\{ 
  \alpha_{i j} \mathfrak{m}_{i j} |
  \alpha_{i j}\in\mathbb R, \mathfrak{m}_{i j}\in\mathfrak{S}_m 
  \Bigr\}
\end{equation}
leads to a simplified multiplication
table for the new basis elements
\begin{equation}
  \label{eq:multiplet-multtable-concl}
  \mathfrak{m}_{i j} \mathfrak{m}_{kl} = \delta_{j k} \mathfrak{m}_{i l},
\end{equation}
significantly simpler than the standard basis of primitive invariants
$\rho_i\in S_m$, $\rho_i \rho_j = \tensor{A}{^k_{i j}} \rho_k$.

The transition operators obtained from Hermitian projectors are
automatically unitary, causing the basis elements to be mutually
orthogonal and normalized to match the dimension of the irreducible
representations
\begin{equation}
 \label{eq:herm-unit-orth-Conclusion}
  \langle\mathfrak m_{i j}, \mathfrak m_{k l}\rangle
  =
              \delta_{i k} \delta_{j l} \text{dim}(\Theta_j)
\ .
\end{equation}
This is an essential prerequisite for a future publication that aims
at constructing an \emph{orthonormal basis} for the space of
\emph{all} global color singlet states for a given Fock-space
configuration; it is important to note
that~\eqref{eq:herm-unit-orth-Conclusion} does \emph{not} hold for the
standard Young projectors and their associated transition operators.

We have used the new form of the multiplication
table~\eqref{eq:multiplet-multtable-concl} to show that the projection
operators in $\API{\SUN,\Pow{m}}$ are only uniquely determined if the
representation occurs precisely once in the decomposition. All
Hermitian projectors onto equivalent representations and their
associated transition operators are only unique up to orthogonal
rotations as described in sec.~\ref{sec:MultiplicationTable}
and~\ref{sec:HermitianYoungProjectorsSection}. Figs.~\ref{fig:YT-hierarchy-4}
and~\ref{fig:hierarchy-4} collect all the examples worked out in this
paper displaying all their relationships in a compact form for
reference.

Our own list of future applications for the tools and insights
presented in this paper are QCD centric: Global singlet state
projections of Wilson-line operators that appear in a myriad of
applications due to factorization of hard and soft contributions help
analyzing the physics content in all of them. We hope that our
presentation is suitable to unify perspectives provided by the various
approaches to representation theory of $\SUN$ and that the results
prove useful beyond these immediate applications.

\paragraph{Acknowledgements:} H.W. is supported by South Africa's
National Research Foundation under CPRR grant nr 90509. J.A-Z. was
supported (in sequence) by the postgraduate funding office of the
University of Cape Town (2014), the National Research Foundation
(2015) and the Science Faculty PhD Fellowship of the University of
Cape Town (2016).

\appendix

\section{Dimensional zeroes}\label{sec:VanishingReps}
\ytableausetup{mathmode, boxsize=normal}

For small enough values of $N$ (and we will define what we mean by
``small enough'' shortly), some of the irreducible representation of
$\SUN$ over $\Pow{m}$ vanish. The reason for this is simple: An
antisymmetrizer over $p$ legs is, viewed as a linear map on $\Pow{p}$,
a null-operator, if $\text{dim}(V)<p$,
\begin{equation}
  \scalebox{0.75}{\FPic{pTotASym}}: \Pow{p} \rightarrow 0
  \quad \text{if} \; \text{dim}(V)<p.
\end{equation}
Thus, if an operator in $\Lin{\Pow{m}}$ with $\text{dim}(V)=N$
contains an antisymmetrizer of length $>N$, the operator will be a
null-operator on the space $\Pow{m}$. For example, the
antisymmetrizer
\begin{equation}
  \FPic{3ASym123SN} 
  = \sfrac{1}{3!} \left(    \FPic{3IdSN} - \FPic{3s12SN} - \FPic{3s13SN} 
    - \FPic{3s23SN} + \FPic{3s123SN} + \FPic{3s132SN} 
  \right)
\end{equation}
acts as a null-operator on the space $\Pow{3}=V\otimes V\otimes V$ if
the dimension of $V$ is $\leq 2$. The reason for this is that the
primitive invariants constituting the antisymmetrizer $\bm{A}_{123}$
as elements of $\Lin{\Pow{m}}$ are not linearly independent if the
vector space $V$ has dimension $\leq 2$. In this situation, the
identity permutation (for example) can be expressed as a linear
combination of the remaining primitive invariants,
\begin{equation}
  \FPic{3IdSN} 
  \xlongequal[]{\text{dim}(V)\leq2} 
  \FPic{3s12SN} 
  + \FPic{3s13SN} + \FPic{3s23SN} 
  - \FPic{3s123SN} - \FPic{3s132SN}
  \ .
\end{equation}

We discussed previously that each irreducible representation of $\SUN$
over $\Pow{m}$ corresponds to a particular Young tableau in
$\mathcal{Y}_m$. From the construction Theorems of (Hermitian) Young
projection operators (eq.~\eqref{eq:YTheta} for Youngs and
sec.~\ref{sec:Three-Hermitian-Ops} for Hermitian) and their transition
operators (sec.~\ref{sec:YoungBasis} for Youngs up to $m=4$ and
sec.~\ref{sec:TransitionOps} for unitary operators for all $m$), it is evident that the
longest antisymmetrizer present in such an operator corresponds to the
longest column of the corresponding Young tableau. In particular, if
$N<m$, there will be at least one Young tableau containing a column
which is longer than $N$, namely
\begin{equation}
  \begin{ytableau}
    1 \\
    2 \\
    \vdots \\
    m
  \end{ytableau}
\ .
\end{equation}
There may be more tableaux with columns longer than $N$, depending by
how much $N$ differs from $m$. If two (non-) Hermitian Young projection
operators $P_{\Theta}$ and $P_{\Phi}$ (resp. $Y_{\Theta}$ and $Y_{\Phi}$)
correspond to equivalent irreducible representations of $\SUN$, they
both will contain antisymmetrizers of equal length ($\mathbf A_\Theta$
or $\mathbf A_{\Phi}$ respectively) and so will their transition
operators by construction (see eq.~\eqref{eq:Ytrans} for Young
operators up to $m=4$ resp. Theorem~\ref{thm:TransitionCompact} for the unitary
operators for all values of $m$). They will therefore all vanish simultaneously if $N$ is
too small.

To summarize, we see that \emph{all} multiplets and transition
operators are only present in $\API{\SUN,\Pow{m}}$ if $N\geq m$. If
$N$ is smaller than $m$, some of them become null-operators. These can
explicitly be identified by their corresponding Young tableau $\Theta$
or directly by $\mathbf{A}_{\Theta}$ in the birdtrack notation.

\section{Illustrating the action of \texorpdfstring{$\rho_{\Theta\Phi}$}{rhoThetaPhi} on Hermitian
  Young projection operators: an example}\label{sec:Illustration-Rho}

In this section, we illustrate why
eq.~\eqref{eq:permute-tableaux4},
\begin{equation}
 \label{eq:Illustrate-Rho-Ex0}
  P_{\Theta} \cdot \rho_{\Theta\Phi} P_{\Phi}
  \rho_{\Phi\Theta} \neq 0
\ ,
\end{equation}
holds by means of an example. In the process, we will show
that eq.~\eqref{eq:permute-tableaux2} (saying that
$Y_{\Theta}=\rho_{\Theta\Phi}Y_{\Phi}\rho_{\Phi\Theta}$) breaks down
for Hermitian projection operators,
\begin{equation}
  \label{eq:Illustrate-Rho-Ex1}
P_{\Theta} \neq \rho_{\Theta\Phi}P_{\Phi}\rho_{\Phi\Theta}
\ .
\end{equation}

Consider two Young tableaux
 \begin{equation}
\label{eq:ThetaPhi-Tableaux}
   \Theta = \begin{ytableau}
     1 & 3 & 5 \\
     2 & 4 \\
     6
   \end{ytableau} \qquad \text{and} \qquad
   \Phi = \begin{ytableau}
     1 & 2 & 6 \\
     3 & 5 \\
     4
   \end{ytableau}
\ .
 \end{equation}
 The permutation $\rho_{\Theta\Phi}$ as defined
 in Definition~\ref{thm:TableauPermutation-birdtrack} is given by
 \begin{equation}
   \label{eq:ThetaPhi-Rho}
   \rho_{\Theta\Phi} = 
\scalebox{0.75}{\FPic{6s23s465}}
 \end{equation}
Let us now construct the MOLD-operators (\emph{c.f.} Theorem~\ref{thm:MOLDConstruction}) corresponding to $\Theta$ and
$\Phi$. To do so, we need to construct their MOLD-ancestries
(\emph{c.f.} Definition~\ref{MOLDDef} for the MOLD of a tableau),
\begin{equation}
  \label{eq:Theta-MOLD-Ancestry}
  \Theta = 
   \begin{ytableau}
     1 & 3 & 5 \\
     2 & 4 \\
     6
   \end{ytableau} \quad \rightarrow \quad
   \begin{ytableau}
     1 & 3 & 5 \\
     2 & 4 
   \end{ytableau} \quad \rightarrow \quad
   \begin{ytableau}
     1 & 3 \\
     2 & 4 
   \end{ytableau}
\end{equation}
and 
\begin{equation}
  \label{eq:Phi-MOLD-Ancestry}
\Phi = 
  \begin{ytableau}
     1 & 2 & 6 \\
     3 & 5 \\
     4
   \end{ytableau} \quad \rightarrow \quad
  \begin{ytableau}
     1 & 2 \\
     3 & 5 \\
     4
   \end{ytableau} \quad \rightarrow \quad
  \begin{ytableau}
     1 & 2 \\
     3 \\
     4
   \end{ytableau} \quad \rightarrow \quad
  \begin{ytableau}
     1 & 2 \\
     3 
   \end{ytableau}
\end{equation}
The MOLD-projectors $P_{\Theta}$ and $P_{\Phi}$ are thus determined by
\begin{align}
  \Bar P_{\Theta} 
= & \; 
\scalebox{0.75}{\FPic{6ASym12ASym34N}\FPic{6s2354N}\FPic{6Sym123Sym45N}\FPic{6s24s36N}\FPic{6ASym123ASym45N}\FPic{6s24s36N}\FPic{6Sym123Sym45N}\FPic{6s24s36N}\FPic{6ASym123ASym45N}\FPic{6s24s36N}\FPic{6Sym123Sym45N}\FPic{6s2453N}\FPic{6ASym12ASym34N}}
\notag \\
= & \; 
\scalebox{0.75}{\FPic{6ASym12ASym34N}\FPic{6s2354N}\FPic{6Sym123Sym45N}\FPic{6s24s36N}\FPic{6ASym123ASym45N}\FPic{6s24s36N}\FPic{6Sym123Sym45N}\FPic{6s2453N}\FPic{6ASym12ASym34N}}
\\
\Bar P_{\Phi}
= & \;
\scalebox{0.75}{\FPic{6Sym12N}\FPic{6s234N}\FPic{6ASym123N}\FPic{6s2453N}\FPic{6Sym12Sym34N}\FPic{6s2354N}\FPic{6ASym123ASym45N}\FPic{6s24s36N}\FPic{6Sym123Sym45N}\FPic{6s24s36N}\FPic{6ASym123ASym45N}\FPic{6s2453N}\FPic{6Sym12Sym34N}\FPic{6s2354N}\FPic{6ASym123N}\FPic{6s243N}\FPic{6Sym12N}}
\ ,
\end{align}
where we simplified $\Bar P_{\Theta}$ according to
Theorem~\ref{thm:CancelMultipleSets} in the second step. The full
projection operators $P_{\Theta}$ and $P_{\Phi}$ require additional
constants $\beta_{\Theta}$ and $\beta_{\Phi}$ respectively to ensure
their idempotency. From the differing lengths of
$P_{\Theta}$ and $P_{\Phi}$ (due to the different MOLD of the tableaux
$\Theta$ and $\Phi$) it is abundantly clear that $P_{\Theta}\neq
\rho_{\Theta\Phi}P_{\Phi}\rho_{\Theta\Phi}^{-1} $, confirming eq.~\eqref{eq:Illustrate-Rho-Ex1}. Let us however take
a closer look at $\rho_{\Theta\Phi}P_{\Phi}\rho_{\Phi\Theta}$,
\begin{equation}
  \label{eq:PTheta-Rho-Transform}
\rho_{\Theta\Phi}P_{\Phi}\rho_{\Phi\Theta} = 
  \scalebox{0.75}{\FPic{6s23s465SN}}\scalebox{0.75}{\FPic{6Sym12N}\FPic{6s234N}\FPic{6ASym123N}\FPic{6s2453N}\FPic{6Sym12Sym34N}\FPic{6s2354N}\FPic{6ASym123ASym45N}\FPic{6s24s36N}\FPic{6Sym123Sym45N}\FPic{6s24s36N}\FPic{6ASym123ASym45N}\FPic{6s2453N}\FPic{6Sym12Sym34N}\FPic{6s2354N}\FPic{6ASym123N}\FPic{6s243N}\FPic{6Sym12N}}\scalebox{0.75}{\FPic{6s23s456SN}}\ .
\end{equation}
By transforming $P_{\Phi}$ with the permutation $\rho_{\Theta\Phi}$,
we have transformed each set of (anti-)symmetrizers into a different
set of the same shape. In particular, the (anti-)symmetrizers
of the ancestor tableaux of $\Phi$ have been transformed
into the (anti-) symmetrizers
of tableaux obtained from $\Theta$ by deleting the
corresponding boxes,
\begin{subequations}
\label{eq:PhiTheta-Transform-Ancestry}
\begin{IEEEeqnarray}{0rCCCCCCCl}
& \Phi && \Phi_{(1)} && \Phi_{(2)} && \Phi_{(3)} & \nonumber \\
    & \overbrace{\begin{ytableau}
     1 & 2 & *(cyan!60) 6 \\
     3 & *(cyan!40) 5 \\
     *(cyan!20) 4
   \end{ytableau}}
& \quad \rightarrow \quad &
  \overbrace{\begin{ytableau}
     1 & 2 \\
     3 & *(cyan!40) 5 \\
     *(cyan!20) 4
   \end{ytableau}} 
& \quad \rightarrow \quad &
  \overbrace{\begin{ytableau}
     1 & 2 \\
     3 \\
     *(cyan!20) 4
   \end{ytableau}}
 & \quad \rightarrow \quad &
  \overbrace{\begin{ytableau}
     1 & 2 \\
     3 
   \end{ytableau}} & \label{eq:Phi-Ancestry} \\
& \FPic{StraightDownArrow} && \FPic{StraightDownArrow} &&
\FPic{StraightDownArrow} && \FPic{StraightDownArrow} & \nonumber \\
    & \underbrace{\begin{ytableau}
     1 & 3 & *(cyan!60) 5 \\
     2 & *(cyan!40) 4 \\
     *(cyan!20) 6
   \end{ytableau}}
& \quad \rightarrow \quad &
  \underbrace{\begin{ytableau}
     1 & 3 \\
     2 & *(cyan!40) 4 \\
     *(cyan!20) 6
   \end{ytableau}} 
& \quad \rightarrow \quad &
  \underbrace{\begin{ytableau}
     1 & 3 \\
     2 \\
     *(cyan!20) 6
   \end{ytableau}}
 & \quad \rightarrow \quad &
  \underbrace{\begin{ytableau}
     1 & 3 \\
     2 
   \end{ytableau}} & \label{eq:ThetaPhi-Ancestry} \\
& \Theta && \Theta_{(\Phi,1)} && \Theta_{(\Phi,2)} &&
\Theta_{(\Phi,3)} & \nonumber
\end{IEEEeqnarray}
\end{subequations}
Each tableau $\Theta_{(\Phi,k)}$ in~\eqref{eq:ThetaPhi-Ancestry}
was obtained from the predecessor $\Theta_{(\Phi,k-1)}$ by removing
the box which is in the same position as the box with the highest
number in $\Phi_{(k-1)}$. We shall refer to the tableaux
in~\eqref{eq:ThetaPhi-Ancestry} as the $\Phi$-MOLD ancestry of
$\Theta$. It should be noted however, that most of the tableaux in the
$\Phi$-MOLD ancestry of $\Theta$ are \emph{not} the ancestor tableaux
of $\Theta$, in fact, most of them are not even Young tableaux. The
$\Theta_{(\Phi,i)}$ emerge by superimposing the $\Phi_{(i)}$ in cookie
cutter fashion over $\Theta$ and thus intrinsically differ from the
ancestry of $\Theta$ itself -- compare
\begin{equation}
   \underbrace{
     \begin{ytableau}
     1 & 3 & 5 \\
     2 & 4 \\
     6
   \end{ytableau}
   }_{\Theta} \quad \rightarrow \quad
   \underbrace{
     \begin{ytableau}
     1 & 3 & 5 \\
     2 & 4 
   \end{ytableau}
   }_{\Theta_{(1)}} \quad \rightarrow \quad
   \underbrace{
     \begin{ytableau}
     1 & 3 \\
     2 & 4 
   \end{ytableau}
    }_{\Theta_{(2)}} \quad \rightarrow \quad
   \underbrace{
     \begin{ytableau}
     1 & 3 \\
     2  \\
   \end{ytableau}
    }_{\Theta_{(3)}}
\end{equation}
with eq.~\eqref{eq:ThetaPhi-Ancestry}.

We now see that the symmetrizers and antisymmetrizers in the
operator~\eqref{eq:PTheta-Rho-Transform} are exactly those
corresponding to the tableaux in the $\Phi$-MOLD ancestry of $\Theta$
eq.~\eqref{eq:ThetaPhi-Ancestry}. This means that, the (anti-)symmetrizers
in~\eqref{eq:PTheta-Rho-Transform} can be obtained from
$\mathbf{S}_{\Theta}$ and $\mathbf{A}_{\Theta}$ by removing index
legs. Thus, all symmetrizers (resp. antisymmetrizers)
of~\eqref{eq:PTheta-Rho-Transform} are contained in
$\mathbf{S}_{\Theta}$ (resp. $\mathbf{A}_{\Theta}$), yielding the
product $P_{\Theta} \cdot \rho_{\Theta\Phi} P_{\Phi}
  \rho_{\Phi\Theta}$ to be non-zero as claimed in~\eqref{eq:Illustrate-Rho-Ex0}.

\section{Consequences of non-Hermiticity -- an example}
\label{sec:cons-non-herm}
\ytableausetup{mathmode, boxsize=normal}

In this appendix, we illustrate the non-unitarity of transition
operators between Young projection operators as given in eq.~\eqref{eq:Herm-Conj-nHTransition-Ops},
\begin{equation}
\label{eq:Herm-Conj-nHTransition-Ops-App1}
\left(T_{\Theta\Phi}\right)^{\dagger} = Y^\dagger_\Phi \rho_{\Phi\Theta} Y^\dagger_{\Theta} \neq T_{\Phi\Theta} \ ,
\end{equation}
by means of an example. Consider the two Young tableaux
\begin{equation}
    \Theta :=
  \begin{ytableau}
    1 & 2 \\
    3
  \end{ytableau}
\qquad \text{and} \qquad \Phi :=
\begin{ytableau}
  1 & 3 \\
  2
\end{ytableau} \ .
\end{equation}
In eq.~\eqref{eq:3-Rho-ThetaThetap} we found that
$\rho_{\Theta\Phi}=\;\FPic{3s23SN}$. Using
eq.~\eqref{eq:Ytrans} we construct $T_{\Theta\Phi}$
\begin{equation}
 \underbrace{\FPic{3s23SN}}_{\rho_{\Theta\Phi}}\underbrace{\FPic{3Sym13ASym12}}_{Y_{\Phi}}
= \underbrace{\FPic{3Sym12s23ASym12}}_{T_{\Theta\Phi}}
= \underbrace{\FPic{3Sym12ASym13}}_{Y_{\Theta}}\underbrace{\FPic{3s23SN}}_{\rho_{\Theta\Phi}}
\end{equation}
 and $T_{\Phi\Theta}$
\begin{equation}
 \underbrace{\FPic{3s23SN}}_{\rho_{\Phi\Theta}}\underbrace{\FPic{3Sym12ASym13}}_{Y_{\Theta}}
= \underbrace{\FPic{3s23Sym12s23ASym12s23}}_{T_{\Phi\Theta}}
=
\underbrace{\FPic{3Sym13ASym12}}_{Y_{\Phi}}\underbrace{\FPic{3s23SN}}_{\rho_{\Phi\Theta}}
\ .
\end{equation}
From this example, it is immediately clear that
$\left(T_{\Theta\Phi}\right)^{\dagger}\neq T_{\Phi\Theta}$, 
\begin{equation}
  \Big(\underbrace{\FPic{3Sym12s23ASym12}}_{T_{\Theta\Phi}}\Big)^{\dagger}
  =
  \underbrace{\FPic{3ASym12s23Sym12}}_{\left(T_{\Theta\Phi}\right)^{\dagger}}
\neq
\underbrace{\FPic{3s23Sym12s23ASym12s23}}_{T_{\Phi\Theta}}
\end{equation}
and vice versa
\begin{equation}
  \Big(\underbrace{\FPic{3s23Sym12s23ASym12s23}}_{T_{\Phi\Theta}}\Big)^{\dagger}
=
\underbrace{\FPic{3s23ASym12s23Sym12s23}}_{\left(T_{\Phi\Theta}\right)^{\dagger}}
\neq
\underbrace{\FPic{3Sym12s23ASym12}}_{T_{\Theta\Phi}} 
\ ,
\end{equation}
confirming eq.~\eqref{eq:Herm-Conj-nHTransition-Ops-App1}.

\section{Proof of Theorem\texorpdfstring{~\ref{thm:TransitionCompact} \emph{``compact
    transition operators''}}{compact transition operators}}\label{sec:ProofsCompactTransition}

\subsection{The significance of the cutting-and-gluing procedure}\label{sec:Cut-Glue-Procedure}

Before we present the proof of Theorem~\ref{thm:TransitionCompact},
we need to make some observations: Let $\mathbf{I}$ be any set of
symmetrizers or antisymmetrizers, and let $\rho$ be a
permutation. Then, using the fact that $\rho^{\dagger}=\rho^{-1}$ for
any permutation,\footnote{This becomes evident in the birdtrack
  formalism, where the inverse of a permutation $\rho$ is obtained by
  flipping $\rho$ about its vertical axis~\cite{Cvitanovic:2008zz}, which is incidentally also
  the process for Hermitian conjugation of a birdtrack~\cite{Cvitanovic:2008zz}.} we have that
\begin{equation}
  \rho \; \mathbf{I} = \rho \; \mathbf{I} \;
  \underbrace{\rho^{\dagger}\rho}_{\mathrm{id}} = \underbrace{\rho \; \mathbf{I} \;
  \rho^{\dagger}}_{=:\mathbf{I'}} \rho = \mathbf{I'} \; \rho,
\end{equation}
where $\mathbf{I'}$ is now a set of symmetrizers, respectively
antisymmetrizers, over a different set of indices.\footnote{We consider this to be self evident, but an example may help diffuse anxiety:
\begin{equation}
  \scalebox{0.75}{$\underbrace{\FPic{5s34N}}_{\mbox{\normalsize $\rho$}}
  \underbrace{\FPic{5Sym123Sym45N}}_{\mbox{\normalsize $\mathbf{I}$}}
  \; = \; \underbrace{\FPic{5s34N}}_{\mbox{\normalsize $\rho$}}
  \underbrace{\FPic{5Sym123Sym45N}}_{\mbox{\normalsize $\mathbf{I}$}}
  \underbrace{\FPic{5s34N}\FPic{5s34N}}_{\mbox{\normalsize
      $\rho^{\dagger} \rho$}} \; = \;
  \underbrace{\FPic{5s34N}\FPic{5Sym123Sym45N}\FPic{5s34N}}_{\mbox{\normalsize
      $\mathbf{I'}$}} \underbrace{\FPic{5s34N}}_{\mbox{\normalsize $\rho$}}$},
\end{equation}
where we had $\mathbf{I}=\lbrace\bm{S}_{123},\bm{S}_{45}\rbrace$ and
$\mathbf{I'}=\lbrace\bm{S}_{124},\bm{S}_{35}\rbrace$.}  

In the proof of Theorem~\ref{thm:TransitionCompact}, we will come
across a particular such case, namely where $\rho$ is the permutation
$\rho_{\Theta\Phi}$ as defined in
Definition~\ref{thm:TableauPermutation-birdtrack}.  The simplest case we
encounter are the products
$\rho_{\Theta\Phi} \mathbf{S}_{{\color{red}\Phi}}$ and
$\rho_{\Theta\Phi} \mathbf{A}_{{\color{red}\Phi}}$. By its very
definition $\rho_{\Theta\Phi}$ explicitly relates $\Theta$ and
$\Phi$ such that
\begin{subequations}
\label{eq:TransElProof4}
\begin{align}
  \rho_{\Theta\Phi} \mathbf{S}_{{\color{red}\Phi}} 
& = \; \mathbf{S}_{{\color{red}\Theta}} \rho_{\Theta\Phi}
\; = \; \mathbf{S}_{{\color{red}\Theta}} \rho_{\Theta\Phi}
  \mathbf{S}_{{\color{red}\Phi}} 
\label{eq:TransElProof4-Sym} \\
\rho_{\Theta\Phi} \mathbf{A}_{{\color{red}\Phi}} 
& = \; \mathbf{A}_{{\color{red}\Theta}} \rho_{\Theta\Phi}
\; = \; \mathbf{A}_{{\color{red}\Theta}} \rho_{\Theta\Phi} \mathbf{A}_{{\color{red}\Phi}} 
\ , \label{eq:TransElProof4-ASym}
\end{align}
\end{subequations}
where the last equality follows from the fact that each (anti-)
symmetrizer individually is idempotent~\eqref{eq:A2AS2S}. Recognizing
the parallel between eq.~\eqref{eq:TransElProof4} and transition
operators eq.~\eqref{eq:TransitionElement} (between Hermitian
projectors, such as symmetrizers $\mathbf{S}_{\Xi}$ and
antisymmetrizers $\mathbf{A}_{\Xi}$),
the objects~\eqref{eq:TransElProof4-Sym}
and~\eqref{eq:TransElProof4-ASym} can be viewed as \emph{transition
  operators} between individual sets of (anti-) symmetrizers.  This
observation extablishes the connection to the graphical
cutting-and-gluing procedure discussed in
Theorem~\ref{thm:TransitionCompact}: cutting antisymmetrizers
$\mathbf{A}_{\Theta}$ and $\mathbf{A}_{\Phi}$ vertically and gluing
them as suggested by the Theorem is
equivalent to forming the product
$\mathbf{A}_{\Theta}\rho_{\Theta\Phi}\mathbf{A}_{\Phi}$ (and similarly
for symmetrizers). This is illustrated in the following example: For the
Young tableaux
\begin{equation}
  \Theta =
  \begin{ytableau}
    1 & 3 \\
    2 \\
    4
  \end{ytableau} \qquad \text{and} \qquad \Phi =
  \begin{ytableau}
    1 & 2 \\
    3 \\
    4
  \end{ytableau},
\end{equation}
we have
\begin{equation}
\underbrace{\FPic{4s34N}\FPic{4ASym123N}\FPic{4s34N}}_{\mathbf{A}_{\Theta}}
\underbrace{\FPic{4s23SN}}_{\rho_{\Theta\Phi}}
\underbrace{\FPic{4s234N}\FPic{4ASym123N}\FPic{4s243N}}_{\mathbf{A}_{\Phi}} 
\;  = \; \FPic{4s34N}\FPic{4ASym123N}\FPic{4ASym123N}\FPic{4s243N}
 \; = \; \FPic{4s34N}\FPic{4ASym123N}\FPic{4s243N}
\ .
\end{equation}
The feature observed in this example is fully general: $\mathbf
\rho_{\Theta\Phi}$ is \emph{defined} to translate the ordering of
the left legs on $\mathbf A_{\Phi}$ into the ordering of the right
legs on $\mathbf A_\Theta$ -- this is precisely what the cutting and
glueing procedure achieves graphically:
\begin{equation}
  \mathbf{A}_{\Theta}\rightarrow \; \FPic{4s34N}\FPic{4ASym123SplitLeftN}\;\cancel{\FPic{4ASym123SplitRightN}\FPic{4s34N}}
\quad \text{and} \quad 
\mathbf{A}_{\Phi}\rightarrow \;  
\cancel{\FPic{4s234N}\FPic{4ASym123SplitLeftN}}\;\FPic{4ASym123SplitRightN}\FPic{4s243N}
  \quad \mapsto \quad 
\FPic{4s34N}\FPic{4ASym123SplitN}\FPic{4s243N}
\ .
\end{equation}
Both procedures lead to the same result (this is a consequence of
relation~\eqref{eq:TransElProof4}). Thus, we will refer to the
algebraic construct~\eqref{eq:TransElProof4-ASym} as the
\emph{cut-antisymmetrizer}
and denote it by
\begin{equation}
  \label{eq:CompactTransProof-Cut-ASym}
\cancel{\mathbf{A}}_{\Theta\Phi} :=
\mathbf{A}_{\Theta}\rho_{\Theta\Phi}\mathbf{A}_{\Phi}
= \mathbf{A}_{\Theta}\rho_{\Theta\Phi}
= \rho_{\Theta\Phi}\mathbf{A}_{\Phi}
\ ,
\end{equation}
and similarly for the \emph{cut-symmetrizer}
$\cancel{\mathbf{S}}_{\Theta\Phi}:=\mathbf{S}_{\Theta}\rho_{\Theta\Phi}\mathbf{S}_{\Phi}$.
For the proof of Theorem~\ref{thm:TransitionCompact}, we will only
concern ourselves with cut-antizymmetrizers, as we already did in the
Theorem. However, all the following arguments hold equally well if
we consider cut-symmetrizers instead.

Before we dive into the proof, we need to notice that
eq.~\eqref{eq:TransElProof4} does not hold for the ancestor sets
$\mathbf{S}_{\Phi_{(k)}}$ and $\mathbf{A}_{\Phi_{(l)}}$ of
$\mathbf{S}_{\Phi}$ and $\mathbf{A}_{\Phi}$, however such ancestor
sets will be transformed (upon commutation with the permutation
$\rho_{\Theta\Phi}$) into sets of the same shape that can be obtained
from $\mathbf{S}_{\Theta}$ resp. $\mathbf{A}_{\Theta}$ by dropping
lines. Thus, the resulting (anti-) symmetrizers can be absorbed into
$\mathbf{S}_{\Theta}$ and $\mathbf{A}_{\Theta}$ respectively,
\begin{subequations}
  \label{eq:TransElProof4b}
\begin{align}
  \rho_{\Theta\Phi} \mathbf{S}_{{\color{red}\Phi_{(k)}}} 
& = \;
  \mathbf{S}_{{\color{red}\Theta_{(\Phi,k)}}}\rho_{\Theta\Phi} 
\quad \text{for $\mathbf{S}_{\Theta_{(\Phi,k)}}\supset\mathbf{S}_{\Theta}$} \\
 \rho_{\Theta\Phi} \mathbf{A}_{{\color{red}\Phi_{(l)}}} 
& = \; \mathbf{A}_{{\color{red}\Theta_{(\Phi,l)}}}
  \rho_{\Theta\Phi}
\quad \text{for
  $\mathbf{A}_{\Theta_{(\Phi,l)}}\supset\mathbf{A}_{\Theta}$}
\ ,
\end{align}
\end{subequations}
the (anti-) symmetrizers $\mathbf{S}_{\Theta_{(\Phi,k)}}$ and
$\mathbf{A}_{\Theta_{(\Phi,l)}}$ correspond to tableaux in the
$\Phi$-MOLD ancestry of $\Theta$, \emph{c.f.}
eq.~\eqref{eq:PhiTheta-Transform-Ancestry} in
app.~\ref{sec:Illustration-Rho}.  For further clarification, we refer
the reader to appendix~\ref{sec:Illustration-Rho} for an explicit
example.

We will now present a proof for the short-hand graphical construction of the
birdtracks of transition operators, Theorem~\ref{thm:TransitionCompact}.

\subsection{Proof of Theorem~\ref{thm:TransitionCompact}}

Let $\Theta,\Phi\in\mathcal{Y}_n$ be two Young tableaux with the
same shape, thus corresponding to equivalent irreducible
representations of $\SUN$, and let the corresponding
Hermitian Young projection operators $P_{\Theta}$ and $P_{\Phi}$ be
constructed according to the MOLD-Theorem
\ref{thm:MOLDConstruction}. Furthermore, let $\mathbf{I}$ denote either a
set of symmetrizers or antisymmetrizers, and $\mathbf{B}$ denote the
other set (that is, if $\mathbf{I}$ denotes a set of symmetrizers then
$\mathbf{B}$ denotes a set of antisymmetrizers and vice versa): we
use these  generalized sets rather than the concrete sets $\mathbf{A}$
and $\mathbf{S}$ in order to discuss all possible forms of
$P_{\Theta}$ and $P_{\Phi}$ in one go. We
then have that $\bar{P}_{\Theta}$ is given by
\begin{equation}
  \bar{P}_{\Theta} 
  = \; 
  \mathcal{C}_{\Theta} \; 
  \mathbf{I}_{\Theta}
  \mathbf{B}_{\Theta} 
  \mathbf{I}_{\Theta} \;
  \mathcal{C}_{\Theta}^{\dagger}
\ ,
\end{equation}
where $\mathcal{C}_{\Theta}$ consists of ancestor sets of (anti-)
symmetrizers of $\Theta$, and the exact structure of
$\mathcal{C}_{\Theta}$ is determined by the MOLD of $\Theta$,
$\mathcal{M}(\Theta)$, and the parity of
$\mathcal{M}(\Theta)$. Similarly, $\bar{P}_{\Phi}$ is of the form
\begin{equation}
\label{eq:CompactTransProof0}
  \bar{P}_{\Phi} 
  = \; 
  \mathcal{D}_{\Phi} \; 
  \mathbf{I}_{\Phi}
  \mathbf{B}_{\Phi} 
  \mathbf{I}_{\Phi} \;
  \mathcal{D}_{\Phi}^{\dagger} 
  \quad \text{or} \quad 
  \bar{P}_{\Phi} 
  = \; 
  \mathcal{D}_{\Phi} \; 
  \mathbf{B}_{\Phi} 
  \mathbf{I}_{\Phi} 
  \mathbf{B}_{\Phi} \;
  \mathcal{D}_{\Phi}^{\dagger}
\ ,
\end{equation}
where, like $\mathcal{C}_{\Theta}$, $\mathcal{D}_{\Phi}$ consists of
ancestor sets of (anti-) symmetrizers of $\Phi$; in
equation~\eqref{eq:CompactTransProof0}, we have taken into account
that the central part of $P_{\Phi}$ can either have the same form as
$P_{\Theta}$ (which is $\mathbf{I}\mathbf{B}\mathbf{I}$), or it may
have symmetrizers and antisymmetrizers exchanged from $P_{\Theta}$. It
should be noted that the set $\mathcal{D}_{\Phi}$ will be different
whether the central part of $\bar{P}_{\Phi}$ is
$\mathbf{I}_{\Phi}\mathbf{B}_{\Phi}\mathbf{I}_{\Phi}$ or
$\mathbf{B}_{\Phi}\mathbf{I}_{\Phi}\mathbf{B}_{\Phi}$, but in both
cases it will consist of ancestor sets of symmetrizers and
antisymmetrizers of $\Theta$. Understanding this, we have chosen not
to introduce different symbols for the set $\mathcal{D}_{\Phi}$ in
order to introduce the following compact notation for
$\bar{P}_{\Phi}$,
\begin{equation}
  \bar{P}_{\Phi} := \; \mathcal{D}_{\Phi} \; 
  \begin{Bmatrix}
    \mathbf{I}_{\Phi} \mathbf{B}_{\Phi} \mathbf{I}_{\Phi}
    \\
    \mathbf{B}_{\Phi} \mathbf{I}_{\Phi} \mathbf{B}_{\Phi}
  \end{Bmatrix}
\; \mathcal{D}_{\Phi}^{\dagger}
\ ,
\end{equation}
which says that the central part of $\bar{P}_{\Phi}$ is either given
by the top row, or by the bottom row in the curly
bracket.\footnote{This notation is convenient, as it will allow us to
  discuss both cases simultaneously.}  According to
Theorem~\ref{thm:TransitionElement}, the birdtrack of the transition
operator $T_{\Theta\Phi}$ is given by
\begin{equation}
\label{eq:CompactTransProof0b} 
  \bar{T}_{\Theta\Phi} 
  = \; 
  \underbrace{
    \mathcal{C}_{\Theta} \; 
    \mathbf{I}_{\Theta}
    \mathbf{B}_{\Theta} 
    \mathbf{I}_{\Theta} \;
    \mathcal{C}_{\Theta}^{\dagger}
  }_{=\bar{P}_{\Theta}} \; 
  \rho_{\Theta\Phi} \; 
  \underbrace{
    \mathcal{D}_{\Phi} \; 
    \begin{Bmatrix}
      \mathbf{I}_{\Phi} \mathbf{B}_{\Phi} \mathbf{I}_{\Phi}
      \\
      \mathbf{B}_{\Phi} \mathbf{I}_{\Phi} \mathbf{B}_{\Phi}
    \end{Bmatrix}
    \; \mathcal{D}_{\Phi}^{\dagger}
  }_{=\bar{P}_{\Phi}}
\ .
\end{equation}
As was discussed in sec.~\ref{sec:Cut-Glue-Procedure}, the permutation $\rho_{\Theta\Phi}$ can be
commuted with $\mathcal{D}_{\Phi}$, in accordance with
relations~\eqref{eq:TransElProof4b}. Furthermore,
equations~\eqref{eq:TransElProof4} tell us that
$\rho_{\Theta\Phi}\mathbf{I}_{\Phi}=\mathbf{I}_{\Theta}\rho_{\Theta\Phi}$
and
$\rho_{\Theta\Phi}\mathbf{B}_{\Phi}=\mathbf{B}_{\Theta}\rho_{\Theta\Phi}$
.

In commuting the $\rho_{\Theta\Phi}$ through the sets
$\mathbf{I}_{\Phi}$ and $\mathbf{B}_{\Phi}$, it will be
convenient to \emph{stop} the commutation in a different place in the
top row than the bottom row of $\bar{T}_{\Theta\Phi}$,
\begin{equation}
\label{eq:CompactTransProof-Stop-Rho}
  \bar{T}_{\Theta\Phi} = \; \mathcal{C}_{\Theta} \; \mathbf{I}_{\Theta}
  \mathbf{B}_{\Theta} \mathbf{I}_{\Theta} \;
  \mathcal{C}_{\Theta}^{\dagger} \; \mathcal{D}_{\Theta} \; 
  \begin{Bmatrix}
    \mathbf{I}_{\Theta} \mathbf{B}_{\Phi} \rho_{\Theta\Phi} \mathbf{I}_{\Phi}
    \\
    \mathbf{B}_{\Theta} \mathbf{I}_{\Theta}  \mathbf{B}_{\Theta} \rho_{\Theta\Phi}
  \end{Bmatrix}
\; \mathcal{D}_{\Phi}^{\dagger}
\ ,
\end{equation}
this choice may seem arbitrary at this point, but the position of
$\rho_{\Theta\Phi}$ in~\eqref{eq:CompactTransProof-Stop-Rho} will turn
out to specify the position of the cut in the cutting-and-gluing
procedure, \emph{c.f.} sec.~\ref{sec:Cut-Glue-Procedure}. 

We may apply the Cancellation-Theorem~\ref{thm:CancelMultipleSets} to
the operator~\eqref{eq:CompactTransProof-Stop-Rho} to simplify
$\bar{T}_{\Theta\Phi}$ as
\begin{equation}\label{CompactTransProof1}
  \bar{T}_{\Theta\Phi} \xlongequal{Thm.~\ref{thm:CancelMultipleSets}} \; \mathcal{C}_{\Theta} \; \mathbf{I}_{\Theta}
  \mathbf{B}_{\Theta} \mathbf{I}_{\Theta} \;
  \begin{Bmatrix}
    \mathbf{I}_{\Theta} \mathbf{B}_{\Theta} \rho_{\Theta\Phi} \mathbf{I}_{\Phi}
    \\
    \mathbf{B}_{\Theta} \mathbf{I}_{\Theta} \mathbf{B}_{\Theta} \rho_{\Theta\Phi}
  \end{Bmatrix}
\; \mathcal{D}_{\Phi}^{\dagger} \; = \; \mathcal{C}_{\Theta} \; 
  \begin{Bmatrix}
    \mathbf{I}_{\Theta}
  \mathbf{B}_{\Theta} \mathbf{I}_{\Theta} \; \mathbf{I}_{\Theta} \mathbf{B}_{\Theta} \rho_{\Theta\Phi} \mathbf{I}_{\Phi}
    \\
    \mathbf{I}_{\Theta}
  \mathbf{B}_{\Theta} \mathbf{I}_{\Theta} \; \mathbf{B}_{\Theta}
  \mathbf{I}_{\Theta} \mathbf{B}_{\Theta} \rho_{\Theta\Phi}
  \end{Bmatrix}
\; \mathcal{D}_{\Phi}^{\dagger}
\ .
\end{equation}
Let us now look at the central part of
$\bar{T}_{\Theta\Phi}$ (the part in the curly brackets)
in more detail: Since $\mathbf{I}_{\Theta}$ denotes either
$\mathbf{A}_{\Theta}$ or $\mathbf{S}_{\Theta}$, and
$\mathbf{B}_{\Theta}$ denotes the other set, then the product
$\mathbf{I}_{\Theta}\mathbf{B}_{\Theta}$ is proportional to either a
Young projection operator or the Hermitian conjugate thereof,
$\mathbf{I}_{\Theta}\mathbf{B}_{\Theta}=\bar{Y}_{\Theta}^{(\dagger)}$. Thus,
if the central part of $\bar{T}_{\Theta\Phi}$ is given by
the top option (implementing that
$\mathbf{I}_{\Theta}\mathbf{I}_{\Theta}=\mathbf{I}_{\Theta}$), we
can use the fact that $\bar{Y}_{\Theta}^{(\dagger)}$ is
quasi-idempotent to obtain
\begin{equation}
  \underbrace{
    \mathbf{I}_{\Theta}
    \mathbf{B}_{\Theta}
  }_{\bar{Y}_{\Theta}^{(\dagger)}}
  \underbrace{
    \mathbf{I}_{\Theta}
    \mathbf{B}_{\Theta}
  }_{\bar{Y}_{\Theta}^{(\dagger)}}
  \rho_{\Theta\Phi} 
  \mathbf{I}_{\Phi} 
  \; \propto \; 
  \mathbf{I}_{\Theta}
  \mathbf{B}_{\Theta} 
  \rho_{\Theta\Phi} 
  \mathbf{I}_{\Phi}
  \ .
\end{equation}
Similarly, if the central part of $\bar{T}_{\Theta\Phi}$
is given by the bottom option of~\eqref{CompactTransProof1}, we may
reduce it to
\begin{equation}
  \underbrace{\mathbf{I}_{\Theta}
    \mathbf{B}_{\Theta}}_{\bar{Y}_{\Theta}^{(\dagger)}}
  \underbrace{\mathbf{I}_{\Theta}
    \mathbf{B}_{\Theta}}_{\bar{Y}_{\Theta}^{(\dagger)}} \underbrace{\mathbf{I}_{\Theta}
    \mathbf{B}_{\Theta}}_{\bar{Y}_{\Theta}^{(\dagger)}}
  \rho_{\Theta\Phi} \; \propto \; \mathbf{I}_{\Theta}
    \mathbf{B}_{\Theta} \rho_{\Theta\Phi}
    \ .
\end{equation}
This turns~\eqref{CompactTransProof1} into (using the bar-notation
introduced in eq.~\eqref{eq:Bar-Notation-Benefit} to retain equality)
\begin{equation}\label{CompactTransProof2}
  \bar{T}_{\Theta\Phi} = \; \mathcal{C}_{\Theta} \; 
  \begin{Bmatrix}
    \mathbf{I}_{\Theta} \mathbf{B}_{\Theta} \rho_{\Theta\Phi} \mathbf{I}_{\Phi}
    \\
    \mathbf{I}_{\Theta} \mathbf{B}_{\Theta} \rho_{\Theta\Phi}
  \end{Bmatrix}
\; \mathcal{D}_{\Phi}^{\dagger}
\ .
\end{equation}

In Theorem~\ref{thm:TransitionCompact}, we discussed three different
cutting-and-gluing procedures, depending on the exact structure of the
projection operators $P_{\Theta}$ and $P_{\Phi}$.
\begin{enumerate}
\item Option~\ref{itm:CompactTrans1} requires both operators
  $\bar{P}_{\Theta}$ and $\bar{P}_{\Phi}$ to contain exactly one
  set of antisymmetrizers $\mathbf{A}_{\Theta}$ and
  $\mathbf{A}_{\Phi}$ respectively. This occurs if we choose the
  top option of $\bar{T}_{\Theta\Phi}$ as given
  in~\eqref{eq:CompactTransProof0b} (and hence the top line
  in~\eqref{CompactTransProof2}) and if $\mathbf{B}$ denotes the set
  of antisymmetrizers and thus $\mathbf{I}_{\Theta}$ denotes the set
  of symmetrizers,
\begin{equation}
\label{eq:CompactTransProof-Option1}
  \eqref{CompactTransProof2}: \qquad \bar{T}_{\Theta\Phi} 
= \; \mathcal{C}_{\Theta} \; 
    \mathbf{I}_{\Theta} \mathbf{B}_{\Theta} \rho_{\Theta\Phi} \mathbf{I}_{\Phi}
\; \mathcal{D}_{\Phi}^{\dagger} \;
\xrightarrow{\mathbf{B}=\mathbf{A}, \;\mathbf{I}=\mathbf{S}} 
\; \mathcal{C}_{\Theta} \; 
    \mathbf{S}_{\Theta} 
\underbrace{
\fcolorbox{red}{white}{$\mathbf{A}_{\Theta}\rho_{\Theta\Phi}$}
}_{=\cancel{\mathbf{A}}_{\Theta\Phi}}
 \mathbf{S}_{\Phi}
\; \mathcal{D}_{\Phi}^{\dagger}
\ ,
\end{equation}
where we marked the cut-antisymmetrizer
$\cancel{\mathbf{A}}_{\Theta\Phi}$ (see
eq.~\eqref{eq:CompactTransProof-Cut-ASym}) in the
above. Clearly,~\eqref{eq:CompactTransProof-Option1} coincides with
the cutting-and-gluing prescription of
Theorem~\ref{thm:TransitionCompact} if each projector $P_{\Theta}$ and
$P_{\Phi}$ contains exactly one set $\mathbf{A}_{\Theta}$ and $
\mathbf{A}_{\Phi}$ respectively.

\item Option~\ref{itm:CompactTrans2} of Theorem~\ref{thm:TransitionCompact}
  requires $\bar{P}_{\Theta}$ and $\bar{P}_{\Phi}$ to have a
  different number of $\mathbf{A}_{\Theta}$ and
  $\mathbf{A}_{\Phi}$. The bottom option of
  operator~\eqref{eq:CompactTransProof0b} (and hence
  operator~\eqref{CompactTransProof2}) corresponds to this case, and
  it does not matter whether $\mathbf{B}$ denotes the set of
  antisymmetrizers and $\mathbf{I}$ the set of symmetrizers or the
  other way around: If $\mathbf{B}$ denotes the set of
  antisymmetrizers, we have
\begin{subequations}
\label{eq:CompactTransProof-Option2}
\begin{equation}
\label{eq:CompactTransProof-Option2a}
\eqref{CompactTransProof2}: \qquad \bar{T}_{\Theta\Phi} 
= \; \mathcal{C}_{\Theta} \; 
    \mathbf{I}_{\Theta} 
    \mathbf{B}_{\Theta} \rho_{\Theta\Phi} 
    \; \mathcal{D}_{\Phi}^{\dagger} \;
\xrightarrow{\mathbf{B}=\mathbf{A},
        \;\mathbf{I}=\mathbf{S}} \; 
    \mathcal{C}_{\Theta} \;
    \mathbf{S}_{\Theta} 
    \underbrace{
\fcolorbox{red}{white}{$\mathbf{A}_{\Theta}\rho_{\Theta\Phi}$}
}_{=\cancel{\mathbf{A}}_{\Theta\Phi}}
    \mathcal{D}_{\Phi}^{\dagger}
    \ .
\end{equation}
The operator~\eqref{eq:CompactTransProof-Option2a}
is the same operator that would have resulted from cutting
$\bar{P}_{\Theta}$ at its left-most set $\mathbf{A}_{\Theta}$ and
$\bar{P}_{\Phi}$ at its right-most set $\mathbf{A}_{\Phi}$, and
gluing the pieces in the appropriate manner as described by the
Theorem~\ref{thm:TransitionCompact}.

Similarly, if $\mathbf{I}$ denotes
the set of antisymmetrizers, then
\begin{equation}
\bar{T}_{\Theta\Phi} 
\xrightarrow{\mathbf{I}=\mathbf{A},\;\mathbf{B}=\mathbf{S}} \; 
    \mathcal{C}_{\Theta} \; 
    \mathbf{A}_{\Theta}
    \mathbf{S}_{\Theta} 
    \rho_{\Theta\Phi} \;
    \mathcal{D}_{\Phi}^{\dagger} 
\;  \xlongequal{\text{eq.~\eqref{eq:TransElProof4-Sym}}} \; 
    \mathcal{C}_{\Theta}
\underbrace{
\fcolorbox{red}{white}{$\mathbf{A}_{\Theta}\rho_{\Theta\Phi}$}
}_{=\cancel{\mathbf{A}}_{\Theta\Phi}}
    \mathbf{S}_{\Phi} 
    \mathcal{D}_{\Phi}^{\dagger} 
\ ,
\end{equation}
\end{subequations}
where we used the commutation relation~\eqref{eq:TransElProof4-Sym} to
commute $\mathbf{S}_{\Theta}$ and $\rho_{\Theta\Phi}$. This again yields the same result as the cutting-and-gluing procedure of
Theorem~\ref{thm:TransitionCompact}.

\item Lastly, suppose that both $\bar{P}_{\Theta}$ and $\bar{P}_{\Phi}$ each
contain two sets of antisymmetrizers $\mathbf{A}_{\Theta}$ and 
$\mathbf{A}_{\Phi}$ respectively. Then, we once again need to
look at the top option of the operator
$\bar{T}_{\Theta\Phi}$ as given
in~\eqref{eq:CompactTransProof0b} (and hence~\eqref{CompactTransProof2}), but this time we require that $\mathbf{I}$
denotes the set of antisymmetrizers. Then,
\begin{subequations}
\label{eq:CompactTransProof-Option3}
\begin{equation}
\label{eq:CompactTransProof-Option3a}
    \eqref{CompactTransProof2}: \qquad 
\bar{T}_{\Theta\Phi} 
= \; \mathcal{C}_{\Theta} \; 
    \mathbf{I}_{\Theta} \mathbf{B}_{\Theta} \rho_{\Theta\Phi} \mathbf{I}_{\Phi}
\; \mathcal{D}_{\Phi}^{\dagger} \;
  \xrightarrow{\mathbf{I}=\mathbf{A}, \;\mathbf{B}=\mathbf{S}} \; \mathcal{C}_{\Theta} \; 
  \mathbf{A}_{\Theta} \mathbf{S}_{\Theta}
\underbrace{
\fcolorbox{red}{white}{$\rho_{\Theta\Phi} \mathbf{A}_{\Phi}$}
}_{=\cancel{\mathbf{A}}_{\Theta\Phi}}
  \; \mathcal{D}_{\Phi}^{\dagger}
\ .
\end{equation}
Equivalently,
\begin{equation}
  \label{eq:CompactTransProof-Option3b}
\Bar{T}_{\Theta\Phi}
 \; = \;    
\mathcal{C}_{\Theta} \; 
    \mathbf{A}_{\Theta}
    \mathbf{S}_{\Theta} 
    \rho_{\Theta\Phi} \;
    \mathbf{A}_{\Phi}
    \mathcal{D}_{\Phi}^{\dagger} 
\;  \xlongequal{\text{eq.~\eqref{eq:TransElProof4-Sym}}} \; 
    \mathcal{C}_{\Theta}
\underbrace{
\fcolorbox{red}{white}{$\mathbf{A}_{\Theta}\rho_{\Theta\Phi}$}
}_{=\cancel{\mathbf{A}}_{\Theta\Phi}}
    \mathbf{S}_{\Phi} 
    \mathbf{A}_{\Phi}
    \mathcal{D}_{\Phi}^{\dagger} 
\ ;
\end{equation}
\end{subequations}
eq.~\eqref{eq:CompactTransProof-Option3a} corresponds to
cutting-and-gluing at the right-most sets of antisymmetrizers
$\mathbf{A}_{\Theta}$ and $\mathbf{A}_{\Phi}$ (respectively) in
both $\bar{P}_{\Theta}$ and $\bar{P}_{\Phi}$, while
eq.~\eqref{eq:CompactTransProof-Option3b} corresponds to
cutting-and-gluing the left-most sets of antisymmetrizers
$\mathbf{A}_{\Theta}$ and $\mathbf{A}_{\Phi}$ in both
$\bar{P}_{\Theta}$ and $\bar{P}_{\Phi}$.
\end{enumerate}

Thus, we have shown that $\bar{T}_{\Theta\Phi}$ can indeed be
obtained by the graphical cutting-and-gluing prescription given in the
Theorem \ref{thm:TransitionCompact}, concluding the proof. \qed

 \bibliographystyle{utphys}
 \bibliography{PaperLibrary,BookLibrary,GroupTheory}

\end{document}